\documentclass[twocolumn]{aastex63}

\usepackage{rotating}
\usepackage{acronym}
\usepackage{xspace}
\usepackage{multirow}
\usepackage{mathtools}
\usepackage{amsmath}
\usepackage{hyperref}
\usepackage{units}

\usepackage{xcolor}

\graphicspath{{./}{figures/}}

\definecolor{chmagenta}{rgb}{0.54, 0.17, 0.88}

\def\q{\ensuremath{q}\xspace}
\def\z{\ensuremath{z}\xspace}

\def\betac{\ensuremath{\beta_\mathrm{c}}\xspace}
\def\Nobs{\ensuremath{N_\mathrm{obs}}\xspace}
\def\Necc{\ensuremath{N_\mathrm{ecc}}\xspace}
\def\pdet{\ensuremath{p_\mathrm{det}}\xspace}
\def\e{\ensuremath{e_{10\,\mathrm{Hz}}\xspace}}

\def\Msun{\ensuremath{\mathit{M_\odot}}\xspace}
\def\Rsun{\ensuremath{\mathit{R_\odot}}\xspace}

\def\ClustersOneEccentricGWTC{\ensuremath{14\%}\xspace}

\def\eccNinety{\ensuremath{0.001}\xspace}
\def\eccNinetyNine{\ensuremath{0.05}\xspace}
\def\TmergeNinety{\ensuremath{32\,\mathrm{yr}}\xspace}
\def\TmergeNinetyNine{\ensuremath{6\,\mathrm{days}}\xspace}

\def\RecoveredPerc{\ensuremath{56\%}\xspace}
\def\EccDetPerc{\ensuremath{4\%}\xspace}
\def\EccDetFrac{\ensuremath{3.9 \times 10^{-2}}\xspace}
\def\EccPerc{\ensuremath{7\%}\xspace}

\def\EccFracZeroPOneDifference{\ensuremath{0.02}\xspace}

\def\NeccZeroPerctileBetaOne{\ensuremath{83^\mathrm{rd}}\xspace}

\def\NeccZeroBetaMaxOneHundredObs{\ensuremath{0.77}\xspace}
\def\NeccZeroBetaMaxThreeHundredObs{\ensuremath{0.26}\xspace}
\def\NobsClustersNotMajority{\ensuremath{150}\xspace}

\def\NeccOneBetaNinetyFive{\ensuremath{0.14}\xspace}
\def\NeccOneBetaNinetyFivePerc{\ensuremath{14\%}\xspace}
\def\NeccTwoBetaNinetyFive{\ensuremath{0.27}\xspace}

\def\NeccExcludedNinetyFive{\ensuremath{4}\xspace}
\def\NeccExcludedNinetyFiveVolume{\ensuremath{96\%}\xspace}

\def\BetaSpreadNeccOne{\ensuremath{1.36}\xspace}
\def\BetaSpreadNeccThree{\ensuremath{0.90}\xspace}
\def\HalfDexPrecision{\ensuremath{10}\xspace}

\def\GCEccRateLow{\ensuremath{0.08}\xspace}
\def\GCEccRateHigh{\ensuremath{0.8}\xspace}

\def\MtotLowModels{\ensuremath{14}\xspace}
\def\MtotHighModels{\ensuremath{135}\xspace}
\def\qLowModels{\ensuremath{0.3}\xspace}
\def\qHighModels{\ensuremath{1.0}\xspace}

\def\ChiEffLowModels{\ensuremath{-0.42}\xspace}
\def\ChiEffHighModels{\ensuremath{0.42}\xspace}

\acrodef{GW}{gravitational-wave}
\acrodef{BH}{black hole}
\acrodef{BBH}{binary black hole}
\acrodef{NS}{black hole}
\acrodef{BNS}{binary neutron star}

\acrodef{LIGO}{Laser Interferometer Gravitational-wave Observatory}
\acrodef{LVC}{LIGO Scientific Collaboration and Virgo Collaboration}
\acrodef{O1}{first observing run}
\acrodef{O2}{second observing run}
\acrodef{O3}{third observing run}
\acrodef{O3a}{first half of the third observing run}

\acrodef{SNR}{signal-to-noise ratio}
\acrodef{PSD}{power spectral density}
\acrodef{NR}{numerical relativity}

\acrodef{GC}{globular cluster}
\acrodef{AGN}{active galactic nucleus}

\acrodef{CDF}{cumulative distribution function}

\newcommand{\changed}[1]{#1}

\newcommand{\KICP}{\affiliation{Kavli Institute for Cosmological Physics, The University of Chicago, 5640 South Ellis Avenue, Chicago, IL 60637, USA}}
\newcommand{\EFI}{\affiliation{Enrico Fermi Institute, The University of Chicago, 933 East 56th Street, Chicago, IL 60637, USA}}
\newcommand{\SPA}{\affiliation{School of Physics and Astronomy, Monash University, Vic 3800, Australia}}
\newcommand{\OzGravMonash}{\affiliation{OzGrav: The ARC Centre of Excellence for Gravitational Wave Discovery, Clayton, VIC 3800, Australia}}

\shorttitle{Eccentric Selection}
\shortauthors{Zevin et al. 2021}

\begin{document}

\title{Implications of Eccentric Observations on Binary Black Hole Formation Channels}

\author[0000-0002-0147-0835]{Michael\,Zevin}\email{michaelzevin@uchicago.edu}\thanks{NASA Hubble Fellow}
\KICP \EFI
\author[0000-0002-4181-8090]{Isobel M. Romero-Shaw}
\SPA\OzGravMonash
\author[0000-0002-4086-3180]{Kyle Kremer}
\affil{TAPIR, California Institute of Technology, Pasadena, CA 91125, USA}
\affil{The Observatories of the Carnegie Institution for Science, Pasadena, CA 91101, USA}
\author[0000-0002-4418-3895]{Eric Thrane}
\SPA\OzGravMonash
\author[0000-0003-3763-1386]{Paul D. Lasky}
\SPA\OzGravMonash

\begin{abstract}
Orbital eccentricity is one of the most robust discriminators for distinguishing between dynamical and isolated formation scenarios of binary black hole mergers using gravitational-wave observatories such as LIGO and Virgo. 
Using state-of-the-art cluster models, we show how selection effects impact the detectable distribution of eccentric mergers from clusters. 
We show that the observation (or lack thereof) of eccentric binary black hole mergers can significantly constrain the fraction of detectable systems that originate from dynamical environments, such as dense star clusters. 
After roughly \NobsClustersNotMajority observations, observing no eccentric binary signals would indicate that clusters cannot make up the majority of the merging binary black hole population in the local universe (95\% credibility). 
However, if dense star clusters dominate the rate of eccentric mergers and a single system is confirmed to be measurably eccentric in the first and second gravitational-wave transient catalogs, clusters must account for at least \ClustersOneEccentricGWTC of detectable binary black hole mergers. 
The constraints on the fraction of detectable systems from dense star clusters become significantly tighter as the number of eccentric observations grows and will be constrained to within 0.5 dex once \HalfDexPrecision eccentric binary black holes are observed. 
\end{abstract}


\section{Introduction}\label{sec:intro}

In the past few years, the dramatic increase in compact binary mergers observed by \ac{GW} detectors has fueled immense interest and debate regarding compact binary formation pathways, particularly for \ac{BBH} systems. 
Over a dozen potential formation scenarios for the \ac{BBH} mergers observed by the LIGO--Virgo detector network~\citep{aLIGO,aVirgo} have been proposed, including isolated massive-star binary progenitors~\citep[e.g.,][]{Bethe1998,Dominik2012,Belczynski2016,Bavera2021}, assembly in dynamical environments~\citep[e.g.,][]{PortegiesZwart2000,OLeary2006,Downing2010,Rodriguez2016a,Banerjee2017,DiCarlo2019}, gas-driven assembly and orbital evolution~\citep[e.g.,][]{Mckernan2014,Bartos2017,Stone2017}, and primordial origins~\citep[e.g.,][]{Bird2016,Sasaki2018,Clesse2020,Franciolini2021}. 
Given the heterogeny of the compact binary coalescences observed to date, a mix of formation channels is currently preferred over a single channel dominating the formation of merging \ac{BBH} systems in the universe~\citep{GWTC2_pops,Bouffanais2021,Wong2021,Zevin2021}. 

Though population-based studies offer insights into the broad features of \ac{BBH} formation, the most efficient means of constraining formation scenarios is to identify features of \ac{BBH} systems unique to particular channels.
One such key feature is orbital eccentricity. 
Compact binary systems that inspiral over long timescales efficiently damp orbital eccentricity through angular momentum loss from \ac{GW} emission~\citep{Peters1964}. 
Therefore, even field binaries born with high eccentricity are nearly circular by the time they enter the LIGO--Virgo sensitive frequency band.
For example, a \ac{BBH} composed of two 20\,\Msun \acp{BH} at an initial orbital separation of 1\,\Rsun and initial eccentricity of 0.9 (0.99) will have an eccentricity at a \ac{GW} frequency of 10 Hz of \eccNinety (\eccNinetyNine) and merge in \TmergeNinety (\TmergeNinetyNine). 
The only means of producing measurably eccentric \ac{BBH} mergers in the LIGO--Virgo band that does not require a high degree of fine-tuning is through strong gravitational encounters in dynamical environments~\citep[e.g.,][]{OLeary2009,Kocsis2012,Samsing2014,Samsing2018c,Samsing2018d,Gondan2018,Rodriguez2018b,Takatsy2019,Zevin2019a,Rasskazov2019,Grobner2020,Samsing2020a,Gondan2021,Tagawa2021a} or through channels that can pump eccentricity into inspiraling binaries, such as the secular evolution of hierarchical systems~\citep[e.g.,][]{Antonini2012,Antognini2014,Silsbee2017,Antonini2017a,Rodriguez2018,Fragione2019a,Fragione2019,Liu2019a,Liu2019} \changed{or through perturbations of wide triples in the Galactic field from flyby encounters~\citep[e.g.][]{Michaely2019,Michaely2020}. }

Orbital eccentricity is arguably the most robust discriminator for distinguishing between isolated and dynamical \ac{BBH} formation scenarios; so long as \acp{BH} form in clusters and are not kicked out of clusters at formation, highly eccentric mergers are an inevitable by-product of two-body relaxation and small-$N$ dynamics~\citep{Samsing2014,Rodriguez2018b}.  
In classical \acp{GC}, which are perhaps the best-studied environment of eccentric mergers from strong gravitational encounters, the presence of \acp{BH} is evidenced both observationally \citep[through detection of BH binary candidates in several Milky Way \acp{GC};][]{Strader2012,Giesers2018,Giesers2019} and computationally through $N$-body modeling~\citep{Mackey2007,Breen2013,Wang2016,Kremer2018,Kremer2019c,ArcaSedda2018a,Antonini2020,Weatherford2020,Kremer2020}. 
Eccentric mergers from strong gravitational encounters have been shown both semianalytically and numerically to account for $\approx\,10\%$ of the underlying population of \ac{BBH} mergers in \acp{GC}, with approximately half of these having eccentricities $\geq 0.1$ at a \ac{GW} frequency of $\unit[10]{Hz}$, the lower edge of the LIGO--Virgo band~\citep{Samsing2017c,Samsing2018c,Samsing2018d,Rodriguez2018b,Zevin2019a,Kremer2020,Antonini2020a}. 

Despite the robust theoretical predictions for generating eccentric \ac{BBH} mergers in clusters, the implications that such a detection would have on constraining \ac{BBH} formation scenarios depends sensitively on the interplay between the measurability and detectability of such systems. 
Eccentricity acts as a double-edged sword: larger eccentricities are easier to distinguish from their circular counterparts using parameter estimation, though there is an inherent selection bias impinging on the detection of eccentric sources due to the template banks used for matched-filter searches for \ac{GW} signals, which typically assume quasi-circular aligned-spin binaries~\citep{Hooper2012,Allen2012,DalCanton2014,Usman2016,Adams2016,Messick2017,Nitz2017,Chu2020,Davies2020,Aubin2021}. 
Though burst searches may be used to detect unmodeled sources, such as highly eccentric mergers~\citep{Tiwari2016,Abbott2019a,Ramos-Buades2020}, they are less capable of digging deep into the noise for signals and their sensitivity is more difficult to quantify~\citep{Klimenko2016}. 
In terms of measurability, \ac{BBH} systems with properties similar to GW150914~\citep{GW150914} can be distinguished as eccentric if their eccentricities are $\gtrsim 0.05$ at a \ac{GW} frequency of 10 Hz~\citep{Lower2018}, consistent with the upper limits on orbital eccentricity~\citep{Romero-Shaw2019} for the events in the first LIGO--Virgo \ac{GW} transient catalog~\citep[GWTC-1;][]{GWTC1}.\footnote{There are claims that eccentric \ac{BBH} mergers have been observed in the most recent observing run~\citep[][]{Romero-Shaw2020b,Gayathri2020,Romero-Shaw2021}, although the interpretation for certain systems, such as GW190521, is speculative. 
We discuss this further in Sections \ref{sec:implications} and \ref{sec:conclusions}. }
However, even if an eccentric \ac{GW} source were detected, there is currently no published selection function for eccentric sources to translate this detection into characteristics of the source population, such as the merger rate. 
This is in part because of the difficulty in modeling eccentric signals~\citep{Loutrel2020} and the computational burden of adding an extra dimension to matched-filter template banks. 

In this Letter, we quantify how the detection of an eccentric \ac{BBH} merger (or lack thereof) impacts the inferred fraction of \ac{BBH} mergers originating from dense stellar clusters\footnote{We use the term ``dense star clusters'' to denote clusters with masses in the range of $\approx\,10^5-10^7\,\Msun$ and virial radii of $\approx0.5-5\,$pc. Critically, however, we do not limit this definition to only old low-metallicity clusters that survive to the present day, like traditional \acp{GC}. We also incorporate high-metallicity clusters born relatively recently \citep[essential for producing lower-mass BH mergers;][]{Chatterjee2017} as well as disrupted clusters that do not survive to the present day~\citep[inclusion of these disrupted clusters may increases the total \ac{BBH} merger rate from clusters by a factor of $\approx\,2$; e.g.,][]{Rodriguez2018a,Fragione2018}.} by approximating selection effects that account for eccentricity in a realistic cluster population. 
With selection effects and measurability of eccentric sources accounted for, \changed{and assuming that dense star clusters dominate the rate of measurably eccentric \ac{BBH} mergers,} the robustness of eccentricity predictions from cluster modeling allows for constraints to be placed on the relative contribution of \ac{BBH} mergers from clusters as a whole. 
In Section \ref{sec:models} we describe the cluster models used as a basis in this analysis. 
Section \ref{sec:detectability_measurability} covers the determination of selection effects for generically eccentric systems and our assumptions regarding the ability of eccentric sources to be distinguished from circular. 
Our main results are presented in Section \ref{sec:implications}, which quantifies the constraints that a bona fide eccentric detection will have on the contribution of dynamical channels to the underlying population of \ac{BBH} systems in the context of both the current catalogs of \acp{GW} and future observations. 
We summarize our results and discuss the caveats of our analysis in Section \ref{sec:conclusions}. 
We use a flat $\Lambda$CDM cosmology with Planck 2015 cosmological parameters~\citep{PlanckCollaboration2016} throughout this work.

\section{Cluster Models}\label{sec:models}


\begin{figure*}[t]
\centering
\includegraphics[width=0.95\textwidth]{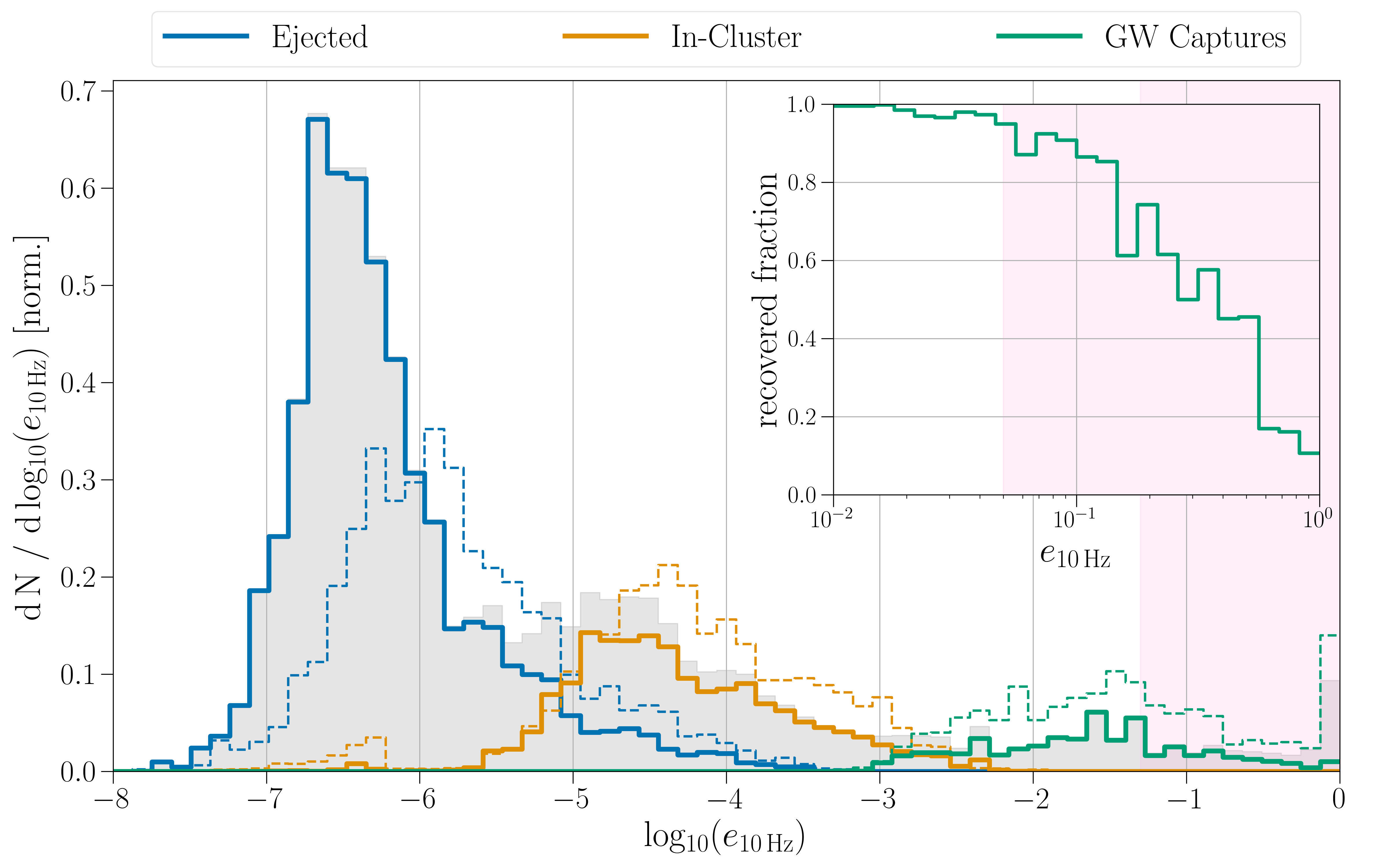}
\caption{Eccentricity distributions for detectable \ac{BBH} mergers assuming perfectly matching templates (gray shading), circular templates (solid lines), and neglecting detection probabilities \pdet altogether (dashed lines).
Colored lines denote whether the \ac{BBH} system was ejected from the cluster (blue), merged inside the cluster between strong dynamical interactions (orange), or merged as a capture during a strong gravitational encounter (green). 
The difference between the background gray histogram and the solid-line histogram for the \ac{GW} capture population shows the systems that are ``missed'' by searching with circular templates; this recovered fraction is also shown in the inset panel as a function of eccentricity. 
The detectable distribution assuming circular templates is normalized to the detectable distribution assuming perfectly matching templates to better visualize this difference; thus, the solid-line histograms integrate to slightly less than unity. 
The pink shaded region marks systems with $\e > 0.05$, the approximate eccentricity requirement for distinguishing GW150914-like systems from circular. 
Ejected mergers are more prevalent in the detectable distribution because more massive systems are ejected earlier in the history of the cluster and have longer inspiral timescales so that they can readily merge in the local universe. 
}
\label{fig:GC_pop}
\end{figure*}

We use dense star cluster models from the \texttt{CMC Cluster Catalog}~\citep{Kremer2020} to construct our population of dynamical \ac{BBH} mergers. 
Collectively, these models span roughly the full parameter space of the Milky Way \acp{GC} (in cluster mass, core/half-light radius, metallicity, and position within the Galactic potential) and include state-of-the-art prescriptions for stellar evolution and \ac{BH} formation \citep[see][for more details]{Kremer2020}. 
Importantly, strong binary-mediated gravitational encounters are modeled with direct integration using \texttt{Fewbody}~\citep{Fregeau2004,Fregeau2007} with updates to include post-Newtonian terms for dynamical encounters involving \acp{BH}~\citep{Rodriguez2018c,Rodriguez2018b}.

The gravitational radiation reaction, which enters at the 2.5 post-Newtonian order, is crucial for the formation of high-eccentricity \ac{BBH} mergers in clusters~\citep{Samsing2014,Samsing2017c,Rodriguez2018b,Zevin2019a}. 
During strong gravitational encounters between single and/or binary \ac{BH} systems, component \acp{BH} can undergo dozens of partner swaps, forming hardened temporary binaries that have eccentricities $e$ drawn from a ``quasi-thermal'' distribution (the probability density is proportional to $e$). 
If this intermediate-state binary has a large enough orbital eccentricity, the efficient loss of orbital energy from \acp{GW} during close periapse passages leads to a rapid merger on a timescale of days with significant eccentricity in the sensitive frequency ranges of ground-based \ac{GW} detectors. 
In addition, classically unbound \acp{BH} can lose enough orbital energy from close passages during these encounters to become bound and rapidly merge. 
This can also occur between two single \acp{BH} in the cluster if the impact parameter is small enough, though this subchannel produces \ac{BBH} mergers with eccentricities more accessible in the decihertz regime~\citep{Samsing2020b}. 

These merger channels from strong gravitational encounters are collectively referred to as \ac{GW} captures and have significantly larger eccentricities compared to \ac{BBH} mergers that were ejected from the cluster due to a prior dynamical interaction or \acp{BBH} that merge between strong gravitational encounters~\citep{Samsing2018d,Rodriguez2018c,Zevin2019a}. 
Figure \ref{fig:GC_pop} shows normalized eccentricity distributions \changed{for \acp{BBH} in our astrophysically weighted cluster models} at a reference \ac{GW} frequency of 10\,Hz, \e, both with and without the inclusion of selection effects (described in detail in the next section). 
\changed{The local \ac{BBH} merger rate from all subchannels in this model is $20\,\mathrm{Gpc}^{-3}\,\mathrm{yr}^{-1}$. }
Further details regarding the cluster population and determination of \ac{BBH} eccentricities can be found in Appendix~\ref{app:cluster_models}.

\section{Selection Effects and Measurability of Eccentric Sources}\label{sec:detectability_measurability}

Due to the lack of large-scale injection campaigns that have the ability to determine a sensitive space-time volume for eccentric sources, we instead estimate detection probabilities using a fixed \ac{SNR} threshold required for detection. 
For all \ac{SNR} calculations, we use the waveform approximant \texttt{TEOBResumS}~\citep{Damour2014,Nagar2018}, which is a time-domain effective one-body approximant~\citep{Buonanno1999,Buonanno2000,Damour2000,Damour2001,Damour2015} that can account for orbital eccentricity in the inspiral. 
\texttt{TEOBResumS} agrees well with \ac{NR} for eccentricities of $e \lesssim 0.3$ at apastron frequency $\omega^\mathrm{EOB}_a \sim 0.03$~\citep[which corresponds to a \ac{GW} frequency of $\approx 10\,\mathrm{Hz}$ for a \ac{BBH} with a total mass of $60\,\Msun$;][]{Chiaramello2020,Albanesi2021,Nagar2021}, and, though untested for larger eccentricities due to a lack of available \ac{NR} waveforms for comparison, it is physically valid even at high eccentricities of $\e \simeq 0.9$.\footnote{\changed{Different waveform models use different definitions of eccentricity. When comparing predictions made with \texttt{TEOBResumS} to detections made with other approximants, care should be taken to ensure that the definition of eccentricity is consistent between models. One element of an eccentric waveform model that can be varied is the point at which the reference frequency is defined, since this varies from apoastron to periastron. By default, \texttt{TEOBResumS} defines the eccentricity at a reference frequency equivalent to that of a circular Keplerian orbit with radius equal to the semimajor axis of the eccentric binary. At low frequencies and eccentricities, this is similar to the eccentricity-dependent peak frequency from ~\cite{Wen2003} used in the \texttt{CMC Cluster Catalog}: with eccentricity defined at a Keplerian reference frequency of \unit[10]{Hz}, the difference between the two frequencies is $\lesssim \unit[3]{Hz}$ ($\unit[0.6]{Hz}$) for $e_{10}\lesssim0.2$ ($0.05$).}}


\begin{figure}[t]
\includegraphics[width=0.47\textwidth]{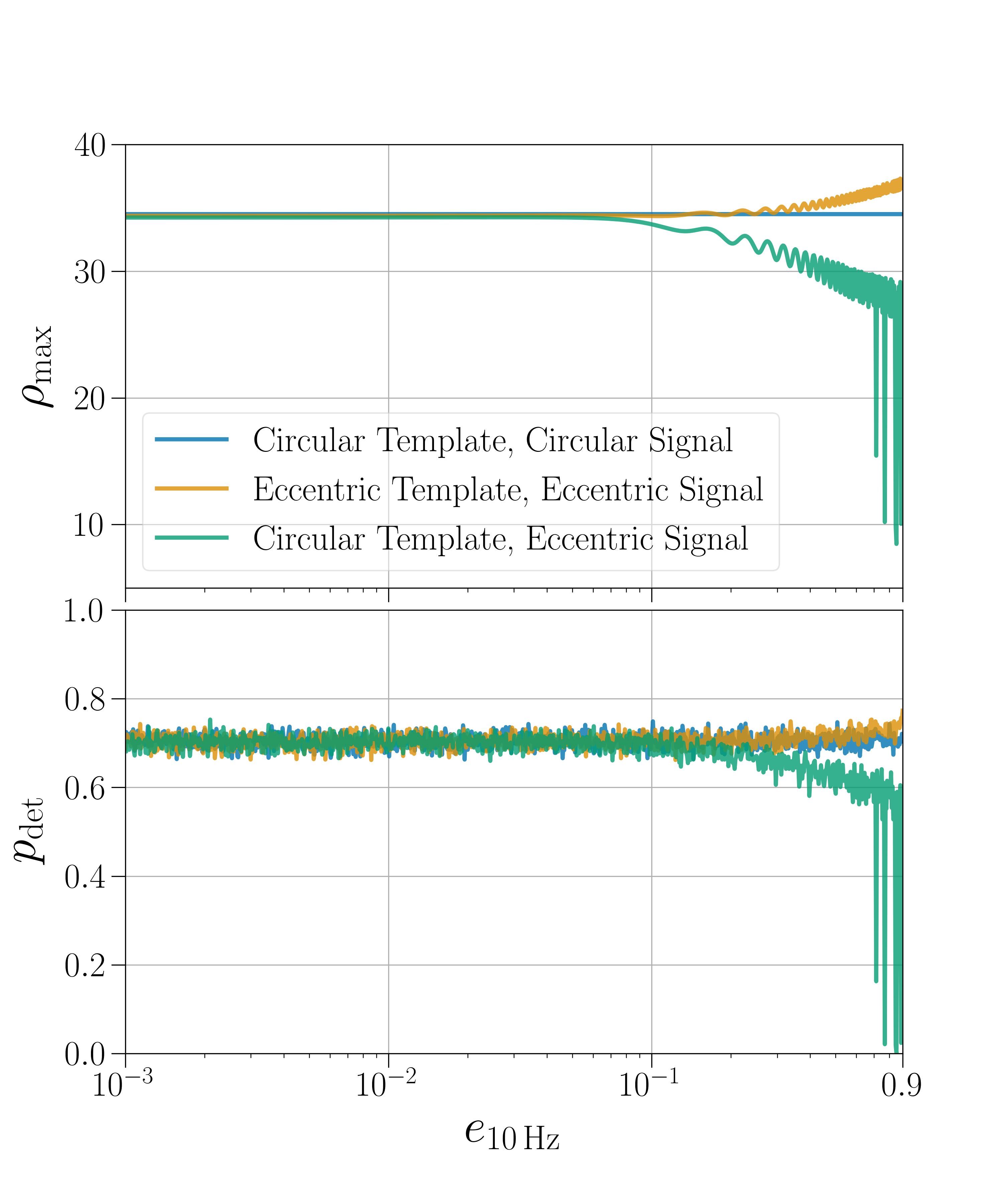}
\caption{Time- and phase-maximized matched-filter \ac{SNR} $\rho_\mathrm{max}$ (top panel) and detection probability \pdet (bottom panel) over a range of \e. 
For this demonstration, we use a fiducial zero-spin system with mass $m_1 = 36 \Msun$ and $m_2 = 29 \Msun$, a luminosity distance $d_\mathrm{L} = 1$\,Gpc, and a \texttt{midhighlatelow} \ac{PSD} (power spectral density) noise curve. 
The three colored curves represent the three signal--template combinations described in Section~\ref{sec:detectability_measurability}: a circular signal with a circular template (blue), an eccentric signal with an eccentric template (yellow), and an eccentric signal with a circular template (green). 
At high eccentricities ($\e\,\gtrsim 0.1$) the time- and phase-maximized \ac{SNR} slightly increases for the eccentric template/eccentric signal case, since the template can perfectly match the amplitude and phase modulations caused by eccentricity in the inspiral. 
It decreases for the circular template/eccentric signal case because of the mismatch between the template and the eccentric signal. 
}
\label{fig:SNR}
\end{figure}

For each system, we calculate three separate time- and phase-maximized \acp{SNR} with different assumptions: (1) the system is circular ($\e=0$) and the template is a perfectly matching circular template, (2) the system is eccentric with $\e > 0$ and the template is a perfectly matching eccentric template, and (3) the system is eccentric with $\e > 0$ and the template is a circular template with all other source properties identical. 
The first two of these correspond to the ``optimal \ac{SNR}'' in the circular and eccentric cases, respectively, while the third characterizes the loss of signal from waveform mismatch.\footnote{A higher maximal \ac{SNR} may be achieved in the third case through marginalization over intrinsic parameters of the template, making our maximal matched-filter \acp{SNR} a conservative lower limit. However, additional factors that may increase our estimated \acp{SNR} will alter the results by less than a factor of 2, as discussed in Section~\ref{sec:conclusions}. } 
In Figure \ref{fig:SNR}, we show both the maximized matched-filter \ac{SNR} and the detection probability for an exemplary \ac{BBH} system over a range of eccentricities when entering the LIGO--Virgo band. 
Further details on \ac{SNR} calculations and determination of detection probabilities are in Appendix~\ref{app:detectability}.


\begin{figure*}[t]
\centering
\includegraphics[width=0.95\textwidth]{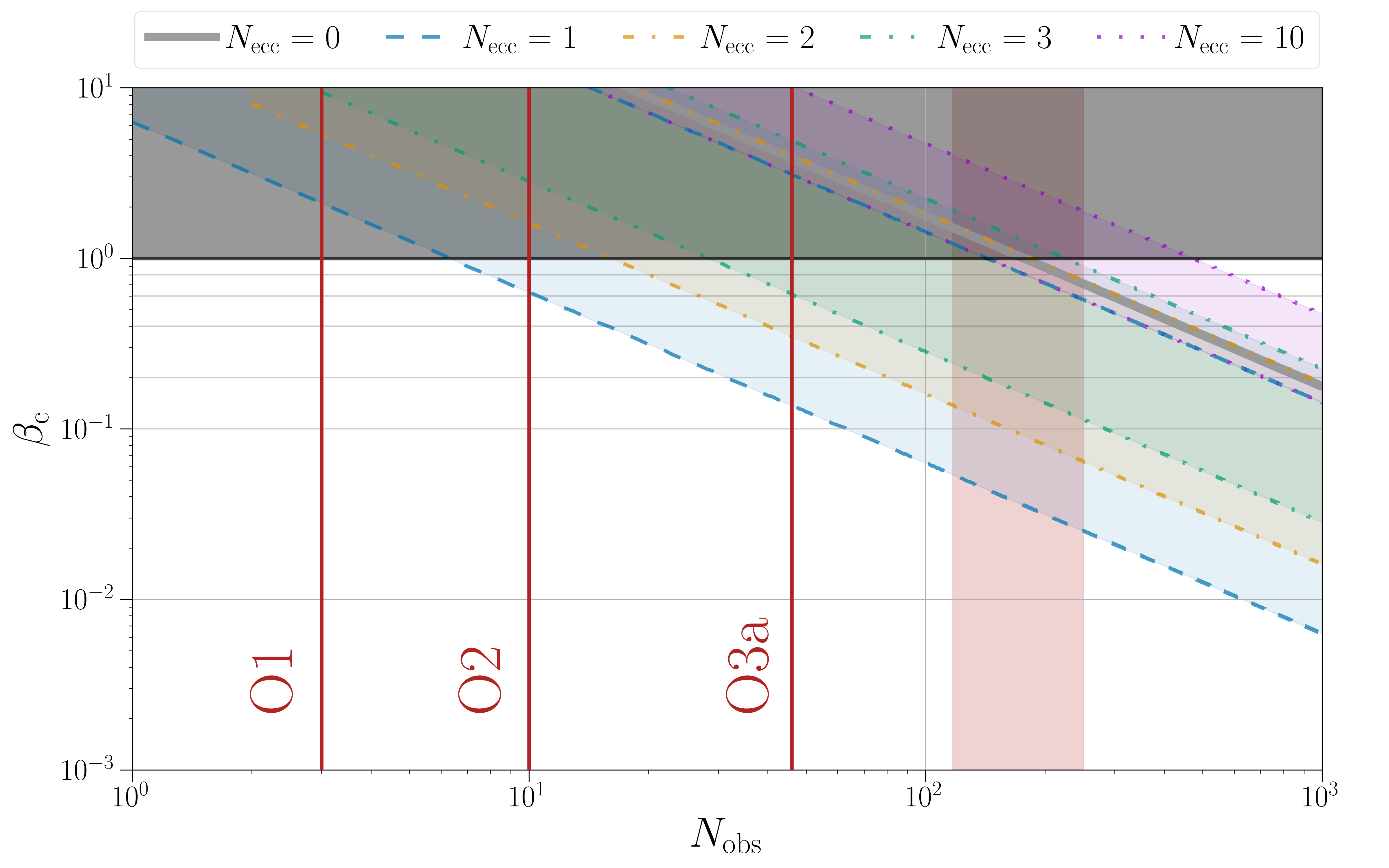}
\caption{Constraints on the detectable branching fraction of dense star clusters, \betac, as a function of the number of \ac{BBH} observations, \Nobs, under the condition of \Necc eccentric observations. 
For $\Necc=0$, the solid gray diagonal line marks the upper $95^{\rm th}$ percentile of the $\betac$ likelihood, and for $\Necc>0$, colored bands encompass the 95\% symmetric credible region of the $\betac$ likelihood. 
We allow for branching fractions above the physical limit of $\betac=1$ in our likelihood (gray shaded region) to display combinations of \Necc and \Nobs that are unphysical, indicating issues with the underlying formation channel model or significant contamination from other eccentric channels. 
Red vertical lines mark the number of confident \ac{BBH} observations reported so far in the first (O1), second (O2), and first half of the third (O3a) observing runs by the \ac{LVC}. 
The vertical red shaded region marks the predicted number of \ac{BBH} observations after the fourth observing run (O4) from \cite{LVC_ObservingScenarios}, assuming the second half of the third observing run observes the same number of \ac{BBH} systems as O3a. 
}
\label{fig:BF_Nobs}
\end{figure*}

In addition to the detection probability of eccentric sources, another important aspect of this analysis is determining the minimum eccentricity that is required for parameter estimation routines to be able to confidently distinguish a system as eccentric. 
For systems with properties similar to GW150914, \cite{Lower2018} found this threshold eccentricity to be $e_\mathrm{thresh}\,\simeq 0.05$. 
This is consistent with the eccentricity upper limits for GWTC-1 events from \cite{Romero-Shaw2019}. 
Due to the computational expense of performing eccentric parameter estimation over a wide range of source parameters, we choose to adopt a fixed threshold eccentricity of $e_\mathrm{thresh} = 0.05$ for this analysis, where $e_\mathrm{thresh}$ is likewise defined at a reference \ac{GW} frequency of 10 Hz. 
At this threshold, $\simeq\,\EccPerc$ of the potentially detectable distribution of cluster binaries have $\e > e_\mathrm{thresh}$. 
\changed{Our chosen value for $e_\mathrm{thresh}$ can be considered a lower limit, since weaker signals may require a higher threshold eccentricity to be distinguished from circular. 
However, small adjustments to this parameter do not impact our analysis significantly. 
For example, increasing $e_\mathrm{thresh}$ from $0.05$ to $0.1$ decreases the fraction of systems above this threshold by only $\lesssim\,\EccFracZeroPOneDifference$. 
As described in the following section, increasing this threshold will decrease the fraction of detectable eccentric systems from dense star clusters, thereby increasing the branching fraction from clusters in the presence of an eccentric detection. 
}

\section{Implications of Eccentric Detections}\label{sec:implications}

Accounting for eccentricity in the selection effects afflicting the cluster population alters the detectable population. 
Under the assumption that matched-filter searches are the only means of detection (i.e., using circular templates for determining the detection probabilities, as shown with the green line in Figure~\ref{fig:SNR}), we find a steep decrease in recovered systems for $\e \gtrsim\,0.1$; see the inset of Figure \ref{fig:GC_pop}. 
Using $e_\mathrm{thresh}=0.05$ as the characteristic threshold for systems that will be measurably eccentric, we find that only \RecoveredPerc of potentially detectable systems with $\e \geq e_\mathrm{thresh}$ will be recovered using matched-filter searches. 
Thus, measurably eccentric systems from clusters make up $\simeq\,\EccDetPerc$ of the detectable distribution of \ac{BBH} mergers from clusters. 
We refer to the fraction of measurably eccentric and detectable systems relative to the full detectable population as the detectable eccentric fraction $\xi_\mathrm{ecc}$, which allows us to translate the presence or absence of eccentric detections to information about the fraction of mergers taking place in dense clusters. 
For the following analyses, we use our calculated value of $\xi_\mathrm{ecc} = \EccDetFrac$. 

Assuming dense star clusters account for a branching fraction $\betac$ of the total number of observed \ac{BBH} mergers and are the dominant contributor of eccentric sources, the probability of detecting \Necc measurably eccentric systems given \Nobs total \ac{BBH} detections follows a homogeneous Poisson process, 
\begin{equation}\label{eq:poisson}
    p(\Necc | \lambda) = e^{- \lambda} \lambda^{\Necc} / \Necc !,
\end{equation}
where 
\begin{equation}
\lambda \equiv \xi_\mathrm{ecc} \betac \Nobs ,
\end{equation}
is the expected number of measurably eccentric detections after \Nobs total observations. 
At a given \Nobs and \Necc, we can determine the likelihood for the cluster branching fraction \betac by evaluating a grid of \betac in the \ac{CDF} of Eq.~\ref{eq:poisson}. 

Constraints on \betac as a function of \Nobs conditioned on various assumed measurably eccentric observations \Necc are shown in Figure~\ref{fig:BF_Nobs}, with colored bands marking the symmetric 95\% credible interval on \betac. 
We plot unphysical values of $\betac > 1$ as a means of a posterior predictive check; though clusters cannot account for more than the entire observed population of \acp{BBH}, significant support for \betac in this region at a given \Nobs and \Necc indicates issues in the underlying cluster models due to the observation of too many eccentric systems. 

We first consider the implications of not detecting measurably eccentric \ac{BBH} mergers on the cluster branching fraction. 
At the number of \ac{BBH} observations through the first half of the third observing run (O3a), the nondetection of a bona fide eccentric signal does not yet place tension on clusters dominating the detectable rate of \acp{BBH}; $\betac = 1$ lies at the \NeccZeroPerctileBetaOne percentile of the likelihood. 
However, the nondetection of an eccentric signal will soon place interesting upper limits on the cluster branching fraction. 
At 100 (300) \ac{BBH} observations, the lack of a bona fide eccentric signal will indicate $\betac < \NeccZeroBetaMaxOneHundredObs$ ($\NeccZeroBetaMaxThreeHundredObs$) at 95\% credibility. 
Once $\simeq\,\NobsClustersNotMajority$ \ac{BBH} observations have been made, the lack of a bona fide eccentric signal will indicate that clusters do not account for the majority of the detectable population ($\betac < 0.5$). 

No unambiguous eccentric detections have been found in the \ac{GW} catalogs to date. 
However, multiple studies have found marginal to strong evidence that GW190521, one of the most massive \ac{BBH} systems observed with \acp{GW} so far~\citep{GWTC2}, is more consistent with being eccentric~\citep{Gayathri2020,Romero-Shaw2020b,Romero-Shaw2021} or from a hyperbolic encounter~\citep{Bustillo2021,Gamba2021} than a quasi-circular inspiral. 
Scattering experiments simulating the strong gravitational encounters typical of \acp{BH} in clusters indicate that GW190521-like binaries can enter the LIGO--Virgo band with appreciable eccentricity~\citep[though not at extreme eccentricities of $\e\,\gtrsim 0.7$; see][]{Holgado2021a}. 
A detailed analysis investigating eccentricity in the \ac{BBH} mergers from the second \ac{GW} catalog~\citep[GWTC-2;][]{GWTC2} also found evidence for eccentricity in the system GW190620A~\citep{Romero-Shaw2021}. 

\begin{figure}[t]
\includegraphics[width=0.47\textwidth]{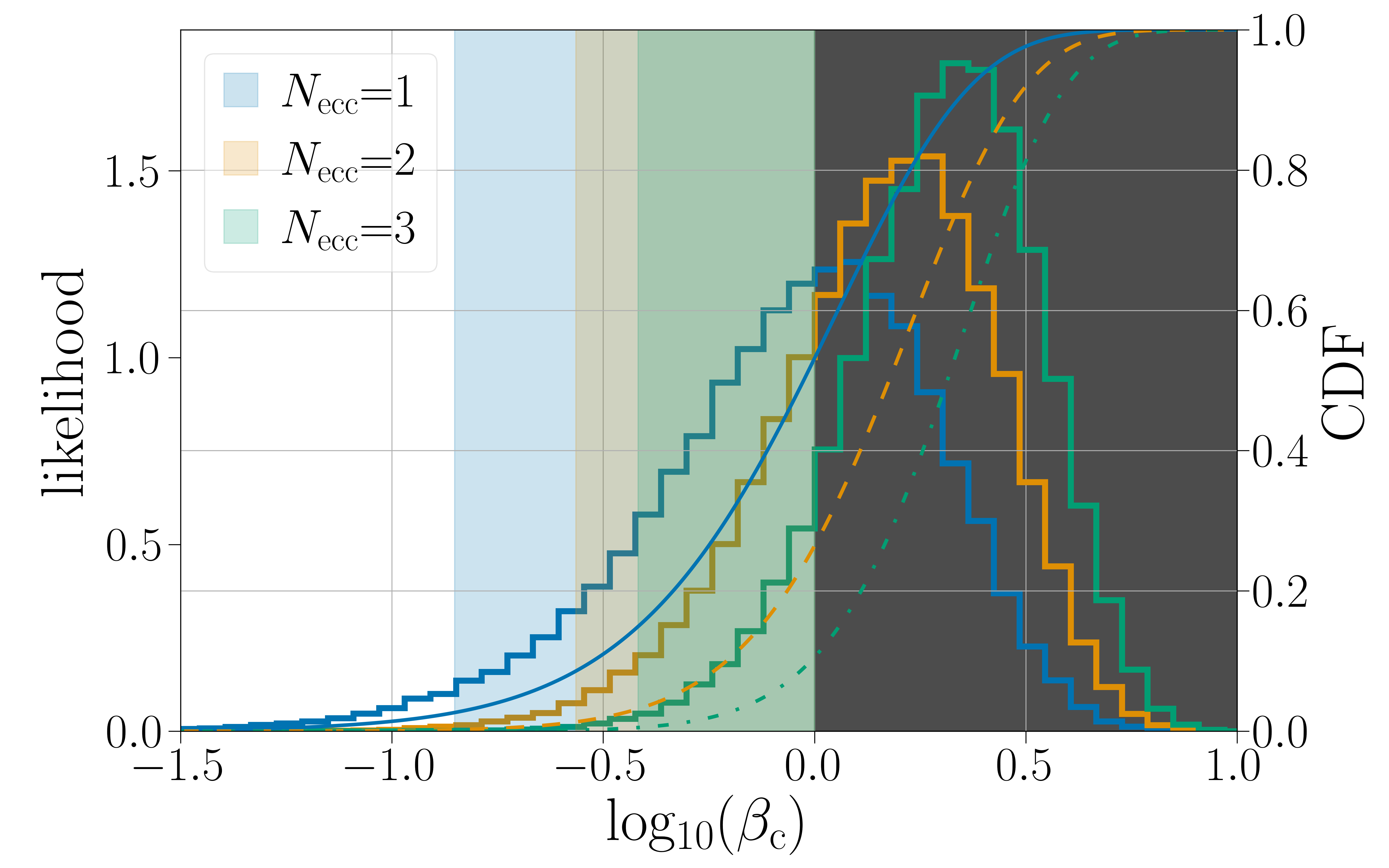}
\caption{Constraints on the detectable branching fraction of the dense star cluster channel, $\betac$, conditioned on \Necc eccentric observations out of $\Nobs=46$ total \ac{BBH} observations. 
As with Figure~\ref{fig:BF_Nobs}, we show the likelihood \changed{(arbitrary units)} without imposing the physical constraint of $\betac \leq 1$ with solid lines and the corresponding \ac{CDF} with dashed lines; the unphysical region $\betac > 1$ is shaded in gray. 
Colored shaded regions mark the $>5\%$ quantile of the posterior distribution, in which we impose a flat-in-log prior on $\betac$ with no support for $\betac > 1$, and can be interpreted as a posteriori lower bounds on the cluster branching fraction given \Necc eccentric observations. 
The relative location where the \ac{CDF} crosses $\betac = 1$ corresponds to the evidence of models in which \Necc eccentric \ac{BBH} systems are observed given $46$ total \ac{BBH} observations. 
}
\label{fig:GWTC2_BF}
\end{figure}

Operating under the assumption that \Necc out of the $\Nobs = 46$ \ac{BBH} observations in GWTC-2 are measurably eccentric, we now focus on the constraints that can be placed on \betac given the presence of eccentric observations. 
Figure \ref{fig:GWTC2_BF} shows the likelihood for \betac assuming different values of \Necc at $\Nobs = 46$. 
Though the likelihood can have support at $\betac > 1$, we also mark the $> 5\%$ quantile of the posterior distribution with shaded regions, where we assume a flat-in-log prior with zero support above $\betac > 1$, thus constraining \betac to its physically valid range. 
Given one (two) eccentric signal(s) in GWTC-2, one can place a lower limit on the detectable branching fraction of \acp{BBH} originating from clusters of \NeccOneBetaNinetyFive (\NeccTwoBetaNinetyFive) at the 95\% credible level. 

An important diagnostic for determining the fidelity of the cluster models considered in this work is identifying the number of eccentric observations that would be infeasible at a given \Nobs. 
We quantify this by the likelihood volume that satisfies $\betac \leq 1$, i.e. where there is nonzero prior probability. 
Given the number of \ac{BBH} observations in GWTC-2, we find that there would be significant tension with eccentricity predictions from cluster models if $\Necc \geq \NeccExcludedNinetyFive$, as $\gtrsim\,\NeccExcludedNinetyFiveVolume$ of the likelihood volume would be in the prior-excluded region. 
Observing this many eccentric mergers in the current catalog of \ac{BBH} observations may also be an indication that a formation channel other than dense stellar clusters is significantly contributing to the eccentric \ac{BBH} population. 

As \Necc grows, tighter constraints on \betac are achieved, since the width of the $\log_{10}(\betac)$ likelihood is independent of \Nobs. 
At $\Necc = 1$ ($3$), the symmetric 95\% credible region for \betac is constrained to \BetaSpreadNeccOne (\BetaSpreadNeccThree) dex. 
Precision of $\approx\,0.5$ dex for the 95\% credible range on \betac will be achieved at $\Necc \gtrsim \HalfDexPrecision$.


\section{Discussion and Conclusions}\label{sec:conclusions}

In this work, we show how the detection or nondetection of measurably eccentric \ac{GW} sources can act to significantly constrain the contribution of certain dynamical channels to the detectable \ac{BBH} population. 
Our main results are. 
\begin{enumerate}
    \item Dense star cluster models robustly predict that ${\approx\,\EccPerc}$ of potentially detectable sources will be measurably eccentric, though matched-filter searches will only recover ${\approx\,\RecoveredPerc}$ of this cluster subpopulation. 
    
    \item Though the nondetection of a measurably eccentric source in GWTC-2 would not rule out the hypothesis that clusters are the dominant contributor of detectable sources, once \NobsClustersNotMajority \ac{BBH} observations have been made, the lack of a bona fide eccentric observation would indicate that clusters do not make up more than $50\%$ of the detectable population of \acp{BBH}. 
    Even with pessimistic predictions, this number of \ac{BBH} mergers will be reached in O4. 
    
    \item Assuming that it originated in a dense star cluster, a single measurably eccentric source in GWTC-2 would indicate that such clusters account for ${>\,\NeccOneBetaNinetyFivePerc}$ of the detectable \ac{BBH} population. 
    
    \item Once \HalfDexPrecision eccentric observations from clusters have been made, the detectable branching fraction from clusters will be constrained at the $0.5$ dex level. 
\end{enumerate}

Eccentricity may be the most robust indicator of a dynamical \ac{BBH} formation pathway, as the presence of eccentric mergers in clusters is relatively insensitive to uncertain physical processes that impact other dynamical indicators, such as binary evolution physics, the efficiency of angular momentum transport in massive stars, and the location of the pair instability mass gap. 
Instead, eccentric mergers in dynamical environments are the result of well-understood physics, and if \acp{BH} are indeed present in clusters, eccentric mergers are an inevitable by-product. 

Current state-of-the-art models of dense star clusters predict local \ac{BBH} merger rates of ${\sim\,2-20\,\mathrm{Gpc}^{-3}\,\mathrm{yr}^{-1}}$~\citep{Kremer2020,Rodriguez2021}. 
Based on this work, this would translate to a merger rate of measurably eccentric sources of ${\sim\,\GCEccRateLow - \GCEccRateHigh\,\mathrm{Gpc}^{-3}\,\mathrm{yr}^{-1}}$. 
However, this translation depends on the fraction of \ac{GW} captures relative to all \ac{BBH} mergers in clusters, which in the local universe is sensitive to the initial density distribution of clusters. 
Though our cluster models have initial densities tuned to reproducing the density profiles of clusters observed in the Milky Way today, the Milky Way's cluster population may not be representative of the local universe as a whole~\citep[e.g.,][]{Jordan2015}, and for high initial cluster densities ($\gtrsim\,10^5\,\Msun\,\mathrm{pc}^{-3}$) \changed{the fraction of local \ac{GW} captures may reduce by a factor of a $\sim\,2-3$~\citep{Antonini2020}. }
\changed{Other model variations, such as metallicity distributions, cluster formation times, binary physics, and remnant formation prescriptions, have a much more minor effect on the fraction of in-cluster mergers and local \ac{GW} captures~\citep[see Table 1 of][]{Antonini2020}. }
Reducing the local \ac{GW} capture fraction (i.e., reducing $\xi_\mathrm{ecc}$) would act to increase our recovered values of \betac, thereby making our lower limits on cluster branching fractions conservative. 
\changed{However, it is important to note that possible systematics can only vary the value of $\xi_\mathrm{ecc}$ by a factor of a few, and no reasonable model variations can cause this eccentric population to disappear entirely, which makes eccentric signals in clusters a robust prediction compared to most other predictions from population modeling. }

The predicted merger rate from dense star clusters is comparable to the empirically measured rate by the LIGO Scientific Collaboration and Virgo Collaboration (LVC) following GWTC-2 of ${15-39\,\mathrm{Gpc}^{-3}\,\mathrm{yr}^{-1}}$~\citep[90\% credibility;][]{GWTC2_pops}. 
However, it is likely that \acp{GC} only account for a portion of this rate, and the diversity of \ac{BBH} observations to date hint at multiple formation channels contributing significantly to the observed population of \acp{BBH}~\citep{GWTC2_pops,Zevin2021}. 
Since eccentric sources will be extremely useful for constraining the merger rate from clusters, they will also assist in limiting the rate contribution of other formation channels. 

This analysis estimates eccentric detectability assuming only matched-filter searches with circular, aligned-spin templates. 
Though these searches are the main workhorse for the detection of compact binary coalescences, a number of other search techniques are used to identify \ac{GW} signals. 
Of particular pertinence to this study are burst searches, which are unmodeled searches that identify coincident and coherent signals in the \ac{GW} data stream. 
For nearby and loud mergers, burst searches would excel over quasi-circular template-based searches at detecting highly eccentric systems~\citep{Klimenko2016}. 
A possible example of this in action is GW190521, which was found by a burst search to have a much lower false-alarm rate ($< 1$ per $28$ yr) than reported for the same candidate by matched-filter compact binary pipelines~\citep[$\geq 1$ per $8$ yr;][]{GW190521}.
In addition, higher matched-filter \acp{SNR} may be achieved by marginalizing over the intrinsic parameters of the template rather than considering a single circular template with analogous source properties. 
Thus, the detectable eccentric fraction $\xi_\mathrm{ecc}$ we report can be considered a lower limit. 
However, if one were to assume the limiting case, where eccentric mergers are equally as likely to be detected as their circular counterparts, the maximum detectable eccentric fraction would be less than twice our reported value. 
Therefore, even in this best-case detectability scenario, our constraints on branching fractions as a function of the number of \ac{BBH} observations are reduced by less than a factor of 2. 

We also assume that strong gravitational encounters in clusters are the only means of generating measurably eccentric mergers in the LIGO--Virgo sensitive frequency range. 
\changed{Other formation channels have been proposed for generating systems with measurable eccentricities, though the predicted local merger rates and/or eccentricity distributions for these channels are typically subject to more uncertainties, as they are not solely a simple by-product of two-body relaxation and small-$N$ dynamics. 
\begin{enumerate}
\item In the secular evolution of isolated hierarchical systems, predicted merger rates for measurably eccentric systems in the LIGO--Virgo band are at most ${\sim\,0.1\,\mathrm{Gpc}^{-3}\,\mathrm{yr}^{-1}}$~\citep{Antonini2017a}, but they are sensitive to initial orbital properties and rely on \acp{BH} having no natal kicks, dropping 1--2 orders of magnitude if even weak natal kicks are assumed~\citep{Silsbee2017,Rodriguez2018}. 
\item Perturbations of field triples by flyby encounters may also contribute significantly with a measurable eccentric merger rate of up to ${\sim\,1-10\,\mathrm{Gpc}^{-3}\,\mathrm{yr}^{-1}}$~\citep{Michaely2020}, though this channel is also highly sensitive to \ac{BH} natal kicks and initial binary properties, and the impact of these parameterizations on the predicted merger rate has yet to be explored. 
\item For \ac{BBH} systems in binaries or triples orbiting supermassive \acp{BH}, the predicted merger rates for measurably eccentric \acp{BBH} range from ${\sim\,0.03 - 0.16\,\mathrm{Gpc}^{-3}\,\mathrm{yr}^{-1}}$~\citep{Fragione2019b} and are dependent on the supermassive \ac{BH} and stellar mass \ac{BH} mass functions, as well as the initial orbital semimajor axis and eccentricity distribution of \acp{BBH}; these rates can be further enhanced by asphericity in the nuclear cluster itself~\citep{Petrovich2017}. 
\item Single-single captures in galactic nuclei are a promising avenue for detectable eccentric mergers, with $\gtrsim\,90\%$ of systems exhibiting potentially measurable signs of eccentricity~\citep{Gondan2021}. 
Furthermore, this channel is relatively insensitive to initial condition and binary evolution uncertainties such as the \ac{BH} mass function, though the overall merger rate of this channel ranges from ${\sim\,0.002 - 0.04\,\mathrm{Gpc}^{-3}\,\mathrm{yr}^{-1}}$ and is exponentially sensitive to the radial number density profile exponent for galactic nuclei~\citep{Rasskazov2019}. 
\item Lastly, though predicted rates for \ac{BBH} mergers in \ac{AGN} disks span many orders of magnitude~\citep{Grobner2020,McKernan2020}, a substantial number ($\gtrsim\,10\%$) of mergers in \ac{AGN} disks may show measurable signs of eccentricity, with this percentage possibly increasing to $\sim\,70\%$ if \acp{BH} are contained to the disk or can migrate to within $10^{-3}\,\mathrm{pc}$ of the supermassive \ac{BH}~\citep{Tagawa2021a,Samsing2020a}.\footnote{\changed{We note that, based on our analysis, the high efficiency of eccentric mergers predicted from the galactic nuclei and \ac{AGN} disk channels combined with only $\sim$\,zero to two eccentric mergers in GWTC-2~\citep[e.g.,][]{Romero-Shaw2021} may indicate that these channels are a subdominant contributor to the total \ac{BBH} merger rate if the predicted eccentricity distributions are similar to those of dense star clusters.}}
\end{enumerate}
Despite the higher degree of uncertainty inherent to many of these eccentric \ac{BBH} channels, if they contribute significantly to the measurably eccentric merger rate, the inclusion of these channels would impact the branching fraction constraints for the dense star clusters reported in this work. }
Robust rate estimates and eccentricity distributions from these channels could be incorporated into the methodology presented here, allowing for mixture models to be created with multiple eccentric detection efficiencies and joint constraints on the contribution of formation channels that generate eccentric \ac{BBH} mergers. 

The utility of eccentric detections for constraining formation channels will improve as more \ac{BBH} mergers are observed, regardless of whether or not eccentric mergers are observed in this population. 
The eccentric detection efficiency will also be relatively stable as the sensitivity of the detector network improves, as the fraction of measurably eccentric mergers relative to the full cluster population does not evolve significantly with redshift~\citep{Rodriguez2018c}. 
Given current \ac{BBH} rate measurements and the anticipated sensitivity of the \ac{GW} detector network in O4, the lack of a measurably eccentric detection after O4 will indicate that clusters do not account for the majority of \ac{BBH} mergers. 
On the other hand, with future eccentric detections, we will significantly constrain the lower limit of mergers that result from clusters and other dynamical channels. 
Once $\sim\,10$ eccentric \ac{BBH} mergers are observed, constraints on the dynamical branching fraction will reach the 0.5 dex level, potentially making eccentricity the most robust and efficient means for constraining the formation pathways of \ac{BBH} mergers.

\acknowledgments
We thank Rossella Gamba for useful comments on this manuscript, Carl Rodriguez for assistance in computing the relative weights for the \texttt{CMC} cluster models, Fabio Antonini and Mark Gieles for interesting discussions regarding uncertainties in initial cluster densities, Daniel Holz for enlightening discussions, and Alessandro Nagar, Sebastiano Bernuzzi, and Piero Rettegno for help with and development of the TEOBResumS waveform model. 
\changed{We also thank the anonymous referee, whose comments and suggestions improved this manuscript.}
Support for this work and for M.Z. was provided by NASA through the NASA Hubble Fellowship grant HST-HF2-51474.001-A awarded by the Space Telescope Science Institute, which is operated by the Association of Universities for Research in Astronomy, Inc., for NASA, under contract NAS5-26555. 
K.K. is supported by an NSF Astronomy and Astrophysics Postdoctoral Fellowship under award AST-2001751.
E.T. and P.D.L. are supported through Australian Research Council (ARC) Centre of Excellence CE170100004.  
P.D.L. is supported through ARC Future Fellowship FT160100112 and ARC Discovery project DP180103155.
This work used computing resources at CIERA funded by NSF grant No.\ PHY-1726951 and resources and staff provided by the Quest high-performance computing facility at Northwestern University, which is jointly supported by the Office of the Provost, the Office for Research, and Northwestern University Information Technology.

\software{\texttt{Astropy}~\citep{TheAstropyCollaboration2013,TheAstropyCollaboration2018}; 
\texttt{iPython}~\citep{ipython}; 
\texttt{Matplotlib}~\citep{matplotlib}; 
\texttt{NumPy}~\citep{numpy,numpy2}; 
\texttt{Pandas}~\citep{pandas}; 
\texttt{PyCBC}~\citep{PyCBC_v1.14.4}; 
\texttt{SciPy}~\citep{scipy};
\texttt{TEOBResumS}~\citep{Nagar2018}.}

\clearpage
\appendix

\section{Cluster Models and Eccentricity of BBH Sources}\label{app:cluster_models}

\changed{We obtain our sample of \acp{BBH} from the 148 $N$-body cluster simulations in \citet{Kremer2020}.} 
In each cluster model, the component masses, component spin magnitudes, and merger redshifts of \ac{BBH} mergers are recorded. 
The \acp{BH} are assumed to have negligible birth spin, as predicted if angular momentum transport in their massive-star progenitors is highly efficient~\citep{Spruit1999,Spruit2002,Fuller2019a}, though spin can be imparted in cluster \acp{BH} through prior \ac{BBH} merger events that are retained in the cluster. 
Spin tilts are assumed to be isotropically distributed on the sphere. 
The \acp{BBH} in our models have (source-frame) total masses between ${\MtotLowModels - \MtotHighModels~\Msun}$, mass ratios between ${\qLowModels - \qHighModels}$, and effective spins between ${\ChiEffLowModels - \ChiEffHighModels}$, where the quoted ranges represent 99\% of all systems. 

\changed{In order to obtain an appropriate astrophysically weighted sample of \acp{BBH} from the \citet{Kremer2020} models, we follow the approach described in \cite{Rodriguez2018a} and \cite{Martinez2021}. 
To briefly summarize, the 148 cluster models from \citet{Kremer2020} are placed into equally spaced bins in cluster mass and logarithmically spaced bins in metallicity. 
Each cluster model is then assigned a relative astrophysical weight, $\gamma_c$, corresponding to the number of clusters that are thought to form in that 2D bin across cosmic time. 
Initial cluster masses are assumed to follow an $M^{-2}$ distribution \citep[e.g.,][]{Lada2003}, and metallicities are obtained from \citet{ElBadry2019}. 
A formation time for each cluster is also drawn from the metallicity-dependent formation time distributions from \citet{ElBadry2019}. 
The cluster formation time, $t_{\rm{form}}$, drawn for a given cluster model is then added to the merger times, $t_{\rm{merge}}$, for all \ac{BBH} mergers identified in that model. 
Systems with $t_{\rm{form}}+t_{\rm{merge}}$ larger than a Hubble time are excluded from the analysis.}

The orbital integration of merging \acp{BBH} is halted once the component \acp{BH} reach a separation of $10\,\mathrm{M}$, where $\mathrm{M}$ is the total mass of the binary in geometric units, and the semimajor axis and eccentricity of the orbit are recorded at various discrete separations prior to this point~(see \citealt{Rodriguez2018c} for further discussion on halting criteria). 
To acquire the orbital properties at a particular \ac{GW} frequency $f_\mathrm{GW}$ for a binary of total mass $M_\mathrm{tot}$, we use the orbital properties recorded at a separation of $100\,\mathrm{M}$ and numerically solve for the orbital properties at an eccentric peak frequency as in \cite{Wen2003}, 
\begin{equation}
    a(e) = \frac{1}{1-e^2} 
    \left[ \frac{G M_\mathrm{tot}}{\pi} 
    \frac{(1+e)^{1.1954}}{f_\mathrm{GW}} \right]^{2/3}, 
\end{equation}
which is coupled to the differential equation governing the coevolution of semimajor axis and eccentricity from \cite{Peters1964}: 
\begin{equation}
    \left \langle \frac{da}{de} \right \rangle = 
    \frac{12}{19}
    \frac{a}{e}
    \frac{[1 + (73/24)e^2 + (37/96)e^4}
    {(1-e^2)[1 + (121/304)e^2]}. 
\end{equation}
In certain cases, the binary forms through a hyperbolic capture at frequencies above $f_\mathrm{GW}$, and we assign its eccentricity at $f_\mathrm{GW}$ to an extremal value of $e_\mathrm{max} = 0.9$. 

As we are interested in the properties of detectable \ac{BBH} mergers rather than the underlying population, we must also account for the larger amount of comoving volume accessible at higher redshifts and the relative sensitivity of \ac{GW} detectors to \acp{BBH} with different properties. 
Each system $i$, parameterized by component masses $m_1$ and $m_2$, component spin vectors $\chi_1$ and $\chi_2$, merger redshift $z$, eccentricity at a reference frequency of $10$ Hz $\e$, and cluster weight $\gamma_\mathrm{c}$ is given a normalized detectability weight of 
\begin{equation}
    w_i = \left[ \gamma_\mathrm{c} \frac{dVc}{dz} \frac{dt_s}{dt_0} p_\mathrm{det}(m_1, m_2, \chi_1, \chi_2, \e, z) \right]_i
\end{equation}
where $\frac{dVc}{dz}$ is the comoving volume element at redshift $z$, $\frac{dt_s}{dt_0} = (1+z)^{-1}$ is the time dilation between clocks at the merger redshift and on Earth, and \pdet is the detection probability of a system with a given set of intrinsic parameters, defined in the following section.

\section{Detectability of Eccentric Sources}\label{app:detectability}

We estimate detection probabilities using a fixed \ac{SNR} threshold required for detection of $\rho_\mathrm{thresh} = 8.0$. 
Given the source parameters of each system, we calculate a matched-filter \ac{SNR} assuming the optimal orientation of face-on and directly overhead, 
\begin{equation}
    \rho_\mathrm{max}^2 = \frac{1}{\langle h|h \rangle} |\langle s|h \rangle|^2, 
\end{equation}
where we maximize over the phase and time of coalescence, and $\langle s|h \rangle$ is the noise-weighted inner product, 
\begin{equation}
    \langle s|h \rangle = 4 \int_0^\infty \frac{\tilde{s}(f) \tilde{h}^\star(f)}{S_n(f)} df, 
\end{equation}
with $\tilde{s}(f)$ and $\tilde{h}(f)$ being the Fourier transform of the time-domain signal and template, respectively, and $S_n(f)$ the one-sided average power spectral density (PSD) of the detector noise. 
The noise-weighted inner product of the template with itself $\langle h|h \rangle$ is the optimal matched-filter \ac{SNR} and defined similarly. 
We assume a stationary, single-detector LIGO \ac{PSD} with \texttt{midhighlatelow} sensitivity from \cite{LVC_ObservingScenarios}, which has been shown to be a decent approximation to the more sophisticated approach of injection/recovery campaigns in search pipelines~\citep[e.g.,][]{LVC_ObservingScenarios,Nitz2020b}. 

For the three matched-filter cases described in Section~\ref{sec:detectability_measurability}, we window the waveforms with a Tukey filter, and in cases where the signal and template are not identical, we zero-pad the shorter waveform, time-align the maximum strain of the template and signal, and apply a frequency-domain time and phase shift determined by maximizing the overlap integral. 
If $\rho_\mathrm{max} < \rho_\mathrm{thresh}$, the detection probabilities are assigned to be zero. 
Otherwise, we determine the detection probabilities by Monte Carlo sampling uniformly over the extrinsic parameters $\psi$ and multiplying the maximized \ac{SNR} with the detector projection factor $\Theta(\psi) \in [0,1]$~\citep{Finn1993}. 
The detection probability is thus determined as
\begin{equation}
    \pdet = \sum_{j=1}^N \mathcal{H} \left[ \Theta(\psi_j) \rho_\mathrm{max} - \rho_\mathrm{thresh} \right]
\end{equation}
where $\mathcal{H}$ is the Heaviside step function, and we draw $N=10^3$ sets of extrinsic parameters.

\bibliography{library}{}

\begin{thebibliography}{}
\expandafter\ifx\csname natexlab\endcsname\relax\def\natexlab#1{#1}\fi
\providecommand{\url}[1]{\href{#1}{#1}}
\providecommand{\dodoi}[1]{doi:~\href{http://doi.org/#1}{\nolinkurl{#1}}}
\providecommand{\doeprint}[1]{\href{http://ascl.net/#1}{\nolinkurl{http://ascl.net/#1}}}
\providecommand{\doarXiv}[1]{\href{https://arxiv.org/abs/#1}{\nolinkurl{https://arxiv.org/abs/#1}}}

\bibitem[{Aasi {et~al.}(2015)Aasi, Abbott, Abbott, Abbott, Abernathy, Ackley,
  Adams, Adams, Addesso, Adhikari, Adya, Affeldt, Aggarwal, Aguiar, Ain, Ajith,
  Alemic, Allen, Amariutei, Anderson, Anderson, Arai, Araya, Arceneaux, Areeda,
  Ashton, Ast, Aston, Aufmuth, Aulbert, Aylott, Babak, Baker, Ballmer,
  Barayoga, Barbet, Barclay, Barish, Barker, Barr, Barsotti, Bartlett, Barton,
  Bartos, Bassiri, Batch, Baune, Behnke, Bell, Bell, Benacquista, Bergman,
  Bergmann, Berry, Betzwieser, Bhagwat, Bhandare, Bilenko, Billingsley, Birch,
  Biscans, Biwer, Blackburn, Blackburn, Blair, Blair, Bock, Bodiya, Bojtos,
  Bond, Bork, Born, Bose, Brady, Braginsky, Brau, Bridges, Brinkmann, Brooks,
  Brown, Brown, Brown, Buchman, Buikema, Buonanno, Cadonati, {Calder{\'{o}}n
  Bustillo}, Camp, Cannon, Cao, Capano, Caride, Caudill, Cavagli{\`{a}},
  Cepeda, Chakraborty, Chalermsongsak, Chamberlin, Chao, Charlton, Chen, Cho,
  Cho, Chow, Christensen, Chu, Chung, Ciani, Clara, Clark, Collette, Cominsky,
  Constancio, Cook, Corbitt, Cornish, Corsi, Costa, Coughlin, Countryman,
  Couvares, Coward, Cowart, Coyne, Coyne, Craig, Creighton, Creighton, Cripe,
  Crowder, Cumming, Cunningham, Cutler, Dahl, {Dal Canton}, Damjanic,
  Danilishin, Danzmann, Dartez, Dave, Daveloza, Davies, Daw, DeBra, {Del
  Pozzo}, Denker, Dent, Dergachev, DeRosa, DeSalvo, Dhurandhar, Diaz, {Di
  Palma}, Dojcinoski, Dominguez, Donovan, Dooley, Doravari, Douglas, Downes,
  Driggers, Du, Dwyer, Eberle, Edo, Edwards, Edwards, Effler, Eggenstein,
  Ehrens, Eichholz, Eikenberry, Essick, Etzel, Evans, Evans, Factourovich,
  Fairhurst, Fan, Fang, Farr, Farr, Favata, Fays, Fehrmann, Fejer, Feldbaum,
  Ferreira, Fisher, Frei, Freise, Frey, Fricke, Fritschel, Frolov,
  Fuentes-Tapia, Fulda, Fyffe, Gair, Gaonkar, Gehrels, Gergely, Giaime,
  Giardina, Gleason, Goetz, Goetz, Gondan, Gonz{\'{a}}lez, Gordon, Gorodetsky,
  Gossan, Go{\ss}ler, Gr{\"{a}}f, Graff, Grant, Gras, Gray, Greenhalgh,
  Gretarsson, Grote, Grunewald, Guido, Guo, Gushwa, Gustafson, Gustafson,
  Hacker, Hall, Hammond, Hanke, Hanks, Hanna, Hannam, Hanson, Hardwick, Harry,
  Harry, Hart, Hartman, Haster, Haughian, Hee, Heintze, Heinzel, Hendry, Heng,
  Heptonstall, Heurs, Hewitson, Hild, Hoak, Hodge, Hollitt, Holt, Hopkins,
  Hosken, Hough, Houston, Howell, Hu, Huerta, Hughey, Husa, Huttner, Huynh,
  Huynh-Dinh, Idrisy, Indik, Ingram, Inta, Islas, Isler, Isogai, Iyer, Izumi,
  Jacobson, Jang, Jawahar, Ji, Jim{\'{e}}nez-Forteza, Johnson, Jones, Jones,
  Ju, Haris, Kalogera, Kandhasamy, Kang, Kanner, Katsavounidis, Katzman,
  Kaufer, Kaufer, Kaur, Kawabe, Kawazoe, Keiser, Keitel, Kelley, Kells, Keppel,
  Key, Khalaidovski, Khalili, Khazanov, Kim, Kim, Kim, Kim, Kim, King, King,
  Kinzel, Kissel, Klimenko, Kline, Koehlenbeck, Kokeyama, Kondrashov, Korobko,
  Korth, Kozak, Kringel, Krishnan, Krueger, Kuehn, Kumar, Kumar, Kuo, Landry,
  Lantz, Larson, Lasky, Lazzarini, Lazzaro, Le, Leaci, Leavey, Lebigot, Lee,
  Lee, Lee, Leong, Levin, Levine, Lewis, Li, Libbrecht, Libson, Lin,
  Littenberg, Lockerbie, Lockett, Logue, Lombardi, Lormand, Lough, Lubinski,
  L{\"{u}}ck, Lundgren, Lynch, Ma, MacArthur, MacDonald, MacHenschalk,
  MacInnis, MacLeod, Maga{\~{n}}a-Sandoval, Magee, Mageswaran, Maglione,
  Mailand, Mandel, Mandic, Mangano, Mansell, M{\'{a}}rka, M{\'{a}}rka,
  Markosyan, Maros, Martin, Martin, Martynov, Marx, Mason, Massinger,
  Matichard, Matone, Mavalvala, Mazumder, Mazzolo, McCarthy, McClelland,
  McCormick, McGuire, McIntyre, McIver, McLin, McWilliams, Meadors, Meinders,
  Melatos, Mendell, Mercer, Meshkov, Messenger, Meyers, Miao, Middleton,
  Mikhailov, Miller, Miller, Millhouse, Ming, Mirshekari, Mishra, Mitra,
  Mitrofanov, Mitselmakher, Mittleman, Moe, Mohanty, Mohapatra, Moore, Moraru,
  Moreno, Morriss, Mossavi, Mow-Lowry, Mueller, Mueller, Mukherjee, Mullavey,
  Munch, Murphy, Murray, Mytidis, Nash, Nayak, Necula, Nedkova, Newton, Nguyen,
  Nielsen, Nissanke, Nitz, Nolting, Normandin, Nuttall, Ochsner, O'Dell,
  Oelker, Ogin, Oh, Oh, Ohme, Oppermann, Oram, O'Reilly, Ortega, O'Shaughnessy,
  Osthelder, Ott, Ottaway, Ottens, Overmier, Owen, Padilla, Pai, Pai, Palashov,
  Pal-Singh, Pan, Pankow, Pannarale, Pant, Papa, Paris, Patrick, Pedraza,
  Pekowsky, Pele, Penn, Perreca, Phelps, Pierro, Pinto, Pitkin, Poeld, Post,
  Poteomkin, Powell, Prasad, Predoi, Premachandra, Prestegard, Price, Principe,
  Privitera, Prix, Prokhorov, Puncken, P{\"{u}}rrer, Qin, Quetschke, Quintero,
  Quiroga, Quitzow-James, Raab, Rabeling, Radkins, Raffai, Raja, Rajalakshmi,
  Rakhmanov, Ramirez, Raymond, Reed, Reid, Reitze, Reula, Riles, Robertson,
  Robie, Rollins, Roma, Romano, Romanov, Romie, Rowan, R{\"{u}}diger, Ryan,
  Sachdev, Sadecki, Sadeghian, Saleem, Salemi, Sammut, Sandberg, Sanders,
  Sannibale, Santiago-Prieto, Sathyaprakash, Saulson, Savage, Sawadsky,
  Scheuer, Schilling, Schmidt, Schnabel, Schofield, Schreiber, Schuette,
  Schutz, Scott, Scott, Sellers, Sengupta, Sergeev, Serna, Sevigny, Shaddock,
  Shahriar, Shaltev, Shao, Shapiro, Shawhan, Shoemaker, Sidery, Siemens, Sigg,
  Silva, Simakov, Singer, Singer, Singh, Sintes, Slagmolen, Smith, Smith,
  Smith, Smith-Lefebvre, Son, Sorazu, Souradeep, Staley, Stebbins, Steinke,
  Steinlechner, Steinlechner, Steinmeyer, Stephens, Steplewski, Stevenson,
  Stone, Strain, Strigin, Sturani, Stuver, Summerscales, Sutton, Szczepanczyk,
  Szeifert, Talukder, Tanner, T{\'{a}}pai, Tarabrin, Taracchini, Taylor,
  Tellez, Theeg, Thirugnanasambandam, Thomas, Thomas, Thorne, Thorne, Thrane,
  Tiwari, Tomlinson, Torres, Torrie, Traylor, Tse, Tshilumba, Ugolini,
  Unnikrishnan, Urban, Usman, Vahlbruch, Vajente, Valdes, Vallisneri, {Van
  Veggel}, Vass, Vaulin, Vecchio, Veitch, Veitch, Venkateswara, Vincent-Finley,
  Vitale, Vo, Vorvick, Vousden, Vyatchanin, Wade, Wade, Wade, Walker, Wallace,
  Walsh, Wang, Wang, Wang, Ward, Warner, Was, Weaver, Weinert, Weinstein,
  Weiss, Welborn, Wen, Wessels, Westphal, Wette, Whelan, Whitcomb, White,
  Whiting, Wilkinson, Williams, Williams, Williamson, Willis, Willke, Wimmer,
  Winkler, Wipf, Wittel, Woan, Worden, Xie, Yablon, Yakushin, Yam, Yamamoto,
  Yancey, Yang, Zanolin, Zhang, Zhang, Zhang, Zhang, Zhao, Zhou, Zhu, Zucker,
  Zuraw, \& Zweizig}]{aLIGO}
Aasi, J., Abbott, B.~P., Abbott, R., {et~al.} 2015, Classical and Quantum
  Gravity, 32, 074001, \dodoi{10.1088/0264-9381/32/7/074001}

\bibitem[{Abbott {et~al.}(2016)Abbott, Abbott, Abbott, Abernathy, Acernese,
  Ackley, Adams, Adams, Addesso, Adhikari, Adya, Affeldt, Agathos, Agatsuma,
  Aggarwal, Aguiar, Aiello, Ain, Ajith, Allen, Allocca, Altin, Anderson,
  Anderson, Arai, Arain, Araya, Arceneaux, Areeda, Arnaud, Arun, Ascenzi,
  Ashton, Ast, Aston, Astone, Aufmuth, Aulbert, Babak, Bacon, Bader, Baker,
  Baldaccini, Ballardin, Ballmer, Barayoga, Barclay, Barish, Barker, Barone,
  Barr, Barsotti, Barsuglia, Barta, Bartlett, Barton, Bartos, Bassiri, Basti,
  Batch, Baune, Bavigadda, Bazzan, Behnke, Bejger, Belczynski, Bell, Bell,
  Berger, Bergman, Bergmann, Berry, Bersanetti, Bertolini, Betzwieser, Bhagwat,
  Bhandare, Bilenko, Billingsley, Birch, Birney, Birnholtz, Biscans, Bisht,
  Bitossi, Biwer, Bizouard, Blackburn, Blair, Blair, Blair, Bloemen, Bock,
  Bodiya, Boer, Bogaert, Bogan, Bohe, Bojtos, Bond, Bondu, Bonnand, Boom, Bork,
  Boschi, Bose, Bouffanais, Bozzi, Bradaschia, Brady, Braginsky, Branchesi,
  Brau, Briant, Brillet, Brinkmann, Brisson, Brockill, Brooks, Brown, Brown,
  Brown, Buchanan, Buikema, Bulik, Bulten, Buonanno, Buskulic, Buy, Byer,
  Cabero, Cadonati, Cagnoli, Cahillane, Bustillo, Callister, Calloni, Camp,
  Cannon, Cao, Capano, Capocasa, Carbognani, Caride, Diaz, Casentini, Caudill,
  Cavagli{\`{a}}, Cavalier, Cavalieri, Cella, Cepeda, Baiardi, Cerretani,
  Cesarini, Chakraborty, Chalermsongsak, Chamberlin, Chan, Chao, Charlton,
  Chassande-Mottin, Chen, Chen, Cheng, Chincarini, Chiummo, Cho, Cho, Chow,
  Christensen, Chu, Chua, Chung, Ciani, Clara, Clark, Cleva, Coccia, Cohadon,
  Colla, Collette, Cominsky, Constancio, Conte, Conti, Cook, Corbitt, Cornish,
  Corsi, Cortese, Costa, Coughlin, Coughlin, Coulon, Countryman, Couvares,
  Cowan, Coward, Cowart, Coyne, Coyne, Craig, Creighton, Creighton, Cripe,
  Crowder, Cruise, Cumming, Cunningham, Cuoco, Canton, Danilishin, D'Antonio,
  Danzmann, Darman, {Da Silva Costa}, Dattilo, Dave, Daveloza, Davier, Davies,
  Daw, Day, De, Debra, Debreczeni, Degallaix, {De Laurentis}, Del{\'{e}}glise,
  {Del Pozzo}, Denker, Dent, Dereli, Dergachev, Derosa, {De Rosa}, Desalvo,
  Dhurandhar, D{\'{i}}az, {Di Fiore}, {Di Giovanni}, {Di Lieto}, {Di Pace}, {Di
  Palma}, {Di Virgilio}, Dojcinoski, Dolique, Donovan, Dooley, Doravari,
  Douglas, Downes, Drago, Drever, Driggers, Du, Ducrot, Dwyer, Edo, Edwards,
  Effler, Eggenstein, Ehrens, Eichholz, Eikenberry, Engels, Essick, Etzel,
  Evans, Evans, Everett, Factourovich, Fafone, Fair, Fairhurst, Fan, Fang,
  Farinon, Farr, Farr, Favata, Fays, Fehrmann, Fejer, Feldbaum, Ferrante,
  Ferreira, Ferrini, Fidecaro, Finn, Fiori, Fiorucci, Fisher, Flaminio,
  Fletcher, Fong, Fournier, Franco, Frasca, Frasconi, Frede, Frei, Freise,
  Frey, Frey, Fricke, Fritschel, Frolov, Fulda, Fyffe, Gabbard, Gair,
  Gammaitoni, Gaonkar, Garufi, Gatto, Gaur, Gehrels, Gemme, Gendre, Genin,
  Gennai, George, Gergely, Germain, Ghosh, Ghosh, Ghosh, Giaime, Giardina,
  Giazotto, Gill, Glaefke, Gleason, Goetz, Goetz, Gondan, Gonz{\'{a}}lez,
  Castro, Gopakumar, Gordon, Gorodetsky, Gossan, Gosselin, Gouaty, Graef,
  Graff, Granata, Grant, Gras, Gray, Greco, Green, Greenhalgh, Groot, Grote,
  Grunewald, Guidi, Guo, Gupta, Gupta, Gushwa, Gustafson, Gustafson, Hacker,
  Hall, Hall, Hammond, Haney, Hanke, Hanks, Hanna, Hannam, Hanson, Hardwick,
  Harms, Harry, Harry, Hart, Hartman, Haster, Haughian, Healy, Heefner,
  Heidmann, Heintze, Heinzel, Heitmann, Hello, Hemming, Hendry, Heng, Hennig,
  Heptonstall, Heurs, Hild, Hoak, Hodge, Hofman, Hollitt, Holt, Holz, Hopkins,
  Hosken, Hough, Houston, Howell, Hu, Huang, Huerta, Huet, Hughey, Husa,
  Huttner, Huynh-Dinh, Idrisy, Indik, Ingram, Inta, Isa, Isac, Isi, Islas,
  Isogai, Iyer, Izumi, Jacobson, Jacqmin, Jang, Jani, Jaranowski, Jawahar,
  Jim{\'{e}}nez-Forteza, Johnson, Johnson-Mcdaniel, Jones, Jones, Jonker, Ju,
  Haris, Kalaghatgi, Kalogera, Kandhasamy, Kang, Kanner, Karki, Kasprzack,
  Katsavounidis, Katzman, Kaufer, Kaur, Kawabe, Kawazoe, K{\'{e}}f{\'{e}}lian,
  Kehl, Keitel, Kelley, Kells, Kennedy, Keppel, Key, Khalaidovski, Khalili,
  Khan, Khan, Khan, Khazanov, Kijbunchoo, Kim, Kim, Kim, Kim, Kim, Kim, King,
  King, Kinzel, Kissel, Kleybolte, Klimenko, Koehlenbeck, Kokeyama, Koley,
  Kondrashov, Kontos, Koranda, Korobko, Korth, Kowalska, Kozak, Kringel,
  Krishnan, Kr{\'{o}}lak, Krueger, Kuehn, Kumar, Kumar, Kuo, Kutynia, Kwee,
  Lackey, Landry, Lange, Lantz, Lasky, Lazzarini, Lazzaro, Leaci, Leavey,
  Lebigot, Lee, Lee, Lee, Lee, Lenon, Leonardi, Leong, Leroy, Letendre, Levin,
  Levine, Li, Libson, Littenberg, Lockerbie, Logue, Lombardi, London, Lord,
  Lorenzini, Loriette, Lormand, Losurdo, Lough, Lousto, Lovelace, L{\"{u}}ck,
  Lundgren, Luo, Lynch, Ma, Macdonald, Machenschalk, Macinnis, Macleod,
  Maga{\~{n}}a-Sandoval, Magee, Mageswaran, Majorana, Maksimovic, Malvezzi,
  Man, Mandel, Mandic, Mangano, Mansell, Manske, Mantovani, Marchesoni, Marion,
  M{\'{a}}rka, M{\'{a}}rka, Markosyan, Maros, Martelli, Martellini, Martin,
  Martin, Martynov, Marx, Mason, Masserot, Massinger, Masso-Reid, Matichard,
  Matone, Mavalvala, Mazumder, Mazzolo, McCarthy, McClelland, McCormick,
  McGuire, McIntyre, McIver, McManus, McWilliams, Meacher, Meadors, Meidam,
  Melatos, Mendell, Mendoza-Gandara, Mercer, Merilh, Merzougui, Meshkov,
  Messenger, Messick, Meyers, Mezzani, Miao, Michel, Middleton, Mikhailov,
  Milano, Miller, Millhouse, Minenkov, Ming, Mirshekari, Mishra, Mitra,
  Mitrofanov, Mitselmakher, Mittleman, Moggi, Mohan, Mohapatra, Montani, Moore,
  Moore, Moraru, Moreno, Morriss, Mossavi, Mours, Mow-Lowry, Mueller, Mueller,
  Muir, Mukherjee, Mukherjee, Mukherjee, Mukund, Mullavey, Munch, Murphy,
  Murray, Mytidis, Nardecchia, Naticchioni, Nayak, Necula, Nedkova, Nelemans,
  Neri, Neunzert, Newton, Nguyen, Nielsen, Nissanke, Nitz, Nocera, Nolting,
  Normandin, Nuttall, Oberling, Ochsner, O'Dell, Oelker, Ogin, Oh, Oh, Ohme,
  Oliver, Oppermann, Oram, O'Reilly, O'Shaughnessy, Ott, Ottaway, Ottens,
  Overmier, Owen, Pai, Pai, Palamos, Palashov, Palomba, Pal-Singh, Pan, Pan,
  Pankow, Pannarale, Pant, Paoletti, Paoli, Papa, Paris, Parker, Pascucci,
  Pasqualetti, Passaquieti, Passuello, Patricelli, Patrick, Pearlstone,
  Pedraza, Pedurand, Pekowsky, Pele, Penn, Perreca, Pfeiffer, Phelps, Piccinni,
  Pichot, Pickenpack, Piergiovanni, Pierro, Pillant, Pinard, Pinto, Pitkin,
  Poeld, Poggiani, Popolizio, Post, Powell, Prasad, Predoi, Premachandra,
  Prestegard, Price, Prijatelj, Principe, Privitera, Prix, Prodi, Prokhorov,
  Puncken, Punturo, Puppo, P{\"{u}}rrer, Qi, Qin, Quetschke, Quintero,
  Quitzow-James, Raab, Rabeling, Radkins, Raffai, Raja, Rakhmanov, Ramet,
  Rapagnani, Raymond, Razzano, Re, Read, Reed, Regimbau, Rei, Reid, Reitze,
  Rew, Reyes, Ricci, Riles, Robertson, Robie, Robinet, Rocchi, Rolland,
  Rollins, Roma, Romano, Romano, Romanov, Romie, Rosi{\'{n}}ska, Rowan,
  R{\"{u}}diger, Ruggi, Ryan, Sachdev, Sadecki, Sadeghian, Salconi, Saleem,
  Salemi, Samajdar, Sammut, Sampson, Sanchez, Sandberg, Sandeen, Sanders,
  Sanders, Sassolas, Sathyaprakash, Saulson, Sauter, Savage, Sawadsky, Schale,
  Schilling, Schmidt, Schmidt, Schnabel, Schofield, Sch{\"{o}}nbeck, Schreiber,
  Schuette, Schutz, Scott, Scott, Sellers, Sengupta, Sentenac, Sequino,
  Sergeev, Serna, Setyawati, Sevigny, Shaddock, Shaffer, Shah, Shahriar,
  Shaltev, Shao, Shapiro, Shawhan, Sheperd, Shoemaker, Shoemaker, Siellez,
  Siemens, Sigg, Silva, Simakov, Singer, Singer, Singh, Singh, Singhal, Sintes,
  Slagmolen, Smith, Smith, Smith, Smith, Son, Sorazu, Sorrentino, Souradeep,
  Srivastava, Staley, Steinke, Steinlechner, Steinlechner, Steinmeyer,
  Stephens, Stevenson, Stone, Strain, Straniero, Stratta, Strauss, Strigin,
  Sturani, Stuver, Summerscales, Sun, Sutton, Swinkels, Szczepa{\'{n}}czyk,
  Tacca, Talukder, Tanner, T{\'{a}}pai, Tarabrin, Taracchini, Taylor, Theeg,
  Thirugnanasambandam, Thomas, Thomas, Thomas, Thorne, Thorne, Thrane, Tiwari,
  Tiwari, Tokmakov, Tomlinson, Tonelli, Torres, Torrie, T{\"{o}}yr{\"{a}},
  Travasso, Traylor, Trifir{\`{o}}, Tringali, Trozzo, Tse, Turconi, Tuyenbayev,
  Ugolini, Unnikrishnan, Urban, Usman, Vahlbruch, Vajente, Valdes, Vallisneri,
  {Van Bakel}, {Van Beuzekom}, {Van Den Brand}, {Van Den Broeck}, Vander-Hyde,
  {Van Der Schaaf}, {Van Heijningen}, {Van Veggel}, Vardaro, Vass,
  Vas{\'{u}}th, Vaulin, Vecchio, Vedovato, Veitch, Veitch, Venkateswara,
  Verkindt, Vetrano, Vicer{\'{e}}, Vinciguerra, Vine, Vinet, Vitale, Vo, Vocca,
  Vorvick, Voss, Vousden, Vyatchanin, Wade, Wade, Wade, Waldman, Walker,
  Wallace, Walsh, Wang, Wang, Wang, Wang, Wang, Ward, Ward, Warner, Was,
  Weaver, Wei, Weinert, Weinstein, Weiss, Welborn, Wen, We{\ss}els, Westphal,
  Wette, Whelan, Whitcomb, White, Whiting, Wiesner, Wilkinson, Willems,
  Williams, Williams, Williamson, Willis, Willke, Wimmer, Winkelmann, Winkler,
  Wipf, Wiseman, Wittel, Woan, Worden, Wright, Wu, Yablon, Yakushin, Yam,
  Yamamoto, Yancey, Yap, Yu, Yvert, Zadro{\.{z}}ny, Zangrando, Zanolin, Zendri,
  Zevin, Zhang, Zhang, Zhang, Zhang, Zhao, Zhou, Zhou, Zhu, Zucker, Zuraw, \&
  Zweizig}]{GW150914}
Abbott, B.~P., Abbott, R., Abbott, T.~D., {et~al.} 2016, Physical Review
  Letters, 116, 061102, \dodoi{10.1103/PhysRevLett.116.061102}

\bibitem[{Abbott {et~al.}(2018)Abbott, Abbott, Abbott, Abernathy, Acernese,
  Ackley, Adams, Adams, Addesso, Adhikari, Adya, Affeldt, Agathos, Agatsuma,
  Aggarwal, Aguiar, Aiello, Ain, Ajith, Akutsu, Allen, Allocca, Altin,
  Ananyeva, Anderson, Anderson, Ando, Appert, Arai, Araya, Araya, Areeda,
  Arnaud, Arun, Asada, Ascenzi, Ashton, Aso, Ast, Aston, Astone, Atsuta,
  Aufmuth, Aulbert, Avila-Alvarez, Awai, Babak, Bacon, Bader, Baiotti, Baker,
  Baldaccini, Ballardin, Ballmer, Barayoga, Barclay, Barish, Barker, Barone,
  Barr, Barsotti, Barsuglia, Barta, Bartlett, Barton, Bartos, Bassiri, Basti,
  Batch, Baune, Bavigadda, Bazzan, B{\'{e}}csy, Beer, Bejger, Belahcene,
  Belgin, Bell, Berger, Bergmann, Berry, Bersanetti, Bertolini, Betzwieser,
  Bhagwat, Bhandare, Bilenko, Billingsley, Billman, Birch, Birney, Birnholtz,
  Biscans, Bisht, Bitossi, Biwer, Bizouard, Blackburn, Blackman, Blair, Blair,
  Blair, Bloemen, Bock, Boer, Bogaert, Bohe, Bondu, Bonnand, Boom, Bork,
  Boschi, Bose, Bouffanais, Bozzi, Bradaschia, Brady, Braginsky, Branchesi,
  Brau, Briant, Brillet, Brinkmann, Brisson, Brockill, Broida, Brooks, Brown,
  Brown, Brown, Brunett, Buchanan, Buikema, Bulik, Bulten, Buonanno, Buskulic,
  Buy, Byer, Cabero, Cadonati, Cagnoli, Cahillane, {Calder{\'{o}}n Bustillo},
  Callister, Calloni, Camp, Cannon, Cao, Cao, Capano, Capocasa, Carbognani,
  Caride, {Casanueva Diaz}, Casentini, Caudill, Cavagli{\`{a}}, Cavalier,
  Cavalieri, Cella, Cepeda, {Cerboni Baiardi}, Cerretani, Cesarini, Chamberlin,
  Chan, Chao, Charlton, Chassande-Mottin, Cheeseboro, Chen, Chen, Cheng,
  Chincarini, Chiummo, Chmiel, Cho, Cho, Chow, Christensen, Chu, Chua, Chua,
  Chung, Ciani, Clara, Clark, Cleva, Cocchieri, Coccia, Cohadon, Colla,
  Collette, Cominsky, Constancio, Conti, Cooper, Corbitt, Cornish, Corsi,
  Cortese, Costa, Coughlin, Coughlin, Coulon, Countryman, Couvares, Covas,
  Cowan, Coward, Cowart, Coyne, Coyne, Creighton, Creighton, Cripe, Crowder,
  Cullen, Cumming, Cunningham, Cuoco, Canton, Danilishin, D'Antonio, Danzmann,
  Dasgupta, {Da Silva Costa}, Dattilo, Dave, Davier, Davies, Davis, Daw, Day,
  Day, De, DeBra, Debreczeni, Degallaix, {De Laurentis}, Del{\'{e}}glise, {Del
  Pozzo}, Denker, Dent, Dergachev, {De Rosa}, DeRosa, DeSalvo, Devine,
  Dhurandhar, D{\'{i}}az, Fiore, Giovanni, Girolamo, Lieto, Pace, Palma,
  Virgilio, Doctor, Doi, Dolique, Donovan, Dooley, Doravari, Dorrington,
  Douglas, {Dovale {\'{A}}lvarez}, Downes, Drago, Drever, Driggers, Du, Ducrot,
  Dwyer, Eda, Edo, Edwards, Effler, Eggenstein, Ehrens, Eichholz, Eikenberry,
  Eisenstein, Essick, Etienne, Etzel, Evans, Evans, Everett, Factourovich,
  Fafone, Fair, Fairhurst, Fan, Farinon, Farr, Farr, Fauchon-Jones, Favata,
  Fays, Fehrmann, Fejer, {Fern{\'{a}}ndez Galiana}, Ferrante, Ferreira,
  Ferrini, Fidecaro, Fiori, Fiorucci, Fisher, Flaminio, Fletcher, Fong,
  Forsyth, Fournier, Frasca, Frasconi, Frei, Freise, Frey, Frey, Fries,
  Fritschel, Frolov, Fujii, Fujimoto, Fulda, Fyffe, Gabbard, Gadre, Gaebel,
  Gair, Gammaitoni, Gaonkar, Garufi, Gaur, Gayathri, Gehrels, Gemme, Genin,
  Gennai, George, Gergely, Germain, Ghonge, Ghosh, Ghosh, Ghosh, Giaime,
  Giardina, Giazotto, Gill, Glaefke, Goetz, Goetz, Gondan, Gonz{\'{a}}lez,
  {Gonzalez Castro}, Gopakumar, Gorodetsky, Gossan, Gosselin, Gouaty, Grado,
  Graef, Granata, Grant, Gras, Gray, Greco, Green, Groot, Grote, Grunewald,
  Guidi, Guo, Gupta, Gupta, Gushwa, Gustafson, Gustafson, Hacker, Hagiwara,
  Hall, Hall, Hammond, Haney, Hanke, Hanks, Hanna, Hannam, Hanson, Hardwick,
  Harms, Harry, Harry, Hart, Hartman, Haster, Haughian, Hayama, Healy,
  Heidmann, Heintze, Heitmann, Hello, Hemming, Hendry, Heng, Hennig, Henry,
  Heptonstall, Heurs, Hild, Hirose, Hoak, Hofman, Holt, Holz, Hopkins, Hough,
  Houston, Howell, Hu, Huerta, Huet, Hughey, Husa, Huttner, Huynh-Dinh, Indik,
  Ingram, Inta, Ioka, Isa, Isac, Isi, Isogai, Itoh, Iyer, Izumi, Jacqmin, Jani,
  Jaranowski, Jawahar, Jim{\'{e}}nez-Forteza, Johnson, Jones, Jones, Jonker,
  Ju, Junker, Kagawa, Kajita, Kakizaki, Kalaghatgi, Kalogera, Kamiizumi, Kanda,
  Kandhasamy, Kanemura, Kaneyama, Kang, Kanner, Karki, Karvinen, Kasprzack,
  Kataoka, Katsavounidis, Katzman, Kaufer, Kaur, Kawabe, Kawai, Kawamura,
  K{\'{e}}f{\'{e}}lian, Keitel, Kelley, Kennedy, Key, Khalili, Khan, Khan,
  Khan, Khazanov, Kijbunchoo, Kim, Kim, Kim, Kim, Kim, Kim, Kimbrell, Kimura,
  King, King, Kirchhoff, Kissel, Klein, Kleybolte, Klimenko, Koch, Koehlenbeck,
  Kojima, Kokeyama, Koley, Komori, Kondrashov, Kontos, Korobko, Korth, Kotake,
  Kowalska, Kozak, Kr{\"{a}}mer, Kringel, Krishnan, Kr{\'{o}}lak, Kuehn, Kumar,
  Kumar, Kumar, Kuo, Kuroda, Kutynia, Kuwahara, Lackey, Landry, Lang, Lange,
  Lantz, Lanza, Lartaux-Vollard, Lasky, Laxen, Lazzarini, Lazzaro, Leaci,
  Leavey, Lebigot, Lee, Lee, Lee, Lee, Lee, Lehmann, Lenon, Leonardi, Leong,
  Leroy, Letendre, Levin, Li, Libson, Littenberg, Liu, Lockerbie, Lombardi,
  London, Lord, Lorenzini, Loriette, Lormand, Losurdo, Lough, Lousto, Lovelace,
  L{\"{u}}ck, Lundgren, Lynch, Ma, Macfoy, Machenschalk, MacInnis, Macleod,
  Maga{\~{n}}a-Sandoval, Majorana, Maksimovic, Malvezzi, Man, Mandic, Mangano,
  Mano, Mansell, Manske, Mantovani, Marchesoni, Marchio, Marion, M{\'{a}}rka,
  M{\'{a}}rka, Markosyan, Maros, Martelli, Martellini, Martin, Martynov, Mason,
  Masserot, Massinger, Masso-Reid, Mastrogiovanni, Matichard, Matone,
  Matsumoto, Matsushima, Mavalvala, Mazumder, McCarthy, McClelland, McCormick,
  McGrath, McGuire, McIntyre, McIver, McManus, McRae, McWilliams, Meacher,
  Meadors, Meidam, Melatos, Mendell, Mendoza-Gandara, Mercer, Merilh,
  Merzougui, Meshkov, Messenger, Messick, Metzdorff, Meyers, Mezzani, Miao,
  Michel, Michimura, Middleton, Mikhailov, Milano, Miller, Miller, Miller,
  Miller, Millhouse, Minenkov, Ming, Mirshekari, Mishra, Mitrofanov,
  Mitselmakher, Mittleman, Miyakawa, Miyamoto, Miyamoto, Miyoki, Moggi, Mohan,
  Mohapatra, Montani, Moore, Moore, Moraru, Moreno, Morii, Morisaki, Moriwaki,
  Morriss, Mours, Mow-Lowry, Mueller, Muir, Mukherjee, Mukherjee, Mukherjee,
  Mukund, Mullavey, Munch, Muniz, Murray, Mytidis, Nagano, Nakamura, Nakamura,
  Nakano, Nakano, Nakano, Nakao, Napier, Nardecchia, Narikawa, Naticchioni,
  Nelemans, Nelson, Neri, Nery, Neunzert, Newport, Newton, Nguyen, Ni, Nielsen,
  Nissanke, Nitz, Noack, Nocera, Nolting, Normandin, Nuttall, Oberling,
  Ochsner, Oelker, Ogin, Oh, Oh, Ohashi, Ohishi, Ohkawa, Ohme, Okutomi, Oliver,
  Ono, Ono, Oohara, Oppermann, Oram, O'Reilly, O'Shaughnessy, Ottaway,
  Overmier, Owen, Pace, Page, Pai, Pai, Palamos, Palashov, Palomba, Pal-Singh,
  Pan, Pankow, Pannarale, Pant, Paoletti, Paoli, Papa, Paris, Parker, Pascucci,
  Pasqualetti, Passaquieti, Passuello, Patricelli, Pearlstone, Pedraza,
  Pedurand, Pekowsky, Pele, {Pe{\~{n}}a Arellano}, Penn, Perez, Perreca, Perri,
  Pfeiffer, Phelps, Piccinni, Pichot, Piergiovanni, Pierro, Pillant, Pinard,
  Pinto, Pitkin, Poe, Poggiani, Popolizio, Post, Powell, Prasad, Pratt, Predoi,
  Prestegard, Prijatelj, Principe, Privitera, Prodi, Prokhorov, Puncken,
  Punturo, Puppo, P{\"{u}}rrer, Qi, Qin, Qiu, Quetschke, Quintero,
  Quitzow-James, Raab, Rabeling, Radkins, Raffai, Raja, Rajan, Rakhmanov,
  Rapagnani, Raymond, Razzano, Re, Read, Regimbau, Rei, Reid, Reitze, Rew,
  Reyes, Rhoades, Ricci, Riles, Rizzo, Robertson, Robie, Robinet, Rocchi,
  Rolland, Rollins, Roma, Romano, Romie, Rosi{\'{n}}ska, Rowan, R{\"{u}}diger,
  Ruggi, Ryan, Sachdev, Sadecki, Sadeghian, Sago, Saijo, Saito, Sakai,
  Sakellariadou, Salconi, Saleem, Salemi, Samajdar, Sammut, Sampson, Sanchez,
  Sandberg, Sanders, Sasaki, Sassolas, Sathyaprakash, Sato, Sato, Saulson,
  Sauter, Savage, Sawadsky, Schale, Scheuer, Schmidt, Schmidt, Schmidt,
  Schnabel, Schofield, Sch{\"{o}}nbeck, Schreiber, Schuette, Schutz, Schwalbe,
  Scott, Scott, Sekiguchi, Sekiguchi, Sellers, Sengupta, Sentenac, Sequino,
  Sergeev, Setyawati, Shaddock, Shaffer, Shahriar, Shapiro, Shawhan, Sheperd,
  Shibata, Shikano, Shimoda, Shoda, Shoemaker, Shoemaker, Siellez, Siemens,
  Sieniawska, Sigg, Silva, Singer, Singer, Singh, Singh, Singhal, Sintes,
  Slagmolen, Smith, Smith, Smith, Somiya, Son, Sorazu, Sorrentino, Souradeep,
  Spencer, Srivastava, Staley, Steinke, Steinlechner, Steinlechner, Steinmeyer,
  Stephens, Stevenson, Stone, Strain, Straniero, Stratta, Strigin, Sturani,
  Stuver, Sugimoto, Summerscales, Sun, Sunil, Sutton, Suzuki, Swinkels,
  Szczepa{\'{n}}czyk, Tacca, Tagoshi, Takada, Takahashi, Takahashi, Takamori,
  Talukder, Tanaka, Tanaka, Tanaka, Tanner, T{\'{a}}pai, Taracchini, Tatsumi,
  Taylor, Telada, Theeg, Thomas, Thomas, Thomas, Thorne, Thrane, Tippens,
  Tiwari, Tiwari, Tokmakov, Toland, Tomaru, Tomlinson, Tonelli, Tornasi,
  Torrie, T{\"{o}}yr{\"{a}}, Travasso, Traylor, Trifir{\`{o}}, Trinastic,
  Tringali, Trozzo, Tse, Tso, Tsubono, Tsuzuki, Turconi, Tuyenbayev, Uchiyama,
  Uehara, Ueki, Ueno, Ugolini, Unnikrishnan, Urban, Ushiba, Usman, Vahlbruch,
  Vajente, Valdes, van Bakel, van Beuzekom, van~den Brand, {Van Den Broeck},
  Vander-Hyde, van~der Schaaf, van Heijningen, van Putten, van Veggel, Vardaro,
  Varma, Vass, Vas{\'{u}}th, Vecchio, Vedovato, Veitch, Veitch, Venkateswara,
  Venugopalan, Verkindt, Vetrano, Vicer{\'{e}}, Viets, Vinciguerra, Vine,
  Vinet, Vitale, Vo, Vocca, Vorvick, Voss, Vousden, Vyatchanin, Wade, Wade,
  Wade, Wakamatsu, Walker, Wallace, Walsh, Wang, Wang, Wang, Wang, Ward,
  Warner, Was, Watchi, Weaver, Wei, Weinert, Weinstein, Weiss, Wen, We{\ss}els,
  Westphal, Wette, Whelan, Whiting, Whittle, Williams, Williams, Williamson,
  Willis, Willke, Wimmer, Winkler, Wipf, Wittel, Woan, Woehler, Worden, Wright,
  Wu, Wu, Yam, Yamamoto, Yamamoto, Yamamoto, Yancey, Yano, Yap, Yokoyama,
  Yokozawa, Yoon, Yu, Yu, Yuzurihara, Yvert, Zadro{\.{z}}ny, Zangrando,
  Zanolin, Zeidler, Zendri, Zevin, Zhang, Zhang, Zhang, Zhang, Zhao, Zhou,
  Zhou, Zhu, Zhu, Zucker, \& Zweizig}]{LVC_ObservingScenarios}
---. 2018, Living Reviews in Relativity, 21, 3,
  \dodoi{10.1007/s41114-018-0012-9}

\bibitem[{Abbott {et~al.}(2019{\natexlab{a}})Abbott, Abbott, Abbott, Abraham,
  Acernese, Ackley, Adams, Adhikari, Adya, Affeldt, Agathos, Agatsuma,
  Aggarwal, Aguiar, Aiello, Ain, Ajith, Allen, Allocca, Aloy, Altin, Amato,
  Anand, Ananyeva, Anderson, Anderson, Angelova, Antier, Appert, Arai, Araya,
  Areeda, Ar{\`{e}}ne, Arnaud, Aronson, Arun, Ascenzi, Ashton, Aston, Astone,
  Aubin, Aufmuth, AultONeal, Austin, Avendano, Avila-Alvarez, Babak, Bacon,
  Badaracco, Bader, Bae, Baird, Baker, Baldaccini, Ballardin, Ballmer, Bals,
  Banagiri, Barayoga, Barbieri, Barclay, Barish, Barker, Barkett, Barnum,
  Barone, Barr, Barsotti, Barsuglia, Barta, Bartlett, Bartos, Bassiri, Basti,
  Bawaj, Bayley, Bazzan, B{\'{e}}csy, Bejger, Belahcene, Bell, Beniwal,
  Benjamin, Bergmann, Bernuzzi, Berry, Bersanetti, Bertolini, Betzwieser,
  Bhandare, Bidler, Biggs, Bilenko, Bilgili, Billingsley, Birney, Birnholtz,
  Biscans, Bischi, Biscoveanu, Bisht, Bitossi, Bizouard, Blackburn, Blackman,
  Blair, Blair, Blair, Bloemen, Bobba, Bode, Boer, Boetzel, Bogaert, Bondu,
  Bonnand, Booker, Boom, Bork, Boschi, Bose, Bossilkov, Bosveld, Bouffanais,
  Bozzi, Bradaschia, Brady, Bramley, Branchesi, Brau, Breschi, Briant, Briggs,
  Brighenti, Brillet, Brinkmann, Brockill, Brooks, Brooks, Brown, Brunett,
  Buikema, Bulik, Bulten, Buonanno, Buskulic, Buy, Byer, Cabero, Cadonati,
  Cagnoli, Cahillane, Bustillo, Callister, Calloni, Camp, Campbell, Canepa,
  Cannon, Cao, Cao, Carapella, Carbognani, Caride, Carney, Carullo, Diaz,
  Casentini, Caudill, Cavagli{\`{a}}, Cavalier, Cavalieri, Cella,
  Cerd{\'{a}}-Dur{\'{a}}n, Cesarini, Chaibi, Chakravarti, Chamberlin, Chan,
  Chao, Charlton, Chase, Chassande-Mottin, Chatterjee, Chaturvedi, Cheeseboro,
  Chen, Chen, Chen, Cheng, Cheong, Chia, Chiadini, Chincarini, Chiummo, Cho,
  Cho, Cho, Christensen, Chu, Chua, Chung, Chung, Ciani, Cie{\'{s}}lar,
  Ciobanu, Ciolfi, Cipriano, Cirone, Clara, Clark, Clearwater, Cleva, Coccia,
  Cohadon, Cohen, Colleoni, Collette, Collins, Colpi, Cominsky, {Constancio
  Jr.}, Conti, Cooper, Corban, Corbitt, Cordero-Carri{\'{o}}n, Corezzi, Corley,
  Cornish, Corre, Corsi, Cortese, Costa, Cotesta, Coughlin, Coughlin, Coulon,
  Countryman, Couvares, Covas, Cowan, Coward, Cowart, Coyne, Coyne, Creighton,
  Creighton, Cripe, Croquette, Crowder, Cullen, Cumming, Cunningham, Cuoco,
  Canton, D{\'{a}}lya, D'Angelo, Danilishin, D'Antonio, Danzmann, Dasgupta,
  Costa, Datrier, Dattilo, Dave, Davier, Davis, Daw, DeBra, Deenadayalan,
  Degallaix, Laurentis, Del{\'{e}}glise, Pozzo, DeMarchi, Demos, Dent, Pietri,
  Rosa, Rossi, DeSalvo, de~Varona, Dhurandhar, D{\'{i}}az, Dietrich, Fiore,
  DiFronzo, Giorgio, Giovanni, Giovanni, Girolamo, Lieto, Ding, Pace, Palma,
  Renzo, Divakarla, Dmitriev, Doctor, Donovan, Dooley, Doravari, Dorrington,
  Downes, Drago, Driggers, Du, Ducoin, Dupej, Durante, Dwyer, Easter, Eddolls,
  Edo, Effler, Ehrens, Eichholz, Eikenberry, Eisenmann, Eisenstein, Errico,
  Essick, Estelles, Estevez, Etienne, Etzel, Evans, Evans, Fafone, Fairhurst,
  Fan, Farinon, Farr, Farr, Fauchon-Jones, Favata, Fays, Fazio, Fee, Feicht,
  Fejer, Feng, Fernandez-Galiana, Ferrante, Ferreira, Ferreira, Fidecaro,
  Fiori, Fiorucci, Fishbach, Fisher, Fishner, Fittipaldi, Fitz-Axen, Fiumara,
  Flaminio, Fletcher, Floden, Flynn, Fong, Font, Forsyth, Fournier, Vivanco,
  Frasca, Frasconi, Frei, Freise, Frey, Frey, Fritschel, Frolov, Fronz{\`{e}},
  Fulda, Fyffe, Gabbard, Gadre, Gaebel, Gair, Gammaitoni, Gaonkar,
  Garc{\'{i}}a-Quir{\'{o}}s, Garufi, Gateley, Gaudio, Gaur, Gayathri, Gemme,
  Genin, Gennai, George, George, Gergely, Ghonge, Ghosh, Ghosh, Ghosh,
  Giacomazzo, Giaime, Giardina, Gibson, Gill, Glover, Gniesmer, Godwin, Goetz,
  Goetz, Goncharov, Gonz{\'{a}}lez, Castro, Gopakumar, Gossan, Gosselin,
  Gouaty, Grace, Grado, Granata, Grant, Gras, Grassia, Gray, Gray, Greco,
  Green, Green, Gretarsson, Grimaldi, Grimm, Groot, Grote, Grunewald, Gruning,
  Guidi, Gulati, Guo, Gupta, Gupta, Gupta, Gustafson, Gustafson, Haegel, Halim,
  Hall, Hall, Hamilton, Hammond, Haney, Hanke, Hanks, Hanna, Hannam,
  Hannuksela, Hansen, Hanson, Harder, Hardwick, Haris, Harms, Harry, Harry,
  Hasskew, Haster, Haughian, Hayes, Healy, Heidmann, Heintze, Heitmann,
  Hellman, Hello, Hemming, Hendry, Heng, Hennig, Heurs, Hild, Hinderer,
  Hochheim, Hofman, Holgado, Holland, Holt, Holz, Hopkins, Horst, Hough,
  Howell, Hoy, Huang, H{\"{u}}bner, Huerta, Huet, Hughey, Hui, Husa, Huttner,
  Huynh-Dinh, Idzkowski, Iess, Inchauspe, Ingram, Inta, Intini, Irwin, Isa,
  Isac, Isi, Iyer, Jacqmin, Jadhav, Jani, Janthalur, Jaranowski, Jariwala,
  Jenkins, Jiang, Johnson, Jones, Jones, Jones, Jones, Jonker, Ju, Junker,
  Kalaghatgi, Kalogera, Kamai, Kandhasamy, Kang, Kanner, Kapadia, Karki,
  Kashyap, Kasprzack, Katsanevas, Katsavounidis, Katzman, Kaufer, Kawabe,
  Keerthana, K{\'{e}}f{\'{e}}lian, Keitel, Kennedy, Key, Khalili, Khan, Khan,
  Khazanov, Khetan, Khursheed, Kijbunchoo, Kim, Kim, Kim, Kim, Kim, Kim,
  Kimball, King, Kinley-Hanlon, Kirchhoff, Kissel, Kleybolte, Klika, Klimenko,
  Knowles, Koch, Koehlenbeck, Koekoek, Koley, Kondrashov, Kontos, Koper,
  Korobko, Korth, Kovalam, Kozak, Kr{\"{a}}mer, Kringel, Krishnendu,
  Kr{\'{o}}lak, Krupinski, Kuehn, Kumar, Kumar, Kumar, Kumar, Kuo, Kutynia,
  Kwang, Lackey, Laghi, Lai, Lam, Landry, Lane, Lang, Lange, Lantz, Lanza,
  Lartaux-Vollard, Lasky, Laxen, Lazzarini, Lazzaro, Leaci, Leavey, Lecoeuche,
  Lee, Lee, Lee, Lee, Lee, Lee, Lehmann, Lenon, Leroy, Letendre, Levin, Li, Li,
  Li, Li, Li, Lin, Linde, Linker, Littenberg, Liu, Liu, Llorens-Monteagudo, Lo,
  London, Longo, Lorenzini, Loriette, Lormand, Losurdo, Lough, Lousto,
  Lovelace, Lower, L{\"{u}}ck, Lumaca, Lundgren, Lynch, Ma, Macas, Macfoy,
  MacInnis, Macleod, Macquet, Hernandez, Maga{\~{n}}a-Sandoval, Magee,
  Majorana, Maksimovic, Malik, Man, Mandic, Mangano, Mansell, Manske,
  Mantovani, Mapelli, Marchesoni, Marion, M{\'{a}}rka, M{\'{a}}rka, Markakis,
  Markosyan, Markowitz, Maros, Marquina, Marsat, Martelli, Martin, Martin,
  Martinez, Martynov, Masalehdan, Mason, Massera, Masserot, Massinger,
  Masso-Reid, Mastrogiovanni, Matas, Matichard, Matone, Mavalvala, McCann,
  McCarthy, McClelland, McCormick, McCuller, McGuire, McIsaac, McIver, McManus,
  McRae, McWilliams, Meacher, Meadors, Mehmet, Mehta, Meidam, Villa, Melatos,
  Mendell, Mercer, Mereni, Merfeld, Merilh, Merzougui, Meshkov, Messenger,
  Messick, Messina, Metzdorff, Meyers, Meylahn, Miani, Miao, Michel, Middleton,
  Milano, Miller, Millhouse, Mills, Milovich-Goff, Minazzoli, Minenkov,
  Mishkin, Mishra, Mistry, Mitra, Mitrofanov, Mitselmakher, Mittleman, Mo,
  Moffa, Mogushi, Mohapatra, Molina-Ruiz, Mondin, Montani, Moore, Moraru,
  Morawski, Moreno, Morisaki, Mours, Mow-Lowry, Muciaccia, Mukherjee,
  Mukherjee, Mukherjee, Mukherjee, Mukund, Mullavey, Munch, Mu{\~{n}}iz,
  Muratore, Murray, Nardecchia, Naticchioni, Nayak, Neil, Neilson, Nelemans,
  Nelson, Nery, Neunzert, Nevin, Ng, Ng, Nguyen, Nguyen, Nichols, Nichols,
  Nissanke, Nocera, North, Nuttall, Obergaulinger, Oberling, O'Brien,
  Oganesyan, Ogin, Oh, Oh, Ohme, Ohta, Okada, Oliver, Oppermann, Oram,
  O'Reilly, Ormiston, Ortega, O'Shaughnessy, Ossokine, Ottaway, Overmier, Owen,
  Pace, Pagano, Page, Pagliaroli, Pai, Pai, Palamos, Palashov, Palomba, Pan,
  Panda, Pang, Pankow, Pannarale, Pant, Paoletti, Paoli, Parida, Parker,
  Pascucci, Pasqualetti, Passaquieti, Passuello, Patil, Patricelli, Payne,
  Pearlstone, Pechsiri, Pedersen, Pedraza, Pedurand, Pele, Penn, Perego, Perez,
  P{\'{e}}rigois, Perreca, Petermann, Pfeiffer, Phelps, Phukon, Piccinni,
  Pichot, Piergiovanni, Pierro, Pillant, Pinard, Pinto, Pirello, Pitkin,
  Plastino, Poggiani, Pong, Ponrathnam, Popolizio, Porter, Powell, Prajapati,
  Prasad, Prasai, Prasanna, Pratten, Prestegard, Principe, Prodi, Prokhorov,
  Punturo, Puppo, P{\"{u}}rrer, Qi, Quetschke, Quinonez, Raab, Raaijmakers,
  Radkins, Radulesco, Raffai, Raja, Rajan, Rajbhandari, Rakhmanov, Ramirez,
  Ramos-Buades, Rana, Rao, Rapagnani, Raymond, Razzano, Read, Regimbau, Rei,
  Reid, Reitze, Rettegno, Ricci, Richardson, Richardson, Ricker,
  Riemenschneider, Riles, Rizzo, Robertson, Robinet, Rocchi, Rolland, Rollins,
  Roma, Romanelli, Romano, Romel, Romie, Rose, Rose, Rose, Rosi{\'{n}}ska,
  Rosofsky, Ross, Rowan, R{\"{u}}diger, Ruggi, Rutins, Ryan, Sachdev, Sadecki,
  Sakellariadou, Salafia, Salconi, Saleem, Samajdar, Sammut, Sanchez, Sanchez,
  Sanchis-Gual, Sanders, Santiago, Santos, Sarin, Sassolas, Sauter, Savage,
  Schale, Scheel, Scheuer, Schmidt, Schnabel, Schofield, Sch{\"{o}}nbeck,
  Schreiber, Schulte, Schutz, Scott, Scott, Seidel, Sellers, Sengupta, Sennett,
  Sentenac, Sequino, Sergeev, Setyawati, Shaddock, Shaffer, Shahriar, Shaner,
  Sharma, Sharma, Shawhan, Shen, Shink, Shoemaker, Shoemaker, Shukla,
  ShyamSundar, Siellez, Sieniawska, Sigg, Singer, Singh, Singh, Singhal,
  Sintes, Sitmukhambetov, Skliris, Slagmolen, Slaven-Blair, Smith, Smith,
  Somala, Son, Soni, Sorazu, Sorrentino, Souradeep, Sowell, Spencer, Spera,
  Srivastava, Srivastava, Staats, Stachie, Standke, Steer, Steinke,
  Steinlechner, Steinlechner, Steinmeyer, Stevenson, Stocks, Stone, Stops,
  Strain, Stratta, Strigin, Strunk, Sturani, Stuver, Sudhir, Summerscales, Sun,
  Sunil, Sur, Suresh, Sutton, Swinkels, Szczepa{\'{n}}czyk, Tacca, Tait,
  Talbot, Tanner, Tao, T{\'{a}}pai, Tapia, Tasson, Taylor, Tenorio, Terkowski,
  Thomas, Thomas, Thondapu, Thorne, Thrane, Tiwari, Tiwari, Tiwari, Toland,
  Tonelli, Tornasi, Torres-Forn{\'{e}}, Torrie, T{\"{o}}yr{\"{a}}, Travasso,
  Traylor, Tringali, Tripathee, Trovato, Trozzo, Tsang, Tse, Tso, Tsukada,
  Tsuna, Tsutsui, Tuyenbayev, Ueno, Ugolini, Unnikrishnan, Urban, Usman,
  Vahlbruch, Vajente, Valdes, Valentini, van Bakel, van Beuzekom, van~den
  Brand, Broeck, Vander-Hyde, van~der Schaaf, VanHeijningen, van Veggel,
  Vardaro, Varma, Vass, Vas{\'{u}}th, Vecchio, Vedovato, Veitch, Veitch,
  Venkateswara, Venugopalan, Verkindt, Vetrano, Vicer{\'{e}}, Viets,
  Vinciguerra, Vine, Vinet, Vitale, Vo, Vocca, Vorvick, Vyatchanin, Wade, Wade,
  Wade, Walet, Walker, Wallace, Walsh, Wang, Wang, Wang, Wang, Wang, Ward,
  Warden, Warner, Was, Watchi, Weaver, Wei, Weinert, Weinstein, Weiss,
  Wellmann, Wen, Wessel, We{\ss}els, Westhouse, Wette, Whelan, Whiting,
  Whittle, Wilken, Williams, Williamson, Willis, Willke, Winkler, Wipf, Wittel,
  Woan, Woehler, Wofford, Wright, Wu, Wysocki, Xiao, Xu, Yamamoto, Yancey,
  Yang, Yang, Yang, Yap, Yazback, Yeeles, Yu, Yu, Yuen, Zadro{\.{z}}ny,
  Zadro{\.{z}}ny, Zanolin, Zelenova, Zendri, Zevin, Zhang, Zhang, Zhang, Zhao,
  Zhao, Zhou, Zhou, Zhu, Zucker, Zweizig, \& Salemi}]{Abbott2019a}
---. 2019{\natexlab{a}}, The Astrophysical Journal, 883, 149,
  \dodoi{10.3847/1538-4357/ab3c2d}

\bibitem[{Abbott {et~al.}(2019{\natexlab{b}})Abbott, Abbott, Abbott, Abernathy,
  Acernese, Ackley, Adams, Adams, Addesso, Adhikari, Adya, Affeldt, Agathos,
  Agatsuma, Aggarwal, Aguiar, Aiello, Ain, Ajith, Akutsu, Allen, Allocca,
  Altin, Ananyeva, Anderson, Anderson, Ando, Appert, Arai, Araya, Araya,
  Areeda, Arnaud, Arun, Asada, Ascenzi, Ashton, Aso, Ast, Aston, Astone,
  Atsuta, Aufmuth, Aulbert, Avila-Alvarez, Awai, Babak, Bacon, Bader, Baiotti,
  Baker, Baldaccini, Ballardin, Ballmer, Barayoga, Barclay, Barish, Barker,
  Barone, Barr, Barsotti, Barsuglia, Barta, Bartlett, Barton, Bartos, Bassiri,
  Basti, Batch, Baune, Bavigadda, Bazzan, B{\'{e}}csy, Beer, Bejger, Belahcene,
  Belgin, Bell, Berger, Bergmann, Berry, Bersanetti, Bertolini, Betzwieser,
  Bhagwat, Bhandare, Bilenko, Billingsley, Billman, Birch, Birney, Birnholtz,
  Biscans, Bisht, Bitossi, Biwer, Bizouard, Blackburn, Blackman, Blair, Blair,
  Blair, Bloemen, Bock, Boer, Bogaert, Bohe, Bondu, Bonnand, Boom, Bork,
  Boschi, Bose, Bouffanais, Bozzi, Bradaschia, Brady, Braginsky, Branchesi,
  Brau, Briant, Brillet, Brinkmann, Brisson, Brockill, Broida, Brooks, Brown,
  Brown, Brown, Brunett, Buchanan, Buikema, Bulik, Bulten, Buonanno, Buskulic,
  Buy, Byer, Cabero, Cadonati, Cagnoli, Cahillane, {Calder{\'{o}}n Bustillo},
  Callister, Calloni, Camp, Cannon, Cao, Cao, Capano, Capocasa, Carbognani,
  Caride, {Casanueva Diaz}, Casentini, Caudill, Cavagli{\`{a}}, Cavalier,
  Cavalieri, Cella, Cepeda, {Cerboni Baiardi}, Cerretani, Cesarini, Chamberlin,
  Chan, Chao, Charlton, Chassande-Mottin, Cheeseboro, Chen, Chen, Cheng,
  Chincarini, Chiummo, Chmiel, Cho, Cho, Chow, Christensen, Chu, Chua, Chua,
  Chung, Ciani, Clara, Clark, Cleva, Cocchieri, Coccia, Cohadon, Colla,
  Collette, Cominsky, Constancio, Conti, Cooper, Corbitt, Cornish, Corsi,
  Cortese, Costa, Coughlin, Coughlin, Coulon, Countryman, Couvares, Covas,
  Cowan, Coward, Cowart, Coyne, Coyne, Creighton, Creighton, Cripe, Crowder,
  Cullen, Cumming, Cunningham, Cuoco, Canton, Danilishin, D'Antonio, Danzmann,
  Dasgupta, {Da Silva Costa}, Dattilo, Dave, Davier, Davies, Davis, Daw, Day,
  Day, De, DeBra, Debreczeni, Degallaix, {De Laurentis}, Del{\'{e}}glise, {Del
  Pozzo}, Denker, Dent, Dergachev, {De Rosa}, DeRosa, DeSalvo, Devine,
  Dhurandhar, D{\'{i}}az, Fiore, Giovanni, Girolamo, Lieto, Pace, Palma,
  Virgilio, Doctor, Doi, Dolique, Donovan, Dooley, Doravari, Dorrington,
  Douglas, {Dovale {\'{A}}lvarez}, Downes, Drago, Drever, Driggers, Du, Ducrot,
  Dwyer, Eda, Edo, Edwards, Effler, Eggenstein, Ehrens, Eichholz, Eikenberry,
  Eisenstein, Essick, Etienne, Etzel, Evans, Evans, Everett, Factourovich,
  Fafone, Fair, Fairhurst, Fan, Farinon, Farr, Farr, Fauchon-Jones, Favata,
  Fays, Fehrmann, Fejer, {Fern{\'{a}}ndez Galiana}, Ferrante, Ferreira,
  Ferrini, Fidecaro, Fiori, Fiorucci, Fisher, Flaminio, Fletcher, Fong,
  Forsyth, Fournier, Frasca, Frasconi, Frei, Freise, Frey, Frey, Fries,
  Fritschel, Frolov, Fujii, Fujimoto, Fulda, Fyffe, Gabbard, Gadre, Gaebel,
  Gair, Gammaitoni, Gaonkar, Garufi, Gaur, Gayathri, Gehrels, Gemme, Genin,
  Gennai, George, Gergely, Germain, Ghonge, Ghosh, Ghosh, Ghosh, Giaime,
  Giardina, Giazotto, Gill, Glaefke, Goetz, Goetz, Gondan, Gonz{\'{a}}lez,
  {Gonzalez Castro}, Gopakumar, Gorodetsky, Gossan, Gosselin, Gouaty, Grado,
  Graef, Granata, Grant, Gras, Gray, Greco, Green, Groot, Grote, Grunewald,
  Guidi, Guo, Gupta, Gupta, Gushwa, Gustafson, Gustafson, Hacker, Hagiwara,
  Hall, Hall, Hammond, Haney, Hanke, Hanks, Hanna, Hannam, Hanson, Hardwick,
  Harms, Harry, Harry, Hart, Hartman, Haster, Haughian, Hayama, Healy,
  Heidmann, Heintze, Heitmann, Hello, Hemming, Hendry, Heng, Hennig, Henry,
  Heptonstall, Heurs, Hild, Hirose, Hoak, Hofman, Holt, Holz, Hopkins, Hough,
  Houston, Howell, Hu, Huerta, Huet, Hughey, Husa, Huttner, Huynh-Dinh, Indik,
  Ingram, Inta, Ioka, Isa, Isac, Isi, Isogai, Itoh, Iyer, Izumi, Jacqmin, Jani,
  Jaranowski, Jawahar, Jim{\'{e}}nez-Forteza, Johnson, Jones, Jones, Jonker,
  Ju, Junker, Kagawa, Kajita, Kakizaki, Kalaghatgi, Kalogera, Kamiizumi, Kanda,
  Kandhasamy, Kanemura, Kaneyama, Kang, Kanner, Karki, Karvinen, Kasprzack,
  Kataoka, Katsavounidis, Katzman, Kaufer, Kaur, Kawabe, Kawai, Kawamura,
  K{\'{e}}f{\'{e}}lian, Keitel, Kelley, Kennedy, Key, Khalili, Khan, Khan,
  Khan, Khazanov, Kijbunchoo, Kim, Kim, Kim, Kim, Kim, Kim, Kimbrell, Kimura,
  King, King, Kirchhoff, Kissel, Klein, Kleybolte, Klimenko, Koch, Koehlenbeck,
  Kojima, Kokeyama, Koley, Komori, Kondrashov, Kontos, Korobko, Korth, Kotake,
  Kowalska, Kozak, Kr{\"{a}}mer, Kringel, Krishnan, Kr{\'{o}}lak, Kuehn, Kumar,
  Kumar, Kumar, Kuo, Kuroda, Kutynia, Kuwahara, Lackey, Landry, Lang, Lange,
  Lantz, Lanza, Lartaux-Vollard, Lasky, Laxen, Lazzarini, Lazzaro, Leaci,
  Leavey, Lebigot, Lee, Lee, Lee, Lee, Lee, Lehmann, Lenon, Leonardi, Leong,
  Leroy, Letendre, Levin, Li, Libson, Littenberg, Liu, Lockerbie, Lombardi,
  London, Lord, Lorenzini, Loriette, Lormand, Losurdo, Lough, Lousto, Lovelace,
  L{\"{u}}ck, Lundgren, Lynch, Ma, Macfoy, Machenschalk, MacInnis, Macleod,
  Maga{\~{n}}a-Sandoval, Majorana, Maksimovic, Malvezzi, Man, Mandic, Mangano,
  Mano, Mansell, Manske, Mantovani, Marchesoni, Marchio, Marion, M{\'{a}}rka,
  M{\'{a}}rka, Markosyan, Maros, Martelli, Martellini, Martin, Martynov, Mason,
  Masserot, Massinger, Masso-Reid, Mastrogiovanni, Matichard, Matone,
  Matsumoto, Matsushima, Mavalvala, Mazumder, McCarthy, McClelland, McCormick,
  McGrath, McGuire, McIntyre, McIver, McManus, McRae, McWilliams, Meacher,
  Meadors, Meidam, Melatos, Mendell, Mendoza-Gandara, Mercer, Merilh,
  Merzougui, Meshkov, Messenger, Messick, Metzdorff, Meyers, Mezzani, Miao,
  Michel, Michimura, Middleton, Mikhailov, Milano, Miller, Miller, Miller,
  Miller, Millhouse, Minenkov, Ming, Mirshekari, Mishra, Mitrofanov,
  Mitselmakher, Mittleman, Miyakawa, Miyamoto, Miyamoto, Miyoki, Moggi, Mohan,
  Mohapatra, Montani, Moore, Moore, Moraru, Moreno, Morii, Morisaki, Moriwaki,
  Morriss, Mours, Mow-Lowry, Mueller, Muir, Mukherjee, Mukherjee, Mukherjee,
  Mukund, Mullavey, Munch, Muniz, Murray, Mytidis, Nagano, Nakamura, Nakamura,
  Nakano, Nakano, Nakano, Nakao, Napier, Nardecchia, Narikawa, Naticchioni,
  Nelemans, Nelson, Neri, Nery, Neunzert, Newport, Newton, Nguyen, Ni, Nielsen,
  Nissanke, Nitz, Noack, Nocera, Nolting, Normandin, Nuttall, Oberling,
  Ochsner, Oelker, Ogin, Oh, Oh, Ohashi, Ohishi, Ohkawa, Ohme, Okutomi, Oliver,
  Ono, Ono, Oohara, Oppermann, Oram, O'Reilly, O'Shaughnessy, Ottaway,
  Overmier, Owen, Pace, Page, Pai, Pai, Palamos, Palashov, Palomba, Pal-Singh,
  Pan, Pankow, Pannarale, Pant, Paoletti, Paoli, Papa, Paris, Parker, Pascucci,
  Pasqualetti, Passaquieti, Passuello, Patricelli, Pearlstone, Pedraza,
  Pedurand, Pekowsky, Pele, {Pe{\~{n}}a Arellano}, Penn, Perez, Perreca, Perri,
  Pfeiffer, Phelps, Piccinni, Pichot, Piergiovanni, Pierro, Pillant, Pinard,
  Pinto, Pitkin, Poe, Poggiani, Popolizio, Post, Powell, Prasad, Pratt, Predoi,
  Prestegard, Prijatelj, Principe, Privitera, Prodi, Prokhorov, Puncken,
  Punturo, Puppo, P{\"{u}}rrer, Qi, Qin, Qiu, Quetschke, Quintero,
  Quitzow-James, Raab, Rabeling, Radkins, Raffai, Raja, Rajan, Rakhmanov,
  Rapagnani, Raymond, Razzano, Re, Read, Regimbau, Rei, Reid, Reitze, Rew,
  Reyes, Rhoades, Ricci, Riles, Rizzo, Robertson, Robie, Robinet, Rocchi,
  Rolland, Rollins, Roma, Romano, Romie, Rosi{\'{n}}ska, Rowan, R{\"{u}}diger,
  Ruggi, Ryan, Sachdev, Sadecki, Sadeghian, Sago, Saijo, Saito, Sakai,
  Sakellariadou, Salconi, Saleem, Salemi, Samajdar, Sammut, Sampson, Sanchez,
  Sandberg, Sanders, Sasaki, Sassolas, Sathyaprakash, Sato, Sato, Saulson,
  Sauter, Savage, Sawadsky, Schale, Scheuer, Schmidt, Schmidt, Schmidt,
  Schnabel, Schofield, Sch{\"{o}}nbeck, Schreiber, Schuette, Schutz, Schwalbe,
  Scott, Scott, Sekiguchi, Sekiguchi, Sellers, Sengupta, Sentenac, Sequino,
  Sergeev, Setyawati, Shaddock, Shaffer, Shahriar, Shapiro, Shawhan, Sheperd,
  Shibata, Shikano, Shimoda, Shoda, Shoemaker, Shoemaker, Siellez, Siemens,
  Sieniawska, Sigg, Silva, Singer, Singer, Singh, Singh, Singhal, Sintes,
  Slagmolen, Smith, Smith, Smith, Somiya, Son, Sorazu, Sorrentino, Souradeep,
  Spencer, Srivastava, Staley, Steinke, Steinlechner, Steinlechner, Steinmeyer,
  Stephens, Stevenson, Stone, Strain, Straniero, Stratta, Strigin, Sturani,
  Stuver, Sugimoto, Summerscales, Sun, Sunil, Sutton, Suzuki, Swinkels,
  Szczepa{\'{n}}czyk, Tacca, Tagoshi, Takada, Takahashi, Takahashi, Takamori,
  Talukder, Tanaka, Tanaka, Tanaka, Tanner, T{\'{a}}pai, Taracchini, Tatsumi,
  Taylor, Telada, Theeg, Thomas, Thomas, Thomas, Thorne, Thrane, Tippens,
  Tiwari, Tiwari, Tokmakov, Toland, Tomaru, Tomlinson, Tonelli, Tornasi,
  Torrie, T{\"{o}}yr{\"{a}}, Travasso, Traylor, Trifir{\`{o}}, Trinastic,
  Tringali, Trozzo, Tse, Tso, Tsubono, Tsuzuki, Turconi, Tuyenbayev, Uchiyama,
  Uehara, Ueki, Ueno, Ugolini, Unnikrishnan, Urban, Ushiba, Usman, Vahlbruch,
  Vajente, Valdes, van Bakel, van Beuzekom, van~den Brand, {Van Den Broeck},
  Vander-Hyde, van~der Schaaf, van Heijningen, van Putten, van Veggel, Vardaro,
  Varma, Vass, Vas{\'{u}}th, Vecchio, Vedovato, Veitch, Veitch, Venkateswara,
  Venugopalan, Verkindt, Vetrano, Vicer{\'{e}}, Viets, Vinciguerra, Vine,
  Vinet, Vitale, Vo, Vocca, Vorvick, Voss, Vousden, Vyatchanin, Wade, Wade,
  Wade, Wakamatsu, Walker, Wallace, Walsh, Wang, Wang, Wang, Wang, Ward,
  Warner, Was, Watchi, Weaver, Wei, Weinert, Weinstein, Weiss, Wen, We{\ss}els,
  Westphal, Wette, Whelan, Whiting, Whittle, Williams, Williams, Williamson,
  Willis, Willke, Wimmer, Winkler, Wipf, Wittel, Woan, Woehler, Worden, Wright,
  Wu, Wu, Yam, Yamamoto, Yamamoto, Yamamoto, Yancey, Yano, Yap, Yokoyama,
  Yokozawa, Yoon, Yu, Yu, Yuzurihara, Yvert, Zadro{\.{z}}ny, Zangrando,
  Zanolin, Zeidler, Zendri, Zevin, Zhang, Zhang, Zhang, Zhang, Zhao, Zhou,
  Zhou, Zhu, Zhu, Zucker, \& Zweizig}]{GWTC1}
---. 2019{\natexlab{b}}, Physical Review X, 9, 31040,
  \dodoi{10.1103/PhysRevX.9.031040}

\bibitem[{Abbott {et~al.}(2020)Abbott, Abbott, Abraham, Acernese, Ackley,
  Adams, Adhikari, Adya, Affeldt, Agathos, Agatsuma, Aggarwal, Aguiar, Aich,
  Aiello, Ain, Ajith, Akcay, Allen, Allocca, Altin, Amato, Anand, Ananyeva,
  Anderson, Anderson, Angelova, Ansoldi, Antier, Appert, Arai, Araya, Areeda,
  Ar{\`{e}}ne, Arnaud, Aronson, Arun, Asali, Ascenzi, Ashton, Aston, Astone,
  Aubin, Aufmuth, AultONeal, Austin, Avendano, Babak, Bacon, Badaracco, Bader,
  Bae, Baer, Baird, Baldaccini, Ballardin, Ballmer, Bals, Balsamo, Baltus,
  Banagiri, Bankar, Bankar, Barayoga, Barbieri, Barish, Barker, Barkett,
  Barneo, Barone, Barr, Barsotti, Barsuglia, Barta, Bartlett, Bartos, Bassiri,
  Basti, Bawaj, Bayley, Bazzan, B{\'{e}}csy, Bejger, Belahcene, Bell, Beniwal,
  Benjamin, Bentley, Bergamin, Berger, Bergmann, Bernuzzi, Berry, Bersanetti,
  Bertolini, Betzwieser, Bhandare, Bhandari, Bidler, Biggs, Bilenko,
  Billingsley, Birney, Birnholtz, Biscans, Bischi, Biscoveanu, Bisht,
  Bissenbayeva, Bitossi, Bizouard, Blackburn, Blackman, Blair, Blair, Blair,
  Bobba, Bode, Boer, Boetzel, Bogaert, Bondu, Bonilla, Bonnand, Booker, Boom,
  Bork, Boschi, Bose, Bossilkov, Bosveld, Bouffanais, Bozzi, Bradaschia, Brady,
  Bramley, Branchesi, Brau, Breschi, Briant, Briggs, Brighenti, Brillet,
  Brinkmann, Brockill, Brooks, Brooks, Brown, Brunett, Bruno, Bruntz, Buikema,
  Bulik, Bulten, Buonanno, Buscicchio, Buskulic, Byer, Cabero, Cadonati,
  Cagnoli, Cahillane, {Calder{\'{o}}n Bustillo}, Callaghan, Callister, Calloni,
  Camp, Canepa, Cannon, Cao, Cao, Carapella, Carbognani, Caride, Carney,
  Carullo, {Casanueva Diaz}, Casentini, Casta{\~{n}}eda, Caudill,
  Cavagli{\`{a}}, Cavalier, Cavalieri, Cella, Cerd{\'{a}}-Dur{\'{a}}n,
  Cesarini, Chaibi, Chakravarti, Chan, Chan, Chandra, Chao, Charlton, Chase,
  Chassande-Mottin, Chatterjee, Chaturvedi, Chatziioannou, Chen, Chen, Chen,
  Cheng, Cheong, Chia, Chiadini, Chierici, Chincarini, Chiummo, Cho, Cho, Cho,
  Christensen, Chu, Chua, Chung, Chung, Ciani, Ciecielag, Cie{\'{s}}lar,
  Ciobanu, Ciolfi, Cipriano, Cirone, Clara, Clark, Clearwater, Clesse, Cleva,
  Coccia, Cohadon, Cohen, Colleoni, Collette, Collins, Colpi, Constancio,
  Conti, Cooper, Corban, Corbitt, Cordero-Carri{\'{o}}n, Corezzi, Corley,
  Cornish, Corre, Corsi, Cortese, Costa, Cotesta, Coughlin, Coughlin, Coulon,
  Countryman, Couvares, Covas, Coward, Cowart, Coyne, Coyne, Creighton,
  Creighton, Cripe, Croquette, Crowder, Cudell, Cullen, Cumming, Cummings,
  Cunningham, Cuoco, Curylo, Canton, D{\'{a}}lya, Dana, Daneshgaran-Bajastani,
  D'Angelo, Danilishin, D'Antonio, Danzmann, Darsow-Fromm, Dasgupta, Datrier,
  Dattilo, Dave, Davier, Davies, Davis, Daw, DeBra, Deenadayalan, Degallaix,
  {De Laurentis}, Del{\'{e}}glise, Delfavero, {De Lillo}, {Del Pozzo},
  DeMarchi, D'Emilio, Demos, Dent, {De Pietri}, {De Rosa}, {De Rossi}, DeSalvo,
  de~Varona, Dhurandhar, D{\'{i}}az, Diaz-Ortiz, Dietrich, {Di Fiore}, {Di
  Fronzo}, {Di Giorgio}, {Di Giovanni}, {Di Giovanni}, {Di Girolamo}, {Di
  Lieto}, Ding, {Di Pace}, {Di Palma}, {Di Renzo}, Divakarla, Dmitriev, Doctor,
  Donovan, Dooley, Doravari, Dorrington, Downes, Drago, Driggers, Du, Ducoin,
  Dupej, Durante, D'Urso, Dwyer, Easter, Eddolls, Edelman, Edo, Edy, Effler,
  Ehrens, Eichholz, Eikenberry, Eisenmann, Eisenstein, Ejlli, Errico, Essick,
  Estelles, Estevez, Etienne, Etzel, Evans, Evans, Ewing, Fafone, Fairhurst,
  Fan, Farinon, Farr, Farr, Fauchon-Jones, Favata, Fays, Fazio, Feicht, Fejer,
  Feng, Fenyvesi, Ferguson, Fernandez-Galiana, Ferrante, Ferreira, Ferreira,
  Fidecaro, Fiori, Fiorucci, Fishbach, Fisher, Fittipaldi, Fitz-Axen, Fiumara,
  Flaminio, Floden, Flynn, Fong, Font, Forsyth, Fournier, Frasca, Frasconi,
  Frei, Freise, Frey, Frey, Fritschel, Frolov, Fronz{\`{e}}, Fulda, Fyffe,
  Gabbard, Gadre, Gaebel, Gair, Galaudage, Ganapathy, Ganguly, Gaonkar,
  Garc{\'{i}}a-Quir{\'{o}}s, Garufi, Gateley, Gaudio, Gayathri, Gemme, Genin,
  Gennai, George, George, Gergely, Ghonge, Ghosh, Ghosh, Ghosh, Giacomazzo,
  Giaime, Giardina, Gibson, Gier, Gill, Glanzer, Gniesmer, Godwin, Goetz,
  Goetz, Gohlke, Goncharov, Gonz{\'{a}}lez, Gopakumar, Gossan, Gosselin,
  Gouaty, Grace, Grado, Granata, Grant, Gras, Grassia, Gray, Gray, Greco,
  Green, Green, Gretarsson, Griggs, Grignani, Grimaldi, Grimm, Grote,
  Grunewald, Gruning, Guidi, Guimaraes, Guix{\'{e}}, Gulati, Guo, Gupta, Gupta,
  Gupta, Gustafson, Gustafson, Haegel, Halim, Hall, Hamilton, Hammond, Haney,
  Hanke, Hanks, Hanna, Hannam, Hannuksela, Hansen, Hanson, Harder, Hardwick,
  Haris, Harms, Harry, Harry, Hasskew, Haster, Haughian, Hayes, Healy,
  Heidmann, Heintze, Heinze, Heitmann, Hellman, Hello, Hemming, Hendry, Heng,
  Hennes, Hennig, Heurs, Hild, Hinderer, Hoback, Hochheim, Hofgard, Hofman,
  Holgado, Holland, Holt, Holz, Hopkins, Horst, Hough, Howell, Hoy, Huang,
  H{\"{u}}bner, Huerta, Huet, Hughey, Hui, Husa, Huttner, Huxford, Huynh-Dinh,
  Idzkowski, Iess, Inchauspe, Ingram, Intini, Isac, Isi, Iyer, Jacqmin, Jadhav,
  Jadhav, James, Jani, Janthalur, Jaranowski, Jariwala, Jaume, Jenkins, Jiang,
  Johns, Johnson-McDaniel, Jones, Jones, Jones, Jones, Jones, Jonker, Ju,
  Junker, Kalaghatgi, Kalogera, Kamai, Kandhasamy, Kang, Kanner, Kapadia,
  Karki, Kashyap, Kasprzack, Kastaun, Katsanevas, Katsavounidis, Katzman,
  Kaufer, Kawabe, K{\'{e}}f{\'{e}}lian, Keitel, Keivani, Kennedy, Key, Khadka,
  Khalili, Khan, Khan, Khan, Khazanov, Khetan, Khursheed, Kijbunchoo, Kim, Kim,
  Kim, Kim, Kim, Kim, Kim, Kimball, King, Kinley-Hanlon, Kirchhoff, Kissel,
  Kleybolte, Klimenko, Knowles, Knyazev, Koch, Koehlenbeck, Koekoek, Koley,
  Kondrashov, Kontos, Koper, Korobko, Korth, Kovalam, Kozak, Kringel,
  Krishnendu, Kr{\'{o}}lak, Krupinski, Kuehn, Kumar, Kumar, Kumar, Kumar,
  Kumar, Kuo, Kutynia, Lackey, Laghi, Lalande, Lam, Lamberts, Landry, Lane,
  Lang, Lange, Lantz, Lanza, {La Rosa}, Lartaux-Vollard, Lasky, Laxen,
  Lazzarini, Lazzaro, Leaci, Leavey, Lecoeuche, Lee, Lee, Lee, Lee, Lee,
  Lehmann, Leroy, Letendre, Levin, Li, Li, Li, Li, Li, Linde, Linker, Linley,
  Littenberg, Liu, Liu, Llorens-Monteagudo, Lo, Lockwood, London, Longo,
  Lorenzini, Loriette, Lormand, Losurdo, Lough, Lousto, Lovelace, L{\"{u}}ck,
  Lumaca, Lundgren, Ma, Macas, Macfoy, MacInnis, Macleod, MacMillan, Macquet,
  {Maga{\~{n}}a Hernandez}, Maga{\~{n}}a-Sandoval, Magee, Majorana, Maksimovic,
  Malik, Man, Mandic, Mangano, Mansell, Manske, Mantovani, Mapelli, Marchesoni,
  Marion, M{\'{a}}rka, M{\'{a}}rka, Markakis, Markosyan, Markowitz, Maros,
  Marquina, Marsat, Martelli, Martin, Martin, Martinez, Martynov, Masalehdan,
  Mason, Massera, Masserot, Massinger, Masso-Reid, Mastrogiovanni, Matas,
  Matichard, Mavalvala, Maynard, McCann, McCarthy, McClelland, McCormick,
  McCuller, McGuire, McIsaac, McIver, McManus, McRae, McWilliams, Meacher,
  Meadors, Mehmet, Mehta, {Mejuto Villa}, Melatos, Mendell, Mercer, Mereni,
  Merfeld, Merilh, Merritt, Merzougui, Meshkov, Messenger, Messick, Metzdorff,
  Meyers, Meylahn, Mhaske, Miani, Miao, Michaloliakos, Michel, Middleton,
  Milano, Miller, Millhouse, Mills, Milotti, Milovich-Goff, Minazzoli,
  Minenkov, Mishkin, Mishra, Mistry, Mitra, Mitrofanov, Mitselmakher,
  Mittleman, Mo, Mogushi, Mohapatra, Mohite, Molina-Ruiz, Mondin, Montani,
  Moore, Moraru, Morawski, Moreno, Morisaki, Mours, Mow-Lowry, Mozzon,
  Muciaccia, Mukherjee, Mukherjee, Mukherjee, Mukherjee, Mukund, Mullavey,
  Munch, Mu{\~{n}}iz, Murray, Nagar, Nardecchia, Naticchioni, Nayak, Neil,
  Neilson, Nelemans, Nelson, Nery, Neunzert, Ng, Ng, Nguyen, Nguyen, Nichols,
  Nichols, Nissanke, Nitz, Nocera, Noh, North, Nothard, Nuttall, Oberling,
  O'Brien, Oganesyan, Ogin, Oh, Oh, Ohme, Ohta, Okada, Oliver, Olivetto,
  Oppermann, Oram, O'Reilly, Ormiston, Ortega, O'Shaughnessy, Ossokine,
  Osthelder, Ottaway, Overmier, Owen, Pace, Pagano, Page, Pagliaroli, Pai, Pai,
  Palamos, Palashov, Palomba, Pan, Panda, Pang, Pankow, Pannarale, Pant,
  Paoletti, Paoli, Parida, Parker, Pascucci, Pasqualetti, Passaquieti,
  Passuello, Patricelli, Payne, Pearlstone, Pechsiri, Pedersen, Pedraza, Pele,
  Penn, Perego, Perez, P{\'{e}}rigois, Perreca, Perri{\`{e}}s, Petermann,
  Pfeiffer, Phelps, Phukon, Piccinni, Pichot, Piendibene, Piergiovanni, Pierro,
  Pillant, Pinard, Pinto, Piotrzkowski, Pirello, Pitkin, Plastino, Poggiani,
  Pong, Ponrathnam, Popolizio, Porter, Powell, Prajapati, Prasai, Prasanna,
  Pratten, Prestegard, Principe, Prodi, Prokhorov, Punturo, Puppo,
  P{\"{u}}rrer, Qi, Quetschke, Quinonez, Raab, Raaijmakers, Radkins, Radulesco,
  Raffai, Rafferty, Raja, Rajan, Rajbhandari, Rakhmanov, Ramirez, Ramos-Buades,
  Rana, Rao, Rapagnani, Raymond, Razzano, Read, Regimbau, Rei, Reid, Reitze,
  Rettegno, Ricci, Richardson, Richardson, Ricker, Riemenschneider, Riles,
  Rizzo, Robertson, Robinet, Rocchi, Rodriguez-Soto, Rolland, Rollins, Roma,
  Romanelli, Romano, Romel, Romero-Shaw, Romie, Rose, Rose, Rose,
  Rosi{\'{n}}ska, Rosofsky, Ross, Rowan, Rowlinson, Roy, Roy, Roy, Ruggi,
  Rutins, Ryan, Sachdev, Sadecki, Sakellariadou, Salafia, Salconi, Saleem,
  Salemi, Samajdar, Sanchez, Sanchez, Sanchis-Gual, Sanders, Santiago, Santos,
  Sarin, Sassolas, Sathyaprakash, Sauter, Savage, Savant, Sawant, Sayah,
  Schaetzl, Schale, Scheel, Scheuer, Schmidt, Schnabel, Schofield,
  Sch{\"{o}}nbeck, Schreiber, Schulte, Schutz, Schwarm, Schwartz, Scott, Scott,
  Seidel, Sellers, Sengupta, Sennett, Sentenac, Sequino, Sergeev, Setyawati,
  Shaddock, Shaffer, Sharifi, Shahriar, Sharma, Sharma, Shawhan, Shen,
  Shikauchi, Shink, Shoemaker, Shoemaker, Shukla, ShyamSundar, Siellez,
  Sieniawska, Sigg, Singer, Singh, Singh, Singha, Singhal, Sintes, Sipala,
  Skliris, Slagmolen, Slaven-Blair, Smetana, Smith, Smith, Somala, Son, Soni,
  Sorazu, Sordini, Sorrentino, Souradeep, Sowell, Spencer, Spera, Srivastava,
  Srivastava, Staats, Stachie, Standke, Steer, Steinke, Steinlechner,
  Steinlechner, Steinmeyer, Stevenson, Stocks, Stops, Stover, Strain, Stratta,
  Strunk, Sturani, Stuver, Sudhagar, Sudhir, Summerscales, Sun, Sunil, Sur,
  Suresh, Sutton, Swinkels, Szczepa{\'{n}}czyk, Tacca, Tait, Talbot,
  Tanasijczuk, Tanner, Tao, T{\'{a}}pai, Tapia, {Tapia San Martin}, Tasson,
  Taylor, Tenorio, Terkowski, Thirugnanasambandam, Thomas, Thomas, Thompson,
  Thondapu, Thorne, Thrane, Tinsman, Saravanan, Tiwari, Tiwari, Tiwari, Toland,
  Tonelli, Tornasi, Torres-Forn{\'{e}}, Torrie, {Tosta e Melo},
  T{\"{o}}yr{\"{a}}, Travasso, Traylor, Tringali, Tripathee, Trovato, Trudeau,
  Tsang, Tse, Tso, Tsukada, Tsuna, Tsutsui, Turconi, Ubhi, Udall, Ueno,
  Ugolini, Unnikrishnan, Urban, Usman, Utina, Vahlbruch, Vajente, Valdes,
  Valentini, van Bakel, van Beuzekom, van~den Brand, {Van Den Broeck},
  Vander-Hyde, van~der Schaaf, {Van Heijningen}, van Veggel, Vardaro, Varma,
  Vass, Vas{\'{u}}th, Vecchio, Vedovato, Veitch, Veitch, Venkateswara,
  Venugopalan, Verkindt, Veske, Vetrano, Vicer{\'{e}}, Viets, Vinciguerra,
  Vine, Vinet, Vitale, Vivanco, Vo, Vocca, Vorvick, Vyatchanin, Wade, Wade,
  Wade, Walet, Walker, Wallace, Wallace, Walsh, Wang, Wang, Wang, Ward, Warden,
  Warner, Was, Watchi, Weaver, Wei, Weinert, Weinstein, Weiss, Wellmann, Wen,
  We{\ss}els, Westhouse, Wette, Whelan, Whiting, Whittle, Wilken, Williams,
  Willis, Willke, Winkler, Wipf, Wittel, Woan, Woehler, Wofford, Wong, Wright,
  Wu, Wysocki, Xiao, Yamamoto, Yang, Yang, Yang, Yap, Yazback, Yeeles, Yu, Yu,
  Yuen, Zadro{\.{z}}ny, Zadro{\.{z}}ny, Zanolin, Zelenova, Zendri, Zevin,
  Zhang, Zhang, Zhang, Zhao, Zhao, Zhou, Zhou, Zhu, Zimmerman, Zucker, \&
  Zweizig}]{GW190521}
Abbott, R., Abbott, T.~D., Abraham, S., {et~al.} 2020, Physical Review Letters,
  125, 101102, \dodoi{10.1103/PhysRevLett.125.101102}

\bibitem[{Abbott {et~al.}(2021{\natexlab{a}})Abbott, Abbott, Abraham, Acernese,
  Ackley, Adams, Adams, Adhikari, Adya, Affeldt, Agathos, Agatsuma, Aggarwal,
  Aguiar, Aiello, Ain, Ajith, Allen, Allocca, Altin, Amato, Anand, Ananyeva,
  Anderson, Anderson, Angelova, Ansoldi, Antelis, Antier, Appert, Arai, Araya,
  Areeda, Ar{\`{e}}ne, Arnaud, Aronson, Arun, Asali, Ascenzi, Ashton, Aston,
  Astone, Aubin, Aufmuth, AultONeal, Austin, Avendano, Babak, Badaracco, Bader,
  Bae, Baer, Bagnasco, Baird, Ball, Ballardin, Ballmer, Bals, Balsamo, Baltus,
  Banagiri, Bankar, Bankar, Barayoga, Barbieri, Barish, Barker, Barneo, Barnum,
  Barone, Barr, Barsotti, Barsuglia, Barta, Bartlett, Bartos, Bassiri, Basti,
  Bawaj, Bayley, Bazzan, Becher, B{\'{e}}csy, Bedakihale, Bejger, Belahcene,
  Beniwal, Benjamin, Bennett, Bentley, Bergamin, Berger, Bergmann, Bernuzzi,
  Berry, Bersanetti, Bertolini, Betzwieser, Bhandare, Bhandari, Bhattacharjee,
  Bidler, Bilenko, Billingsley, Birney, Birnholtz, Biscans, Bischi, Biscoveanu,
  Bisht, Bitossi, Bizouard, Blackburn, Blackman, Blair, Blair, Blair, Blanch,
  Bobba, Bode, Boer, Boetzel, Bogaert, Boldrini, Bondu, Bonilla, Bonnand,
  Booker, Boom, Bork, Boschi, Bose, Bossilkov, Boudart, Bouffanais, Bozzi,
  Bradaschia, Brady, Bramley, Branchesi, Brau, Breschi, Briant, Briggs,
  Brighenti, Brillet, Brinkmann, Brockill, Brooks, Brooks, Brown, Brunett,
  Bruno, Bruntz, Buikema, Bulik, Bulten, Buonanno, Buscicchio, Buskulic, Byer,
  Cabero, Cadonati, Caesar, Cagnoli, Cahillane, {Calder{\'{o}}n Bustillo},
  Callaghan, Callister, Calloni, Camp, Canepa, Cannon, Cao, Cao, Carapella,
  Carbognani, Carney, Carpinelli, Carullo, Carver, {Casanueva Diaz}, Casentini,
  Caudill, Cavagli{\`{a}}, Cavalier, Cavalieri, Cella, Cerd{\'{a}}-Dur{\'{a}}n,
  Cesarini, Chaibi, Chakravarti, Chan, Chan, Chandra, Chanial, Chao, Charlton,
  Chase, Chassande-Mottin, Chatterjee, Chattopadhyay, Chaturvedi,
  Chatziioannou, Chen, Chen, Chen, Chen, Cheng, Cheong, Chia, Chiadini,
  Chierici, Chincarini, Chiummo, Cho, Cho, Cho, Choate, Christensen, Chu, Chua,
  Chung, Chung, Ciani, Ciecielag, Cie{\'{s}}lar, Cifaldi, Ciobanu, Ciolfi,
  Cipriano, Cirone, Clara, Clark, Clark, Clarke, Clearwater, Clesse, Cleva,
  Coccia, Cohadon, Cohen, Colleoni, Collette, Collins, Colpi, Constancio,
  Conti, Cooper, Corban, Corbitt, Cordero-Carri{\'{o}}n, Corezzi, Corley,
  Cornish, Corre, Corsi, Cortese, Costa, Cotesta, Coughlin, Coughlin, Coulon,
  Countryman, Couvares, Covas, Coward, Cowart, Coyne, Coyne, Creighton,
  Creighton, Croquette, Crowder, Cudell, Cullen, Cumming, Cummings, Cunningham,
  Cuoco, Curylo, {Dal Canton}, D{\'{a}}lya, Dana, DaneshgaranBajastani,
  D'Angelo, Danilishin, D'Antonio, Danzmann, Darsow-Fromm, Dasgupta, Datrier,
  Dattilo, Dave, Davier, Davies, Davis, Daw, Dean, DeBra, Deenadayalan,
  Degallaix, {De Laurentis}, Del{\'{e}}glise, {Del Favero}, {De Lillo}, {De
  Lillo}, {Del Pozzo}, DeMarchi, {De Matteis}, D'Emilio, Demos, Denker, Dent,
  Depasse, {De Pietri}, {De Rosa}, {De Rossi}, DeSalvo, de~Varona, Dhurandhar,
  D{\'{i}}az, Diaz-Ortiz, Didio, Dietrich, {Di Fiore}, DiFronzo, {Di Giorgio},
  {Di Giovanni}, {Di Giovanni}, {Di Girolamo}, {Di Lieto}, Ding, {Di Pace}, {Di
  Palma}, {Di Renzo}, Divakarla, Dmitriev, Doctor, D'Onofrio, Donovan, Dooley,
  Doravari, Dorrington, Downes, Drago, Driggers, Du, Ducoin, Dupej, Durante,
  D'Urso, Duverne, Dwyer, Easter, Eddolls, Edelman, Edo, Edy, Effler, Eichholz,
  Eikenberry, Eisenmann, Eisenstein, Ejlli, Errico, Essick, Estell{\'{e}}s,
  Estevez, Etienne, Etzel, Evans, Evans, Ewing, Fafone, Fair, Fairhurst, Fan,
  Farah, Farinon, Farr, Farr, Fauchon-Jones, Favata, Fays, Fazio, Feicht,
  Fejer, Feng, Fenyvesi, Ferguson, Fernandez-Galiana, Ferrante, Ferreira,
  Fidecaro, Figura, Fiori, Fiorucci, Fishbach, Fisher, Fishner, Fittipaldi,
  Fitz-Axen, Fiumara, Flaminio, Floden, Flynn, Fong, Font, Forsyth, Fournier,
  Frasca, Frasconi, Frei, Freise, Frey, Frey, Fritschel, Frolov, Fronz{\'{e}},
  Fulda, Fyffe, Gabbard, Gadre, Gaebel, Gair, Gais, Galaudage, Gamba,
  Ganapathy, Ganguly, Gaonkar, Garaventa, Garc{\'{i}}a-Quir{\'{o}}s, Garufi,
  Gateley, Gaudio, Gayathri, Gemme, Gennai, George, George, Gergely, Ghonge,
  Ghosh, Ghosh, Ghosh, Giacomazzo, Giacoppo, Giaime, Giardina, Gibson, Gier,
  Gill, Giri, Glanzer, Gleckl, Godwin, Goetz, Goetz, Gohlke, Goncharov,
  Gonz{\'{a}}lez, Gopakumar, Gossan, Gosselin, Gouaty, Grace, Grado, Granata,
  Granata, Grant, Gras, Grassia, Gray, Gray, Greco, Green, Green, Gretarsson,
  Griggs, Grignani, Grimaldi, Grimes, Grimm, Grote, Grunewald, Gruning,
  Guerrero, Guidi, Guimaraes, Guix{\'{e}}, Gulati, Guo, Gupta, Gupta, Gupta,
  Gustafson, Gustafson, Guzman, Haegel, Halim, Hall, Hamilton, Hammond, Haney,
  Hanke, Hanks, Hanna, Hannuksela, Hannuksela, Hansen, Hansen, Hanson, Harder,
  Hardwick, Haris, Harms, Harry, Harry, Hartwig, Hasskew, Haster, Haughian,
  Hayes, Healy, Heidmann, Heintze, Heinze, Heinzel, Heitmann, Hellman, Hello,
  Helmling-Cornell, Hemming, Hendry, Heng, Hennes, Hennig, Hennig, {Hernandez
  Vivanco}, Heurs, Hild, Hill, Hines, Hochheim, Hofgard, Hofman, Hohmann,
  Holgado, Holland, Hollows, Holmes, Holt, Holz, Hopkins, Horst, Hough, Howell,
  Hoy, Hoyland, Huang, H{\"{u}}bner, Huddart, Huerta, Hughey, Hui, Husa,
  Huttner, Hutzler, Huxford, Huynh-Dinh, Idzkowski, Iess, Imperato, Inchauspe,
  Ingram, Intini, Isi, Iyer, JaberianHamedan, Jacqmin, Jadhav, Jadhav, James,
  Jani, Janssens, Janthalur, Jaranowski, Jariwala, Jaume, Jenkins, Jeunon,
  Jiang, Johns, Jones, Jones, Jones, Jones, Jones, Jonker, Ju, Junker,
  Kalaghatgi, Kalogera, Kamai, Kandhasamy, Kang, Kanner, Kapadia, Kapasi,
  Karathanasis, Karki, Kashyap, Kasprzack, Kastaun, Katsanevas, Katsavounidis,
  Katzman, Kawabe, K{\'{e}}f{\'{e}}lian, Keitel, Key, Khadka, Khalili, Khan,
  Khan, Khazanov, Khetan, Khursheed, Kijbunchoo, Kim, Kim, Kim, Kim, Kim, Kim,
  Kimball, King, Kinley-Hanlon, Kirchhoff, Kissel, Kleybolte, Klimenko,
  Knowles, Knyazev, Koch, Koehlenbeck, Koekoek, Koley, Kolstein, Komori,
  Kondrashov, Kontos, Koper, Korobko, Korth, Kovalam, Kozak, Kr{\"{a}}mer,
  Kringel, Krishnendu, Kr{\'{o}}lak, Kuehn, Kumar, Kumar, Kumar, Kumar, Kuns,
  Kwang, Lackey, Laghi, Lalande, Lam, Lamberts, Landry, Lane, Lang, Lange,
  Lantz, Lanza, {La Rosa}, Lartaux-Vollard, Lasky, Laxen, Lazzarini, Lazzaro,
  Leaci, Leavey, Lecoeuche, Lee, Lee, Lee, Lee, Lehmann, Leon, Leroy, Letendre,
  Levin, Li, Li, Li, Li, Li, Linde, Linker, Linley, Littenberg, Liu, Liu,
  Llorens-Monteagudo, Lo, Lockwood, London, Longo, Lorenzini, Loriette,
  Lormand, Losurdo, Lough, Lousto, Lovelace, L{\"{u}}ck, Lumaca, Lundgren, Ma,
  Macas, MacInnis, Macleod, MacMillan, Macquet, {Maga{\~{n}}a Hernandez},
  Maga{\~{n}}a-Sandoval, Magazz{\`{u}}, Magee, Majorana, Maksimovic, Maliakal,
  Malik, Man, Mandic, Mangano, Mansell, Manske, Mantovani, Mapelli, Marchesoni,
  Marion, M{\'{a}}rka, M{\'{a}}rka, Markakis, Markosyan, Markowitz, Maros,
  Marquina, Marsat, Martelli, Martin, Martin, Martinez, Martinez, Martynov,
  Masalehdan, Mason, Massera, Masserot, Massinger, Masso-Reid, Mastrogiovanni,
  Matas, Mateu-Lucena, Matichard, Matiushechkina, Mavalvala, Maynard, McCann,
  McCarthy, McClelland, McCormick, McCuller, McGuire, McIsaac, McIver, McManus,
  McRae, McWilliams, Meacher, Meadors, Mehmet, Mehta, Melatos, Melchor,
  Mendell, Menendez-Vazquez, Mercer, Mereni, Merfeld, Merilh, Merritt,
  Merzougui, Meshkov, Messenger, Messick, Metzdorff, Meyers, Meylahn, Mhaske,
  Miani, Miao, Michaloliakos, Michel, Middleton, Milano, Miller, Miller,
  Millhouse, Mills, Milotti, Milovich-Goff, Minazzoli, Minenkov, Mir, Mishkin,
  Mishra, Mistry, Mitra, Mitrofanov, Mitselmakher, Mittleman, Mo, Mogushi,
  Mohapatra, Mohite, Molina, Molina-Ruiz, Mondin, Montani, Moore, Moraru,
  Morawski, Moreno, Morisaki, Mours, Mow-Lowry, Mozzon, Muciaccia, Mukherjee,
  Mukherjee, Mukherjee, Mukherjee, Mukund, Mullavey, Munch, Mu{\~{n}}iz,
  Murray, Nadji, Nagar, Nardecchia, Naticchioni, Nayak, Neil, Neilson,
  Nelemans, Nelson, Nery, Neunzert, Ng, Ng, Nguyen, Nguyen, Nguyen, Nichols,
  Nissanke, Nocera, Noh, North, Nothard, Nuttall, Oberling, O'Brien, O'Dell,
  Oganesyan, Ogin, Oh, Oh, Ohme, Ohta, Okada, Olivetto, Oppermann, Oram,
  O'Reilly, Ormiston, Ormsby, Ortega, O'Shaughnessy, Ossokine, Osthelder,
  Ottaway, Overmier, Owen, Pace, Pagano, Page, Pagliaroli, Pai, Pai, Palamos,
  Palashov, Palomba, Pan, Panda, Pang, Pankow, Pannarale, Pant, Paoletti,
  Paoli, Paolone, Parker, Pascucci, Pasqualetti, Passaquieti, Passuello, Patel,
  Patricelli, Payne, Pechsiri, Pedraza, Pegoraro, Pele, Penn, Perego, Perez,
  P{\'{e}}rigois, Perreca, Perri{\`{e}}s, Petermann, Petterson, Pfeiffer, Pham,
  Phukon, Piccinni, Pichot, Piendibene, Piergiovanni, Pierini, Pierro, Pillant,
  Pilo, Pinard, Pinto, Piotrzkowski, Pirello, Pitkin, Placidi, Plastino,
  Pluchar, Poggiani, Polini, Pong, Ponrathnam, Popolizio, Porter, Poverman,
  Powell, Pracchia, Prajapati, Prasai, Prasanna, Pratten, Prestegard, Principe,
  Prodi, Prokhorov, Prosposito, Puecher, Punturo, Puosi, Puppo, P{\"{u}}rrer,
  Qi, Quetschke, Quinonez, Quitzow-James, Raab, Raaijmakers, Radkins,
  Radulesco, Raffai, Rafferty, Rail, Raja, Rajan, Rajbhandari, Rakhmanov,
  Ramirez, Ramirez, Ramos-Buades, Rana, Rao, Rapagnani, Rapol, Ratto, Raymond,
  Razzano, Read, Regimbau, Rei, Reid, Reitze, Rettegno, Ricci, Richardson,
  Richardson, Richardson, Ricker, Riemenschneider, Riles, Rizzo, Robertson,
  Robinet, Rocchi, Rocha, Rodriguez, Rodriguez-Soto, Rolland, Rollins, Roma,
  Romanelli, Romano, Romel, Romero, Romero-Shaw, Romie, Ronchini, Rose, Rose,
  Rose, Rosell, Rosi{\'{n}}ska, Rosofsky, Ross, Rowan, Rowlinson, Roy, Roy,
  Ruggi, Ryan, Sachdev, Sadecki, Sakellariadou, Salafia, Salconi, Saleem,
  Samajdar, Sanchez, Sanchez, Sanchez, Sanchis-Gual, Sanders, Santiago, Santos,
  Saravanan, Sarin, Sassolas, Sathyaprakash, Sauter, Savage, Savant, Sawant,
  Sayah, Schaetzl, Schale, Scheel, Scheuer, Schindler-Tyka, Schmidt, Schnabel,
  Schofield, Sch{\"{o}}nbeck, Schreiber, Schulte, Schutz, Schwarm, Schwartz,
  Scott, Scott, Seglar-Arroyo, Seidel, Sellers, Sengupta, Sennett, Sentenac,
  Sequino, Sergeev, Setyawati, Shaffer, Shahriar, Sharifi, Sharma, Sharma,
  Shawhan, Shen, Shikauchi, Shink, Shoemaker, Shoemaker, Shukla, ShyamSundar,
  Sieniawska, Sigg, Singer, Singh, Singh, Singha, Singhal, Sintes, Sipala,
  Skliris, Slagmolen, Slaven-Blair, Smetana, Smith, Smith, Somala, Son, Soni,
  Sorazu, Sordini, Sorrentino, Sorrentino, Soulard, Souradeep, Sowell, Spencer,
  Spera, Srivastava, Srivastava, Staats, Stachie, Steer, Steinke, Steinlechner,
  Steinlechner, Steinmeyer, Stevenson, Stolle-McAllister, Stops, Stover,
  Strain, Stratta, Strunk, Sturani, Stuver, S{\"{u}}dbeck, Sudhagar, Sudhir,
  Suh, Summerscales, Sun, Sun, Sunil, Sur, Suresh, Sutton, Swinkels,
  Szczepa{\'{n}}czyk, Tacca, Tait, Talbot, Tanasijczuk, Tanner, Tao, Tapia,
  {Tapia San Martin}, Tasson, Taylor, Tenorio, Terkowski, Thirugnanasambandam,
  Thomas, Thomas, Thomas, Thompson, Thondapu, Thorne, Thrane, Tiwari, Tiwari,
  Tiwari, Toland, Tolley, Tonelli, Tornasi, Torres-Forn{\'{e}}, Torrie, {Tosta
  e Melo}, T{\"{o}}yr{\"{a}}, Tran, Trapananti, Travasso, Traylor, Tringali,
  Tripathee, Trovato, Trudeau, Tsai, Tsang, Tse, Tso, Tsukada, Tsuna, Tsutsui,
  Turconi, Ubhi, Udall, Ueno, Ugolini, Unnikrishnan, Urban, Usman, Utina,
  Vahlbruch, Vajente, Vajpeyi, Valdes, Valentini, Valsan, van Bakel, van
  Beuzekom, van~den Brand, {Van Den Broeck}, Vander-Hyde, van~der Schaaf, van
  Heijningen, Vardaro, Vargas, Varma, Vass, Vas{\'{u}}th, Vecchio, Vedovato,
  Veitch, Veitch, Venkateswara, Venneberg, Venugopalan, Verkindt, Verma, Veske,
  Vetrano, Vicer{\'{e}}, Viets, Villa-Ortega, Vinet, Vitale, Vo, Vocca,
  Vorvick, Vyatchanin, Wade, Wade, Wade, Walet, Walker, Wallace, Wallace,
  Walsh, Wang, Wang, Wang, Wang, Ward, Warner, Was, Washington, Watchi, Weaver,
  Wei, Weinert, Weinstein, Weiss, Wellmann, Wen, We{\ss}els, Westhouse, Wette,
  Whelan, White, White, Whiting, Whittle, Wilken, Williams, Williams,
  Williamson, Willis, Willke, Wilson, Wimmer, Winkler, Wipf, Woan, Woehler,
  Wofford, Wong, Wrangel, Wright, Wu, Wysocki, Xiao, Yamamoto, Yang, Yang,
  Yang, Yap, Yeeles, Yoon, Yu, Yu, Yuen, Zadro{\.{z}}ny, Zanolin, Zelenova,
  Zendri, Zevin, Zhang, Zhang, Zhang, Zhang, Zhao, Zhao, Zhou, Zhou, Zhu,
  Zimmerman, Zucker, \& Zweizig}]{GWTC2_pops}
---. 2021{\natexlab{a}}, The Astrophysical Journal Letters, 913, L7,
  \dodoi{10.3847/2041-8213/abe949}

\bibitem[{Abbott {et~al.}(2021{\natexlab{b}})Abbott, Abbott, Abraham, Acernese,
  Ackley, Adams, Adams, Adhikari, Adya, Affeldt, Agathos, Agatsuma, Aggarwal,
  Aguiar, Aiello, Ain, Ajith, Akcay, Allen, Allocca, Altin, Amato, Anand,
  Ananyeva, Anderson, Anderson, Angelova, Ansoldi, Antelis, Antier, Appert,
  Arai, Araya, Areeda, Ar{\`{e}}ne, Arnaud, Aronson, Arun, Asali, Ascenzi,
  Ashton, Aston, Astone, Aubin, Aufmuth, Aultoneal, Austin, Avendano, Babak,
  Badaracco, Bader, Bae, Baer, Bagnasco, Baird, Ball, Ballardin, Ballmer, Bals,
  Balsamo, Baltus, Banagiri, Bankar, Bankar, Barayoga, Barbieri, Barish,
  Barker, Barneo, Barnum, Barone, Barr, Barsotti, Barsuglia, Barta, Bartlett,
  Bartos, Bassiri, Basti, Bawaj, Bayley, Bazzan, Becher, B{\'{e}}csy,
  Bedakihale, Bejger, Belahcene, Beniwal, Benjamin, Bennett, Bentley, Bergamin,
  Berger, Bergmann, Bernuzzi, Berry, Bersanetti, Bertolini, Betzwieser,
  Bhandare, Bhandari, Bhattacharjee, Bidler, Bilenko, Billingsley, Birney,
  Birnholtz, Biscans, Bischi, Biscoveanu, Bisht, Bitossi, Bizouard, Blackburn,
  Blackman, Blair, Blair, Blair, Blanch, Bobba, Bode, Boer, Boetzel, Bogaert,
  Boldrini, Bondu, Bonilla, Bonnand, Booker, Boom, Bork, Boschi, Bose,
  Bossilkov, Boudart, Bouffanais, Bozzi, Bradaschia, Brady, Bramley, Branchesi,
  Brau, Breschi, Briant, Briggs, Brighenti, Brillet, Brinkmann, Brockill,
  Brooks, Brooks, Brown, Brunett, Bruno, Bruntz, Buikema, Bulik, Bulten,
  Buonanno, Buscicchio, Buskulic, Byer, Cabero, Cadonati, Caesar, Cagnoli,
  Cahillane, {Calder{\'{o}}n Bustillo}, Callaghan, Callister, Calloni, Camp,
  Canepa, Cannon, Cao, Cao, Carapella, Carbognani, Carney, Carpinelli, Carullo,
  Carver, {Casanueva Diaz}, Casentini, Caudill, Cavagli{\`{a}}, Cavalier,
  Cavalieri, Cella, Cerd{\'{a}}-Dur{\'{a}}n, Cesarini, Chaibi, Chakravarti,
  Chan, Chan, Chandra, Chanial, Chao, Charlton, Chase, Chassande-Mottin,
  Chatterjee, Chattopadhyay, Chaturvedi, Chatziioannou, Chen, Chen, Chen, Chen,
  Cheng, Cheong, Chia, Chiadini, Chierici, Chincarini, Chiummo, Cho, Cho, Cho,
  Choate, Christensen, Chu, Chua, Chung, Chung, Ciani, Ciecielag,
  Cie{\'{s}}lar, Cifaldi, Ciobanu, Ciolfi, Cipriano, Cirone, Clara, Clark,
  Clark, Clarke, Clearwater, Clesse, Cleva, Coccia, Cohadon, Cohen, Colleoni,
  Collette, Collins, Colpi, Constancio, Conti, Cooper, Corban, Corbitt,
  Cordero-Carri{\'{o}}n, Corezzi, Corley, Cornish, Corre, Corsi, Cortese,
  Costa, Cotesta, Coughlin, Coughlin, Coulon, Countryman, Cousins, Couvares,
  Covas, Coward, Cowart, Coyne, Coyne, Creighton, Creighton, Croquette,
  Crowder, Cudell, Cullen, Cumming, Cummings, Cunningham, Cuoco, Cury{\l}o,
  Canton, D{\'{a}}lya, Dana, Daneshgaranbajastani, D'Angelo, Danila,
  Danilishin, D'Antonio, Danzmann, Darsow-Fromm, Dasgupta, Datrier, Dattilo,
  Dave, Davier, Davies, Davis, Daw, Dean, Debra, Deenadayalan, Degallaix, {De
  Laurentis}, Del{\'{e}}glise, {Del Favero}, {De Lillo}, {De Lillo}, {Del
  Pozzo}, Demarchi, {De Matteis}, D'Emilio, Demos, Denker, Dent, Depasse, {De
  Pietri}, {De Rosa}, {De Rossi}, Desalvo, {De Varona}, Dhurandhar, D{\'{i}}az,
  Diaz-Ortiz, Didio, Dietrich, {Di Fiore}, Difronzo, {Di Giorgio}, {Di
  Giovanni}, {Di Giovanni}, {Di Girolamo}, {Di Lieto}, Ding, {Di Pace}, {Di
  Palma}, {Di Renzo}, Divakarla, Dmitriev, Doctor, D'Onofrio, Donovan, Dooley,
  Doravari, Dorrington, Downes, Drago, Driggers, Du, Ducoin, Dupej, Durante,
  D'Urso, Duverne, Dwyer, Easter, Eddolls, Edelman, Edo, Edy, Effler, Eichholz,
  Eikenberry, Eisenmann, Eisenstein, Ejlli, Errico, Essick, Estell{\'{e}}s,
  Estevez, Etienne, Etzel, Evans, Evans, Ewing, Fafone, Fair, Fairhurst, Fan,
  Farah, Farinon, Farr, Farr, Fauchon-Jones, Favata, Fays, Fazio, Feicht,
  Fejer, Feng, Fenyvesi, Ferguson, Fernandez-Galiana, Ferrante, Ferreira,
  Fidecaro, Figura, Fiori, Fiorucci, Fishbach, Fisher, Fishner, Fittipaldi,
  Fitz-Axen, Fiumara, Flaminio, Floden, Flynn, Fong, Font, Forsyth, Fournier,
  Frasca, Frasconi, Frei, Freise, Frey, Frey, Fritschel, Frolov, Fronz{\'{e}},
  Fulda, Fyffe, Gabbard, Gadre, Gaebel, Gair, Gais, Galaudage, Gamba,
  Ganapathy, Ganguly, Gaonkar, Garaventa, Garc{\'{i}}a-Quir{\'{o}}s, Garufi,
  Gateley, Gaudio, Gayathri, Gemme, Gennai, George, George, George, Gergely,
  Ghonge, Ghosh, Ghosh, Ghosh, Giacomazzo, Giacoppo, Giaime, Giardina, Gibson,
  Gier, Gill, Giri, Glanzer, Gleckl, Godwin, Goetz, Goetz, Gohlke, Goncharov,
  Gonz{\'{a}}lez, Gopakumar, Gossan, Gosselin, Gouaty, Grace, Grado, Granata,
  Granata, Grant, Gras, Grassia, Gray, Gray, Greco, Green, Green, Gretarsson,
  Griggs, Grignani, Grimaldi, Grimes, Grimm, Grote, Grunewald, Gruning,
  Guerrero, Guidi, Guimaraes, Guix{\'{e}}, Gulati, Guo, Gupta, Gupta, Gupta,
  Gustafson, Gustafson, Guzman, Haegel, Halim, Hall, Hamilton, Hammond, Haney,
  Hanke, Hanks, Hanna, Hannam, Hannuksela, Hannuksela, Hansen, Hansen, Hanson,
  Harder, Hardwick, Haris, Harms, Harry, Harry, Hartwig, Hasskew, Haster,
  Haughian, Hayes, Healy, Heidmann, Heintze, Heinze, Heinzel, Heitmann,
  Hellman, Hello, Helmling-Cornell, Hemming, Hendry, Heng, Hennes, Hennig,
  Hennig, {Hernandez Vivanco}, Heurs, Hild, Hill, Hines, Hochheim, Hofgard,
  Hofman, Hohmann, Holgado, Holland, Hollows, Holmes, Holt, Holz, Hopkins,
  Horst, Hough, Howell, Hoy, Hoyland, Huang, H{\"{u}}bner, Huddart, Huerta,
  Hughey, Hui, Husa, Huttner, Hutzler, Huxford, Huynh-Dinh, Idzkowski, Iess,
  Imperato, Inchauspe, Ingram, Intini, Isi, Iyer, Jaberianhamedan, Jacqmin,
  Jadhav, Jadhav, James, Jani, Janssens, Janthalur, Jaranowski, Jariwala,
  Jaume, Jenkins, Jeunon, Jiang, Johns, Johnson-Mcdaniel, Jones, Jones, Jones,
  Jones, Jones, Jonker, Ju, Junker, Kalaghatgi, Kalogera, Kamai, Kandhasamy,
  Kang, Kanner, Kapadia, Kapasi, Karathanasis, Karki, Kashyap, Kasprzack,
  Kastaun, Katsanevas, Katsavounidis, Katzman, Kawabe, K{\'{e}}f{\'{e}}lian,
  Keitel, Key, Khadka, Khalili, Khan, Khan, Khazanov, Khetan, Khursheed,
  Kijbunchoo, Kim, Kim, Kim, Kim, Kim, Kim, Kimball, King, Kinley-Hanlon,
  Kirchhoff, Kissel, Kleybolte, Klimenko, Knowles, Knyazev, Koch, Koehlenbeck,
  Koekoek, Koley, Kolstein, Komori, Kondrashov, Kontos, Koper, Korobko, Korth,
  Kovalam, Kozak, Kr{\"{a}}mer, Kringel, Krishnendu, Kr{\'{o}}lak, Kuehn,
  Kumar, Kumar, Kumar, Kumar, Kuns, Kwang, Lackey, Laghi, Lalande, Lam,
  Lamberts, Landry, Lane, Lang, Lange, Lantz, Lanza, {La Rosa},
  Lartaux-Vollard, Lasky, Laxen, Lazzarini, Lazzaro, Leaci, Leavey, Lecoeuche,
  Lee, Lee, Lee, Lee, Lehmann, Leon, Leroy, Letendre, Levin, Li, Li, Li, Li,
  Li, Linde, Linker, Linley, Littenberg, Liu, Liu, Llorens-Monteagudo, Lo,
  Lockwood, London, Longo, Lorenzini, Loriette, Lormand, Losurdo, Lough,
  Lousto, Lovelace, L{\"{u}}ck, Lumaca, Lundgren, Ma, Macas, Macinnis, Macleod,
  Macmillan, Macquet, {Maga{\~{n}}a Hernandez}, Maga{\~{n}}a-Sandoval,
  Magazz{\`{u}}, Magee, Majorana, Maksimovic, Maliakal, Malik, Man, Mandic,
  Mangano, Mansell, Manske, Mantovani, Mapelli, Marchesoni, Marion,
  M{\'{a}}rka, M{\'{a}}rka, Markakis, Markosyan, Markowitz, Maros, Marquina,
  Marsat, Martelli, Martin, Martin, Martinez, Martinez, Martynov, Masalehdan,
  Mason, Massera, Masserot, Massinger, Masso-Reid, Mastrogiovanni, Matas,
  Mateu-Lucena, Matichard, Matiushechkina, Mavalvala, Maynard, McCann,
  McCarthy, McClelland, McCormick, McCuller, McGuire, McIsaac, McIver, McManus,
  McRae, McWilliams, Meacher, Meadors, Mehmet, Mehta, Melatos, Melchor,
  Mendell, Menendez-Vazquez, Mercer, Mereni, Merfeld, Merilh, Merritt,
  Merzougui, Meshkov, Messenger, Messick, Metzdorff, Meyers, Meylahn, Mhaske,
  Miani, Miao, Michaloliakos, Michel, Middleton, Milano, Miller, Millhouse,
  Mills, Milotti, Milovich-Goff, Minazzoli, Minenkov, Mir, Mishkin, Mishra,
  Mistry, Mitra, Mitrofanov, Mitselmakher, Mittleman, Mo, Mogushi, Mohapatra,
  Mohite, Molina, Molina-Ruiz, Mondin, Montani, Moore, Moraru, Morawski,
  Moreno, Morisaki, Mours, Mow-Lowry, Mozzon, Muciaccia, Mukherjee, Mukherjee,
  Mukherjee, Mukherjee, Mukund, Mullavey, Munch, Mu{\~{n}}iz, Murray, Nadji,
  Nagar, Nardecchia, Naticchioni, Nayak, Neil, Neilson, Nelemans, Nelson, Nery,
  Neunzert, Nitz, Ng, Ng, Nguyen, Nguyen, Nguyen, Nichols, Nissanke, Nocera,
  Noh, North, Nothard, Nuttall, Oberling, O'Brien, O'Dell, Oganesyan, Ogin, Oh,
  Oh, Ohme, Ohta, Okada, Olivetto, Oppermann, Oram, O'Reilly, Ormiston, Ortega,
  O'Shaughnessy, Ossokine, Osthelder, Ottaway, Overmier, Owen, Pace, Pagano,
  Page, Pagliaroli, Pai, Pai, Palamos, Palashov, Palomba, Pan, Panda, Pang,
  Pankow, Pannarale, Pant, Paoletti, Paoli, Paolone, Parker, Pascucci,
  Pasqualetti, Passaquieti, Passuello, Patel, Patricelli, Payne, Pechsiri,
  Pedraza, Pegoraro, Pele, Penn, Perego, Perez, P{\'{e}}rigois, Perreca,
  Perri{\`{e}}s, Petermann, Petterson, Pfeiffer, Pham, Phukon, Piccinni,
  Pichot, Piendibene, Piergiovanni, Pierini, Pierro, Pillant, Pilo, Pinard,
  Pinto, Piotrzkowski, Pirello, Pitkin, Placidi, Plastino, Pluchar, Poggiani,
  Polini, Pong, Ponrathnam, Popolizio, Porter, Poverman, Powell, Pracchia,
  Prajapati, Prasai, Prasanna, Pratten, Prestegard, Principe, Prodi, Prokhorov,
  Prosposito, Prudenzi, Puecher, Punturo, Puosi, Puppo, P{\"{u}}rrer, Qi,
  Quetschke, Quinonez, Quitzow-James, Raab, Raaijmakers, Radkins, Radulesco,
  Raffai, Rafferty, Rail, Raja, Rajan, Rajbhandari, Rakhmanov, Ramirez,
  Ramirez, Ramos-Buades, Rana, Rao, Rapagnani, Rapol, Ratto, Raymond, Razzano,
  Read, Regimbau, Rei, Reid, Reitze, Rettegno, Ricci, Richardson, Richardson,
  Richardson, Ricker, Riemenschneider, Riles, Rizzo, Robertson, Robinet,
  Rocchi, Rocha, Rodriguez, Rodriguez-Soto, Rolland, Rollins, Roma, Romanelli,
  Romano, Romel, Romero, Romero-Shaw, Romie, Ronchini, Rose, Rose, Rose,
  Rosell, Rosi{\'{n}}ska, Rosofsky, Ross, Rowan, Rowlinson, Roy, Roy, Ruggi,
  Ryan, Sachdev, Sadecki, Sadiq, Sakellariadou, Salafia, Salconi, Saleem,
  Samajdar, Sanchez, Sanchez, Sanchez, Sanchis-Gual, Sanders, Sandles,
  Santiago, Santos, Saravanan, Sarin, Sassolas, Sathyaprakash, Sauter, Savage,
  Savant, Sawant, Sayah, Schaetzl, Schale, Scheel, Scheuer, Schindler-Tyka,
  Schmidt, Schnabel, Schofield, Sch{\"{o}}nbeck, Schreiber, Schulte, Schutz,
  Schwarm, Schwartz, Scott, Scott, Seglar-Arroyo, Seidel, Sellers, Sengupta,
  Sennett, Sentenac, Sequino, Sergeev, Setyawati, Shaffer, Shahriar, Sharifi,
  Sharma, Sharma, Shawhan, Shen, Shikauchi, Shink, Shoemaker, Shoemaker,
  Shukla, Shyamsundar, Sieniawska, Sigg, Singer, Singh, Singh, Singha, Singhal,
  Sintes, Sipala, Skliris, Slagmolen, Slaven-Blair, Smetana, Smith, Smith,
  Somala, Son, Soni, Soni, Sorazu, Sordini, Sorrentino, Sorrentino, Soulard,
  Souradeep, Sowell, Spencer, Spera, Srivastava, Srivastava, Staats, Stachie,
  Steer, Steinhoff, Steinke, Steinlechner, Steinlechner, Steinmeyer, Stevenson,
  Stolle-Mcallister, Stops, Stover, Strain, Stratta, Strunk, Sturani, Stuver,
  S{\"{u}}dbeck, Sudhagar, Sudhir, Suh, Summerscales, Sun, Sun, Sunil, Sur,
  Suresh, Sutton, Swinkels, Szczepa{\'{n}}czyk, Tacca, Tait, Talbot,
  Tanasijczuk, Tanner, Tao, Tapia, {Tapia San Martin}, Tasson, Taylor, Tenorio,
  Terkowski, Thirugnanasambandam, Thomas, Thomas, Thomas, Thompson, Thondapu,
  Thorne, Thrane, Tiwari, Tiwari, Tiwari, Toland, Tolley, Tonelli, Tornasi,
  Torres-Forn{\'{e}}, Torrie, {E Melo}, T{\"{o}}yr{\"{a}}, Tran, Trapananti,
  Travasso, Traylor, Tringali, Tripathee, Trovato, Trudeau, Tsai, Tsang, Tse,
  Tso, Tsukada, Tsuna, Tsutsui, Turconi, Ubhi, Udall, Ueno, Ugolini,
  Unnikrishnan, Urban, Usman, Utina, Vahlbruch, Vajente, Vajpeyi, Valdes,
  Valentini, Valsan, {Van Bakel}, {Van Beuzekom}, {Van Den Brand}, {Van Den
  Broeck}, Vander-Hyde, {Van Der Schaaf}, {Van Heijningen}, Vardaro, Vargas,
  Varma, Vass, Vas{\'{u}}th, Vecchio, Vedovato, Veitch, Veitch, Venkateswara,
  Venneberg, Venugopalan, Verkindt, Verma, Veske, Vetrano, Vicer{\'{e}}, Viets,
  Vijaykumar, Villa-Ortega, Vinet, Vitale, Vo, Vocca, Vorvick, Vyatchanin,
  Wade, Wade, Wade, Walet, Walker, Wallace, Wallace, Walsh, Wang, Wang, Wang,
  Wang, Ward, Warner, Was, Washington, Watchi, Weaver, Wei, Weinert, Weinstein,
  Weiss, Wellmann, Wen, We{\ss}els, Westhouse, Wette, Whelan, White, White,
  Whiting, Whittle, Wilken, Williams, Williams, Williamson, Willis, Willke,
  Wilson, Wimmer, Winkler, Wipf, Woan, Woehler, Wofford, Wong, Wrangel, Wright,
  Wu, Wysocki, Xiao, Yamamoto, Yang, Yang, Yang, Yap, Yeeles, Yoon, Yu, Yu,
  Yuen, Zadro{\.{z}}ny, Zanolin, Zelenova, Zendri, Zevin, Zhang, Zhang, Zhang,
  Zhang, Zhao, Zhao, Zheng, Zhou, Zhou, Zhu, Zimmerman, Zlochower, Zucker, \&
  Zweizig}]{GWTC2}
---. 2021{\natexlab{b}}, Physical Review X, 11, 21053,
  \dodoi{10.1103/PhysRevX.11.021053}

\bibitem[{Acernese {et~al.}(2015)Acernese, Agathos, Agatsuma, Aisa, Allemandou,
  Allocca, Amarni, Astone, Balestri, Ballardin, Barone, Baronick, Barsuglia,
  Basti, Basti, Bauer, Bavigadda, Bejger, Beker, Belczynski, Bersanetti,
  Bertolini, Bitossi, Bizouard, Bloemen, Blom, Boer, Bogaert, Bondi, Bondu,
  Bonelli, Bonnand, Boschi, Bosi, Bouedo, Bradaschia, Branchesi, Briant,
  Brillet, Brisson, Bulik, Bulten, Buskulic, Buy, Cagnoli, Calloni, Campeggi,
  Canuel, Carbognani, Cavalier, Cavalieri, Cella, Cesarini, Mottin, Chincarini,
  Chiummo, Chua, Cleva, Coccia, Cohadon, Colla, Colombini, Conte, Coulon,
  Cuoco, Dalmaz, D'Antonio, Dattilo, Davier, Day, Debreczeni, Degallaix,
  Del{\'{e}}glise, Pozzo, Dereli, Rosa, Fiore, Lieto, Virgilio, Doets, Dolique,
  Drago, Ducrot, Endrczi, Fafone, Farinon, Ferrante, Ferrini, Fidecaro, Fiori,
  Flaminio, Fournier, Franco, Frasca, Frasconi, Gammaitoni, Garufi, Gaspard,
  Gatto, Gemme, Gendre, Genin, Gennai, Ghosh, Giacobone, Giazotto, Gouaty,
  Granata, Greco, Groot, Guidi, Harms, Heidmann, Heitmann, Hello, Hemming,
  Hennes, Hofman, Jaranowski, Jonker, Kasprzack, K{\'{e}}f{\'{e}}lian,
  Kowalska, Kraan, Kr{\'{o}}lak, Kutynia, Lazzaro, Leonardi, Leroy, Letendre,
  Li, Lieunard, Lorenzini, Loriette, Losurdo, Magazz{\'{u}}, Majorana,
  Maksimovic, Malvezzi, Man, Mangano, Mantovani, Marchesoni, Marion, Marque,
  Martelli, Martellini, Masserot, Meacher, Meidam, Mezzani, Michel, Milano,
  Minenkov, Moggi, Mohan, Montani, Morgado, Mours, Mul, Nagy, Nardecchia,
  Naticchioni, Nelemans, Neri, Neri, Nocera, Pacaud, Palomba, Paoletti, Paoli,
  Pasqualetti, Passaquieti, Passuello, Perciballi, Petit, Pichot, Piergiovanni,
  Pillant, Piluso, Pinard, Poggiani, Prijatelj, Prodi, Punturo, Puppo,
  Rabeling, R{\'{a}}cz, Rapagnani, Razzano, Re, Regimbau, Ricci, Robinet,
  Rocchi, Rolland, Romano, Rosi{\'{n}}ska, Ruggi, Saracco, Sassolas, Schimmel,
  Sentenac, Sequino, Shah, Siellez, Straniero, Swinkels, Tacca, Tonelli,
  Travasso, Turconi, Vajente, {Van Bakel}, {Van Beuzekom}, {Van Den Brand},
  {Van Den Broeck}, {Van Der Sluys}, {Van Heijningen}, Vas{\'{u}}th, Vedovato,
  Veitch, Verkindt, Vetrano, Vicer{\'{e}}, Vinet, Visser, Vocca, Ward, Was,
  Wei, Yvert, Zny, \& Zendri}]{aVirgo}
Acernese, F., Agathos, M., Agatsuma, K., {et~al.} 2015, Classical and Quantum
  Gravity, 32, 024001, \dodoi{10.1088/0264-9381/32/2/024001}

\bibitem[{Adams {et~al.}(2016)Adams, Buskulic, Germain, Guidi, Marion, Montani,
  Mours, Piergiovanni, \& Wang}]{Adams2016}
Adams, T., Buskulic, D., Germain, V., {et~al.} 2016, Classical and Quantum
  Gravity, 33, 175012, \dodoi{10.1088/0264-9381/33/17/175012}

\bibitem[{Ade {et~al.}(2016)Ade, Aghanim, Arnaud, Ashdown, Aumont, Baccigalupi,
  Banday, Barreiro, Bartlett, Bartolo, Battaner, Battye, Benabed, Beno{\^{i}}t,
  Benoit-L{\'{e}}vy, Bernard, Bersanelli, Bielewicz, Bock, Bonaldi, Bonavera,
  Bond, Borrill, Bouchet, Boulanger, Bucher, Burigana, Butler, Calabrese,
  Cardoso, Catalano, Challinor, Chamballu, Chary, Chiang, Chluba, Christensen,
  Church, Clements, Colombi, Colombo, Combet, Coulais, Crill, Curto, Cuttaia,
  Danese, Davies, Davis, {De Bernardis}, {De Rosa}, {De Zotti}, Delabrouille,
  D{\'{e}}sert, {Di Valentino}, Dickinson, Diego, Dolag, Dole, Donzelli,
  Dor{\'{e}}, Douspis, Ducout, Dunkley, Dupac, Efstathiou, Elsner, En{\ss}lin,
  Eriksen, Farhang, Fergusson, Finelli, Forni, Frailis, Fraisse, Franceschi,
  Frejsel, Galeotta, Galli, Ganga, Gauthier, Gerbino, Ghosh, Giard,
  Giraud-H{\'{e}}raud, Giusarma, Gjerl{\o}w, Gonz{\'{a}}lez-Nuevo,
  G{\'{o}}rski, Gratton, Gregorio, Gruppuso, Gudmundsson, Hamann, Hansen,
  Hanson, Harrison, Helou, Henrot-Versill{\'{e}}, Hern{\'{a}}ndez-Monteagudo,
  Herranz, Hildebrandt, Hivon, Hobson, Holmes, Hornstrup, Hovest, Huang,
  Huffenberger, Hurier, Jaffe, Jaffe, Jones, Juvela, Keih{\"{a}}nen, Keskitalo,
  Kisner, Kneissl, Knoche, Knox, Kunz, Kurki-Suonio, Lagache,
  L{\"{a}}hteenm{\"{a}}ki, Lamarre, Lasenby, Lattanzi, Lawrence, Leahy,
  Leonardi, Lesgourgues, Levrier, Lewis, Liguori, Lilje, Linden-V{\o}rnle,
  L{\'{o}}pez-Caniego, Lubin, Maci{\'{a}}s-P{\'{e}}rez, Maggio, Maino,
  Mandolesi, Mangilli, Marchini, Maris, Martin, Martinelli,
  Mart{\'{i}}nez-Gonz{\'{a}}lez, Masi, Matarrese, Mcgehee, Meinhold,
  Melchiorri, Melin, Mendes, Mennella, Migliaccio, Millea, Mitra,
  Miville-Desch{\^{e}}nes, Moneti, Montier, Morgante, Mortlock, Moss, Munshi,
  Murphy, Naselsky, Nati, Natoli, Netterfield, N{\o}rgaard-Nielsen, Noviello,
  Novikov, Novikov, Oxborrow, Paci, Pagano, Pajot, Paladini, Paoletti,
  Partridge, Pasian, Patanchon, Pearson, Perdereau, Perotto, Perrotta,
  Pettorino, Piacentini, Piat, Pierpaoli, Pietrobon, Plaszczynski,
  Pointecouteau, Polenta, Popa, Pratt, Pr{\'{e}}zeau, Prunet, Puget, Rachen,
  Reach, Rebolo, Reinecke, Remazeilles, Renault, Renzi, Ristorcelli, Rocha,
  Rosset, Rossetti, Roudier, {Rouill{\'{e}} D'orfeuil}, Rowan-Robinson,
  Rubin{\~{o}}-Mart{\'{i}}n, Rusholme, Said, Salvatelli, Salvati, Sandri,
  Santos, Savelainen, Savini, Scott, Seiffert, Serra, Shellard, Spencer,
  Spinelli, Stolyarov, Stompor, Sudiwala, Sunyaev, Sutton, Suur-Uski, Sygnet,
  Tauber, Terenzi, Toffolatti, Tomasi, Tristram, Trombetti, Tucci, Tuovinen,
  T{\"{u}}rler, Umana, Valenziano, Valiviita, {Van Tent}, Vielva, Villa, Wade,
  Wandelt, Wehus, White, White, Wilkinson, Yvon, Zacchei, \&
  Zonca}]{PlanckCollaboration2016}
Ade, P.~A., Aghanim, N., Arnaud, M., {et~al.} 2016, Astronomy and Astrophysics,
  594, A13, \dodoi{10.1051/0004-6361/201525830}

\bibitem[{Albanesi {et~al.}(2021)Albanesi, Nagar, \& Bernuzzi}]{Albanesi2021}
Albanesi, S., Nagar, A., \& Bernuzzi, S. 2021, arXiv e-prints.
\newblock \doarXiv{2104.10559}

\bibitem[{Allen {et~al.}(2012)Allen, Anderson, Brady, Brown, \&
  Creighton}]{Allen2012}
Allen, B., Anderson, W.~G., Brady, P.~R., Brown, D.~A., \& Creighton, J.~D.
  2012, Physical Review D - Particles, Fields, Gravitation and Cosmology, 85,
  122006, \dodoi{10.1103/PhysRevD.85.122006}

\bibitem[{Antognini {et~al.}(2014)Antognini, Shappee, Thompson, \&
  Amaro-seoane}]{Antognini2014}
Antognini, J.~M., Shappee, B.~J., Thompson, T.~A., \& Amaro-seoane, P. 2014,
  Monthly Notices of the Royal Astronomical Society, 439, 1079,
  \dodoi{10.1093/mnras/stu039}

\bibitem[{Antonini \& Gieles(2020{\natexlab{a}})}]{Antonini2020}
Antonini, F., \& Gieles, M. 2020{\natexlab{a}}, Physical Review D, 102, 123016,
  \dodoi{10.1103/PhysRevD.102.123016}

\bibitem[{Antonini \& Gieles(2020{\natexlab{b}})}]{Antonini2020a}
---. 2020{\natexlab{b}}, Monthly Notices of the Royal Astronomical Society,
  492, 2936, \dodoi{10.1093/mnras/stz3584}

\bibitem[{Antonini \& Perets(2012)}]{Antonini2012}
Antonini, F., \& Perets, H.~B. 2012, The Astrophysical Journal, 757, 27,
  \dodoi{10.1088/0004-637X/757/1/27}

\bibitem[{Antonini {et~al.}(2017)Antonini, Toonen, \& Hamers}]{Antonini2017a}
Antonini, F., Toonen, S., \& Hamers, A.~S. 2017, The Astrophysical Journal,
  841, 77, \dodoi{10.3847/1538-4357/aa6f5e}

\bibitem[{{Arca Sedda} {et~al.}(2018){Arca Sedda}, Askar, \&
  Giersz}]{ArcaSedda2018a}
{Arca Sedda}, M., Askar, A., \& Giersz, M. 2018, Monthly Notices of the Royal
  Astronomical Society, 479, 4652, \dodoi{10.1093/mnras/sty1859}

\bibitem[{Aubin {et~al.}(2021)Aubin, Brighenti, Chierici, Estevez, Greco,
  Guidi, Juste, Marion, Mours, Nitoglia, Sauter, \& Sordini}]{Aubin2021}
Aubin, F., Brighenti, F., Chierici, R., {et~al.} 2021, Classical and Quantum
  Gravity, 38, 095004, \dodoi{10.1088/1361-6382/abe913}

\bibitem[{Banerjee(2017)}]{Banerjee2017}
Banerjee, S. 2017, Monthly Notices of the Royal Astronomical Society, 467, 524,
  \dodoi{10.1093/mnras/stx2347}

\bibitem[{Bartos {et~al.}(2017)Bartos, Kocsis, Haiman, \&
  M{\'{a}}rka}]{Bartos2017}
Bartos, I., Kocsis, B., Haiman, Z., \& M{\'{a}}rka, S. 2017, The Astrophysical
  Journal, 835, 165, \dodoi{10.3847/1538-4357/835/2/165}

\bibitem[{Bavera {et~al.}(2021)Bavera, Fragos, Zevin, Berry, Marchant, Andrews,
  Coughlin, Dotter, Kovlakas, Misra, Serra-Perez, Qin, Rocha,
  Rom{\'{a}}n-Garza, Tran, \& Zapartas}]{Bavera2021}
Bavera, S.~S., Fragos, T., Zevin, M., {et~al.} 2021, Astronomy and
  Astrophysics, 647, A153, \dodoi{10.1051/0004-6361/202039804}

\bibitem[{Belczynski {et~al.}(2016)Belczynski, Holz, Bulik, \&
  O'shaughnessy}]{Belczynski2016}
Belczynski, K., Holz, D.~E., Bulik, T., \& O'shaughnessy, R. 2016, Nature, 534,
  512, \dodoi{10.1038/nature18322}

\bibitem[{Bethe \& Brown(1998)}]{Bethe1998}
Bethe, H.~A., \& Brown, G.~E. 1998, The Astrophysical Journal, 506, 780,
  \dodoi{10.1086/306265}

\bibitem[{Bird {et~al.}(2016)Bird, Cholis, Mu{\~{n}}oz, Ali-Ha{\"{i}}moud,
  Kamionkowski, Kovetz, Raccanelli, \& Riess}]{Bird2016}
Bird, S., Cholis, I., Mu{\~{n}}oz, J.~B., {et~al.} 2016, Physical Review
  Letters, 116, 201301, \dodoi{10.1103/PhysRevLett.116.201301}

\bibitem[{Bouffanais {et~al.}(2021)Bouffanais, Mapelli, Santoliquido, Giacobbo,
  {Di Carlo}, Rastello, Artale, \& Iorio}]{Bouffanais2021}
Bouffanais, Y., Mapelli, M., Santoliquido, F., {et~al.} 2021, Monthly Notices
  of the Royal Astronomical Society, 507, 5224, \dodoi{10.1093/mnras/stab2438}

\bibitem[{Breen \& Heggie(2013)}]{Breen2013}
Breen, P.~G., \& Heggie, D.~C. 2013, Monthly Notices of the Royal Astronomical
  Society, 436, 584, \dodoi{10.1093/mnras/stt1599}

\bibitem[{Buonanno \& Damour(1999)}]{Buonanno1999}
Buonanno, A., \& Damour, T. 1999, Physical Review D, 59, 084006,
  \dodoi{10.1103/PhysRevD.59.084006}

\bibitem[{Buonanno \& Damour(2000)}]{Buonanno2000}
---. 2000, Physical Review D, 62, 064015, \dodoi{10.1103/physrevd.62.064015}

\bibitem[{Bustillo {et~al.}(2021)Bustillo, Sanchis-Gual, Torres-Forn{\'{e}},
  Font, Vajpeyi, Smith, Herdeiro, Radu, \& Leong}]{Bustillo2021}
Bustillo, J.~C., Sanchis-Gual, N., Torres-Forn{\'{e}}, A., {et~al.} 2021,
  Physical Review Letters, 126, 081101, \dodoi{10.1103/PhysRevLett.126.081101}

\bibitem[{Chatterjee {et~al.}(2017)Chatterjee, Rodriguez, Kalogera, \&
  Rasio}]{Chatterjee2017}
Chatterjee, S., Rodriguez, C.~L., Kalogera, V., \& Rasio, F.~A. 2017, The
  Astrophysical Journal, 836, L26, \dodoi{10.3847/2041-8213/aa5caa}

\bibitem[{Chiaramello \& Nagar(2020)}]{Chiaramello2020}
Chiaramello, D., \& Nagar, A. 2020, Physical Review D, 101, 101501,
  \dodoi{10.1103/PhysRevD.101.101501}

\bibitem[{Chu {et~al.}(2020)Chu, Kovalam, Wen, Slaven-Blair, Bosveld, Chen,
  Clearwater, Codoreanu, Du, Guo, Guo, Kim, Li, Oloworaran, Panther, Powell,
  Sengupta, Wette, \& Zhu}]{Chu2020}
Chu, Q., Kovalam, M., Wen, L., {et~al.} 2020, arXiv e-prints.
\newblock \doarXiv{2011.06787}

\bibitem[{Clesse \& Garcia-Bellido(2020)}]{Clesse2020}
Clesse, S., \& Garcia-Bellido, J. 2020, arXiv e-prints.
\newblock \doarXiv{2007.06481}

\bibitem[{{Dal Canton} {et~al.}(2014){Dal Canton}, Nitz, Lundgren, Nielsen,
  Brown, Dent, Harry, Krishnan, Miller, Wette, Wiesner, \&
  Willis}]{DalCanton2014}
{Dal Canton}, T., Nitz, A.~H., Lundgren, A.~P., {et~al.} 2014, Physical Review
  D - Particles, Fields, Gravitation and Cosmology, 90, 082004,
  \dodoi{10.1103/PhysRevD.90.082004}

\bibitem[{Damour(2001)}]{Damour2001}
Damour, T. 2001, Physical Review D, 64, 124013,
  \dodoi{10.1103/PhysRevD.64.124013}

\bibitem[{Damour {et~al.}(2000)Damour, Jaranowski, \&
  Sch{\"{a}}fer}]{Damour2000}
Damour, T., Jaranowski, P., \& Sch{\"{a}}fer, G. 2000, Physical Review D, 62,
  084011, \dodoi{10.1103/PhysRevD.62.084011}

\bibitem[{Damour {et~al.}(2015)Damour, Jaranowski, \&
  Sch{\"{a}}fer}]{Damour2015}
---. 2015, Physical Review D, 91, 084024, \dodoi{10.1103/PhysRevD.91.084024}

\bibitem[{Damour \& Nagar(2014)}]{Damour2014}
Damour, T., \& Nagar, A. 2014, Physical Review D, 90, 044018,
  \dodoi{10.1103/PhysRevD.90.044018}

\bibitem[{Davies {et~al.}(2020)Davies, Dent, T{\'{a}}pai, Harry, McIsaac, \&
  Nitz}]{Davies2020}
Davies, G.~S., Dent, T., T{\'{a}}pai, M., {et~al.} 2020, Physical Review D,
  102, 22004, \dodoi{10.1103/PhysRevD.102.022004}

\bibitem[{di~Carlo {et~al.}(2019)di~Carlo, Giacobbo, Mapelli, Pasquato, Spera,
  Wang, \& Haardt}]{DiCarlo2019}
di~Carlo, U.~N., Giacobbo, N., Mapelli, M., {et~al.} 2019, Monthly Notices of
  the Royal Astronomical Society, 487, 2947, \dodoi{10.1093/mnras/stz1453}

\bibitem[{Dominik {et~al.}(2012)Dominik, Belczynski, Fryer, Holz, Berti, Bulik,
  Mandel, \& O'Shaughnessy}]{Dominik2012}
Dominik, M., Belczynski, K., Fryer, C., {et~al.} 2012, The Astrophysical
  Journal, 759, 52, \dodoi{10.1088/0004-637X/759/1/52}

\bibitem[{Downing {et~al.}(2010)Downing, Benacquista, Giersz, \&
  Spurzem}]{Downing2010}
Downing, J. M.~B., Benacquista, M.~J., Giersz, M., \& Spurzem, R. 2010, Monthly
  Notices of the Royal Astronomical Society, 407, 1946,
  \dodoi{10.1111/j.1365-2966.2010.17040.x}

\bibitem[{El-Badry {et~al.}(2019)El-Badry, Quataert, Weisz, Choksi, \&
  Boylan-Kolchin}]{ElBadry2019}
El-Badry, K., Quataert, E., Weisz, D.~R., Choksi, N., \& Boylan-Kolchin, M.
  2019, Monthly Notices of the Royal Astronomical Society, 482, 4528,
  \dodoi{10.1093/mnras/sty3007}

\bibitem[{Finn \& Chernoff(1993)}]{Finn1993}
Finn, L.~S., \& Chernoff, D.~F. 1993, Physical Review D, 47, 2198,
  \dodoi{10.1103/PhysRevD.47.2198}

\bibitem[{Fragione \& Bromberg(2019)}]{Fragione2019a}
Fragione, G., \& Bromberg, O. 2019, Monthly Notices of the Royal Astronomical
  Society, 488, 4370, \dodoi{10.1093/mnras/stz2024}

\bibitem[{Fragione {et~al.}(2019)Fragione, Grishin, Leigh, Perets, \&
  Perna}]{Fragione2019b}
Fragione, G., Grishin, E., Leigh, N. W.~C., Perets, H.~B., \& Perna, R. 2019,
  Monthly Notices of the Royal Astronomical Society, 488, 2825,
  \dodoi{10.1093/mnras/stz1803}

\bibitem[{Fragione \& Kocsis(2018)}]{Fragione2018}
Fragione, G., \& Kocsis, B. 2018, Physical Review Letters, 121, 161103,
  \dodoi{10.1103/PhysRevLett.121.161103}

\bibitem[{Fragione \& Kocsis(2019)}]{Fragione2019}
---. 2019, Monthly Notices of the Royal Astronomical Society, 486, 4781,
  \dodoi{10.1093/mnras/stz1175}

\bibitem[{Franciolini {et~al.}(2021)Franciolini, Baibhav, {De Luca}, Ng, Wong,
  Berti, Pani, Riotto, \& Vitale}]{Franciolini2021}
Franciolini, G., Baibhav, V., {De Luca}, V., {et~al.} 2021, arXiv e-prints.
\newblock \doarXiv{2105.03349}

\bibitem[{Fregeau {et~al.}(2004)Fregeau, Cheung, Zwart, \& Rasio}]{Fregeau2004}
Fregeau, J.~M., Cheung, P., Zwart, S. F.~P., \& Rasio, F.~A. 2004, Monthly
  Notices of the Royal Astronomical Society, 352, 1,
  \dodoi{10.1111/j.1365-2966.2004.07914.x}

\bibitem[{Fregeau \& Rasio(2007)}]{Fregeau2007}
Fregeau, J.~M., \& Rasio, F.~A. 2007, The Astrophysical Journal, 658, 1047,
  \dodoi{10.1086/511809}

\bibitem[{Fuller {et~al.}(2019)Fuller, Piro, \& Jermyn}]{Fuller2019a}
Fuller, J., Piro, A.~L., \& Jermyn, A.~S. 2019, Monthly Notices of the Royal
  Astronomical Society, 485, 3661, \dodoi{10.1093/mnras/stz514}

\bibitem[{Gamba {et~al.}(2021)Gamba, Breschi, Carullo, Rettegno, Albanesi,
  Bernuzzi, \& Nagar}]{Gamba2021}
Gamba, R., Breschi, M., Carullo, G., {et~al.} 2021, arXiv e-prints.
\newblock \doarXiv{2106.05575}

\bibitem[{Gayathri {et~al.}(2020)Gayathri, Healy, Lange, O'Brien, Szczepanczyk,
  Bartos, Campanelli, Klimenko, Lousto, \& O'Shaughnessy}]{Gayathri2020}
Gayathri, V., Healy, J., Lange, J., {et~al.} 2020, arXiv e-prints.
\newblock \doarXiv{2009.05461}

\bibitem[{Giesers {et~al.}(2018)Giesers, Dreizler, Husser, Kamann,
  Escud{\'{e}}, Brinchmann, {Marcella Carollo}, Roth, Weilbacher, \&
  Wisotzki}]{Giesers2018}
Giesers, B., Dreizler, S., Husser, T.~O., {et~al.} 2018, Monthly Notices of the
  Royal Astronomical Society: Letters, 475, L15, \dodoi{10.1093/mnrasl/slx203}

\bibitem[{Giesers {et~al.}(2019)Giesers, Kamann, Dreizler, Husser, Askar,
  G{\"{o}}ttgens, Brinchmann, Latour, Weilbacher, Wendt, \& Roth}]{Giesers2019}
Giesers, B., Kamann, S., Dreizler, S., {et~al.} 2019, Astronomy and
  Astrophysics, 632, A3, \dodoi{10.1051/0004-6361/201936203}

\bibitem[{Gond{\'{a}}n \& Kocsis(2021)}]{Gondan2021}
Gond{\'{a}}n, L., \& Kocsis, B. 2021, Monthly Notices of the Royal Astronomical
  Society, 506, 1665, \dodoi{10.1093/mnras/stab1722}

\bibitem[{Gond{\'{a}}n {et~al.}(2018)Gond{\'{a}}n, Kocsis, Raffai, \&
  Frei}]{Gondan2018}
Gond{\'{a}}n, L., Kocsis, B., Raffai, P., \& Frei, Z. 2018, The Astrophysical
  Journal, 860, 5, \dodoi{10.3847/1538-4357/aabfee}

\bibitem[{Gr{\"{o}}bner {et~al.}(2020)Gr{\"{o}}bner, Ishibashi, Tiwari, Haney,
  Jetzer, \& Gr{\"{o}}bner}]{Grobner2020}
Gr{\"{o}}bner, M., Ishibashi, W., Tiwari, S., {et~al.} 2020, Astronomy and
  Astrophysics, 638, A119, \dodoi{10.1051/0004-6361/202037681}

\bibitem[{Holgado {et~al.}(2021)Holgado, Ortega, \& Rodriguez}]{Holgado2021a}
Holgado, A.~M., Ortega, A., \& Rodriguez, C.~L. 2021, The Astrophysical Journal
  Letters, 909, L24, \dodoi{10.3847/2041-8213/abe7f5}

\bibitem[{Hooper {et~al.}(2012)Hooper, Chung, Luan, Blair, Chen, \&
  Wen}]{Hooper2012}
Hooper, S., Chung, S.~K., Luan, J., {et~al.} 2012, Physical Review D -
  Particles, Fields, Gravitation and Cosmology, 86, 024012,
  \dodoi{10.1103/PhysRevD.86.024012}

\bibitem[{Hunter(2007)}]{matplotlib}
Hunter, J.~D. 2007, Computing in Science and Engineering, 9, 99,
  \dodoi{10.1109/MCSE.2007.55}

\bibitem[{Jord{\'{a}}n {et~al.}(2015)Jord{\'{a}}n, Peng, Blakeslee,
  C{\^{o}}t{\'{e}}, Eyheramendy, \& Ferrarese}]{Jordan2015}
Jord{\'{a}}n, A., Peng, E.~W., Blakeslee, J.~P., {et~al.} 2015, Astrophysical
  Journal, Supplement Series, 221, 13, \dodoi{10.1088/0067-0049/221/1/13}

\bibitem[{Klimenko {et~al.}(2016)Klimenko, Vedovato, Drago, Salemi, Tiwari,
  Prodi, Lazzaro, Ackley, Tiwari, {Da Silva}, \& Mitselmakher}]{Klimenko2016}
Klimenko, S., Vedovato, G., Drago, M., {et~al.} 2016, Physical Review D, 93,
  042004, \dodoi{10.1103/PhysRevD.93.042004}

\bibitem[{Kocsis \& Levin(2012)}]{Kocsis2012}
Kocsis, B., \& Levin, J. 2012, Physical Review D, 85, 123005,
  \dodoi{10.1103/PhysRevD.85.123005}

\bibitem[{Kremer {et~al.}(2019)Kremer, Chatterjee, Ye, Rodriguez, \&
  Rasio}]{Kremer2019c}
Kremer, K., Chatterjee, S., Ye, C.~S., Rodriguez, C.~L., \& Rasio, F.~A. 2019,
  The Astrophysical Journal, 871, 38, \dodoi{10.3847/1538-4357/aaf646}

\bibitem[{Kremer {et~al.}(2018)Kremer, Ye, Chatterjee, Rodriguez, \&
  Rasio}]{Kremer2018}
Kremer, K., Ye, C.~S., Chatterjee, S., Rodriguez, C.~L., \& Rasio, F.~A. 2018,
  The Astrophysical Journal Letters, 855, L15, \dodoi{10.3847/2041-8213/aab26c}

\bibitem[{Kremer {et~al.}(2020)Kremer, Ye, Rui, Weatherford, Chatterjee,
  Fragione, Rodriguez, Spera, \& Rasio}]{Kremer2020}
Kremer, K., Ye, C.~S., Rui, N.~Z., {et~al.} 2020, The Astrophysical Journal
  Supplement Series, 247, 48, \dodoi{10.3847/1538-4365/ab7919}

\bibitem[{Lada \& Lada(2003)}]{Lada2003}
Lada, C.~J., \& Lada, E.~A. 2003, Annual Review of Astronomy and Astrophysics,
  41, 57, \dodoi{10.1146/annurev.astro.41.011802.094844}

\bibitem[{Liu \& Lai(2019)}]{Liu2019}
Liu, B., \& Lai, D. 2019, Monthly Notices of the Royal Astronomical Society,
  483, 4060, \dodoi{10.1093/mnras/sty3432}

\bibitem[{Liu {et~al.}(2019)Liu, Lai, \& Wang}]{Liu2019a}
Liu, B., Lai, D., \& Wang, Y.-H. 2019, The Astrophysical Journal, 881, 41,
  \dodoi{10.3847/1538-4357/ab2dfb}

\bibitem[{Loutrel(2020)}]{Loutrel2020}
Loutrel, N. 2020, arXiv e-prints.
\newblock \doarXiv{2009.11332}

\bibitem[{Lower {et~al.}(2018)Lower, Thrane, Lasky, \& Smith}]{Lower2018}
Lower, M.~E., Thrane, E., Lasky, P.~D., \& Smith, R. 2018, Physical Review D,
  98, 083028, \dodoi{10.1103/PhysRevD.98.083028}

\bibitem[{Mackey {et~al.}(2007)Mackey, Wilkinson, Davies, \&
  Gilmore}]{Mackey2007}
Mackey, A.~D., Wilkinson, M.~I., Davies, M.~B., \& Gilmore, G.~F. 2007, Monthly
  Notices of the Royal Astronomical Society: Letters, 379, 40,
  \dodoi{10.1111/j.1745-3933.2007.00330.x}

\bibitem[{Martinez {et~al.}(2021)Martinez, Rodriguez, \&
  Fragione}]{Martinez2021}
Martinez, M. A.~S., Rodriguez, C.~L., \& Fragione, G. 2021, arXiv e-prints.
\newblock \doarXiv{2105.01671}

\bibitem[{McKernan {et~al.}(2014)McKernan, Ford, Kocsis, Lyra, \&
  Winter}]{Mckernan2014}
McKernan, B., Ford, K.~E., Kocsis, B., Lyra, W., \& Winter, L.~M. 2014, Monthly
  Notices of the Royal Astronomical Society, 441, 900,
  \dodoi{10.1093/mnras/stu553}

\bibitem[{McKernan {et~al.}(2020)McKernan, Ford, \&
  O'Shaughnessy}]{McKernan2020}
McKernan, B., Ford, K.~E., \& O'Shaughnessy, R. 2020, Monthly Notices of the
  Royal Astronomical Society, 498, 4088, \dodoi{10.1093/mnras/staa2681}

\bibitem[{McKinney(2010)}]{pandas}
McKinney, W. 2010, in Proceedings of the 9th Python in Science Conference, ed.
  S.~van~der Walt \& J.~Millman, 51--56, \dodoi{10.25080/Majora-92bf1922-00a}

\bibitem[{Messick {et~al.}(2017)Messick, Blackburn, Brady, Brockill, Cannon,
  Cariou, Caudill, Chamberlin, Creighton, Everett, Hanna, Keppel, Lang, Li,
  Meacher, Nielsen, Pankow, Privitera, Qi, Sachdev, Sadeghian, Singer, Thomas,
  Wade, Wade, Weinstein, \& Wiesner}]{Messick2017}
Messick, C., Blackburn, K., Brady, P., {et~al.} 2017, Physical Review D, 95,
  042001, \dodoi{10.1103/PhysRevD.95.042001}

\bibitem[{Michaely \& Perets(2019)}]{Michaely2019}
Michaely, E., \& Perets, H.~B. 2019, The Astrophysical Journal, 887, L36,
  \dodoi{10.3847/2041-8213/ab5b9b}

\bibitem[{Michaely \& Perets(2020)}]{Michaely2020}
---. 2020, Monthly Notices of the Royal Astronomical Society, 498, 4924,
  \dodoi{10.1093/mnras/staa2720}

\bibitem[{Nagar {et~al.}(2021)Nagar, Bonino, \& Rettegno}]{Nagar2021}
Nagar, A., Bonino, A., \& Rettegno, P. 2021, Physical Review D, 103, 104021,
  \dodoi{10.1103/physrevd.103.104021}

\bibitem[{Nagar {et~al.}(2018)Nagar, Bernuzzi, {Del Pozzo}, Riemenschneider,
  Akcay, Carullo, Fleig, Babak, Tsang, Colleoni, Messina, Pratten, Radice,
  Rettegno, Agathos, Fauchon-Jones, Hannam, Husa, Dietrich, Cerd{\'{a}}-Duran,
  Font, Pannarale, Schmidt, \& Damour}]{Nagar2018}
Nagar, A., Bernuzzi, S., {Del Pozzo}, W., {et~al.} 2018, Physical Review D, 98,
  104052, \dodoi{10.1103/PhysRevD.98.104052}

\bibitem[{Nitz {et~al.}(2019)Nitz, Harry, Brown, Biwer, Willis, {Dal Canton},
  Capano, Pekowsky, Dent, Williamson, Cabero, De, Davies, Macleod,
  Machenschalk, Kumar, Reyes, Massinger, Pannarale, T{\'{a}}pai, Dfinstad,
  Fairhurst, Khan, Nielsen, Shasvath, Singer, Kumar, Idorrington92, Gabbard, \&
  {Varsha Uday Gadre}}]{PyCBC_v1.14.4}
Nitz, A., Harry, I., Brown, D., {et~al.} 2019, {gwastro/pycbc: PyCBC Release
  v1.14.4},  Zenodo, \dodoi{10.5281/zenodo.3546372}

\bibitem[{Nitz {et~al.}(2017)Nitz, Dent, {Dal Canton}, Fairhurst, \&
  Brown}]{Nitz2017}
Nitz, A.~H., Dent, T., {Dal Canton}, T., Fairhurst, S., \& Brown, D.~A. 2017,
  The Astrophysical Journal, 849, 118, \dodoi{10.3847/1538-4357/aa8f50}

\bibitem[{Nitz {et~al.}(2020)Nitz, Sch{\"{a}}fer, \& Canton}]{Nitz2020b}
Nitz, A.~H., Sch{\"{a}}fer, M., \& Canton, T.~D. 2020, The Astrophysical
  Journal Letters, 902, L29, \dodoi{10.3847/2041-8213/abbc10}

\bibitem[{O'Leary {et~al.}(2009)O'Leary, Kocsis, \& Loeb}]{OLeary2009}
O'Leary, R.~M., Kocsis, B., \& Loeb, A. 2009, Monthly Notices of the Royal
  Astronomical Society, 395, 2127, \dodoi{10.1111/j.1365-2966.2009.14653.x}

\bibitem[{O'Leary {et~al.}(2006)O'Leary, Rasio, Fregeau, Ivanova, \&
  O'Shaughnessy}]{OLeary2006}
O'Leary, R.~M., Rasio, F.~A., Fregeau, J.~M., Ivanova, N., \& O'Shaughnessy, R.
  2006, The Astrophysical Journal, 637, 937, \dodoi{10.1086/498446}

\bibitem[{Oliphant(2006)}]{numpy}
Oliphant, T.~E. 2006, {A guide to NumPy} (USA: Trelgol Publishing).
\newblock \url{https://web.mit.edu/dvp/Public/numpybook.pdf}

\bibitem[{P{\'{e}}rez \& Granger(2007)}]{ipython}
P{\'{e}}rez, F., \& Granger, B.~E. 2007, IEEE Journals {\&} Magazines, 9, 21,
  \dodoi{10.1109/MCSE.2007.53}

\bibitem[{Peters(1964)}]{Peters1964}
Peters, P.~C. 1964, Physical Review, 136, 1224,
  \dodoi{10.1103/PhysRev.136.B1224}

\bibitem[{Petrovich \& Antonini(2017)}]{Petrovich2017}
Petrovich, C., \& Antonini, F. 2017, The Astrophysical Journal, 846, 146,
  \dodoi{10.3847/1538-4357/aa8628}

\bibitem[{{Portegies Zwart} \& McMillan(2000)}]{PortegiesZwart2000}
{Portegies Zwart}, S.~F., \& McMillan, S. L.~W. 2000, The Astrophysical Journal
  Letters, 528, 17, \dodoi{10.1086/312422}

\bibitem[{Price-Whelan {et~al.}(2018)Price-Whelan, Sipocz, G{\"{u}}nther, Lim,
  Crawford, Conseil, Shupe, Craig, Dencheva, Ginsburg, VanderPlas, Bradley,
  P{\'{e}}rez-Su{\'{a}}rez, de~Val-Borro, Aldcroft, Cruz, Robitaille, Tollerud,
  Ardelean, Babej, Bachetti, Bakanov, Bamford, Barentsen, Barmby, Baumbach,
  Berry, Biscani, Boquien, Bostroem, Bouma, Brammer, Bray, Breytenbach,
  Buddelmeijer, Burke, Calderone, Rodr{\'{i}}guez, Cara, Cardoso, Cheedella,
  Copin, Crichton, D{\'{A}}vella, Deil, Depagne, Dietrich, Donath, Droettboom,
  Earl, Erben, Fabbro, Ferreira, Finethy, Fox, Garrison, Gibbons, Goldstein,
  Gommers, Greco, Greenfield, Groener, Grollier, Hagen, Hirst, Homeier, Horton,
  Hosseinzadeh, Hu, Hunkeler, Ivezi{\'{c}}, Jain, Jenness, Kanarek, Kendrew,
  Kern, Kerzendorf, Khvalko, King, Kirkby, Kulkarni, Kumar, Lee, Lenz,
  Littlefair, Ma, Macleod, Mastropietro, McCully, Montagnac, Morris, Mueller,
  Mumford, Muna, Murphy, Nelson, Nguyen, Ninan, N{\"{o}}the, Ogaz, Oh, Parejko,
  Parley, Pascual, Patil, Patil, Plunkett, Prochaska, Rastogi, Janga, Sabater,
  Sakurikar, Seifert, Sherbert, Sherwood-Taylor, Shih, Sick, Silbiger,
  Singanamalla, Singer, Sladen, Sooley, Sornarajah, Streicher, Teuben, Thomas,
  Tremblay, Turner, Terr{\'{o}}n, van Kerkwijk, de~la Vega, Watkins, Weaver,
  Whitmore, Woillez, \& Zabalza}]{TheAstropyCollaboration2018}
Price-Whelan, A.~M., Sipocz, B.~M., G{\"{u}}nther, H.~M., {et~al.} 2018, The
  Astronomical Journal, 156, 123, \dodoi{10.3847/1538-3881/aabc4f}

\bibitem[{Ramos-Buades {et~al.}(2020)Ramos-Buades, Tiwari, Haney, \&
  Husa}]{Ramos-Buades2020}
Ramos-Buades, A., Tiwari, S., Haney, M., \& Husa, S. 2020, Physical Review D,
  102, 043005, \dodoi{10.1103/PhysRevD.102.043005}

\bibitem[{Rasskazov \& Kocsis(2019)}]{Rasskazov2019}
Rasskazov, A., \& Kocsis, B. 2019, The Astrophysical Journal, 881, 20,
  \dodoi{10.3847/1538-4357/ab2c74}

\bibitem[{Robitaille {et~al.}(2013)Robitaille, Tollerud, Greenfield,
  Droettboom, Bray, Aldcroft, Davis, Ginsburg, Price-Whelan, Kerzendorf,
  Conley, Crighton, Barbary, Muna, Ferguson, Grollier, Parikh, Nair,
  G{\"{u}}nther, Deil, Woillez, Conseil, Kramer, Turner, Singer, Fox, Weaver,
  Zabalza, Edwards, Bostroem, Burke, Casey, Crawford, Dencheva, Ely, Jenness,
  Labrie, Lim, Pierfederici, Pontzen, Ptak, Refsdal, Servillat, \&
  Streicher}]{TheAstropyCollaboration2013}
Robitaille, T.~P., Tollerud, E.~J., Greenfield, P., {et~al.} 2013, Astronomy
  {\&} Astrophysics, 558, A33, \dodoi{10.1051/0004-6361/201322068}

\bibitem[{Rodriguez {et~al.}(2018{\natexlab{a}})Rodriguez, Amaro-seoane,
  Chatterjee, Kremer, Rasio, Samsing, Ye, \& Zevin}]{Rodriguez2018c}
Rodriguez, C.~L., Amaro-seoane, P., Chatterjee, S., {et~al.}
  2018{\natexlab{a}}, Physical Review D, 98, 123005,
  \dodoi{10.1103/PhysRevD.98.123005}

\bibitem[{Rodriguez {et~al.}(2018{\natexlab{b}})Rodriguez, Amaro-Seoane,
  Chatterjee, \& Rasio}]{Rodriguez2018b}
Rodriguez, C.~L., Amaro-Seoane, P., Chatterjee, S., \& Rasio, F.~A.
  2018{\natexlab{b}}, Physical Review Letters, 120, 151101,
  \dodoi{10.1103/PhysRevLett.120.151101}

\bibitem[{Rodriguez \& Antonini(2018)}]{Rodriguez2018}
Rodriguez, C.~L., \& Antonini, F. 2018, The Astrophysical Journal, 863, 7,
  \dodoi{10.3847/1538-4357/aacea4}

\bibitem[{Rodriguez {et~al.}(2016)Rodriguez, Chatterjee, \&
  Rasio}]{Rodriguez2016a}
Rodriguez, C.~L., Chatterjee, S., \& Rasio, F.~A. 2016, Physical Review D, 93,
  084029, \dodoi{10.1103/PhysRevD.93.084029}

\bibitem[{Rodriguez {et~al.}(2021)Rodriguez, Kremer, Chatterjee, Fragione,
  Loeb, Rasio, Weatherford, \& Ye}]{Rodriguez2021}
Rodriguez, C.~L., Kremer, K., Chatterjee, S., {et~al.} 2021, Research Notes of
  the AAS, 5, 19, \dodoi{10.3847/2515-5172/abdf54}

\bibitem[{Rodriguez \& Loeb(2018)}]{Rodriguez2018a}
Rodriguez, C.~L., \& Loeb, A. 2018, The Astrophysical Journal Letters, 866, L5,
  \dodoi{10.3847/2041-8213/aae377}

\bibitem[{Romero-Shaw {et~al.}(2020)Romero-Shaw, Lasky, Thrane, \&
  Bustillo}]{Romero-Shaw2020b}
Romero-Shaw, I., Lasky, P.~D., Thrane, E., \& Bustillo, J.~C. 2020, The
  Astrophysical Journal Letters, 903, L5, \dodoi{10.3847/2041-8213/abbe26}

\bibitem[{Romero-Shaw {et~al.}(2019)Romero-Shaw, Lasky, \&
  Thrane}]{Romero-Shaw2019}
Romero-Shaw, I.~M., Lasky, P.~D., \& Thrane, E. 2019, Monthly Notices of the
  Royal Astronomical Society, 490, 5210, \dodoi{10.1093/mnras/stz2996}

\bibitem[{Romero-Shaw {et~al.}(2021)Romero-Shaw, Lasky, \&
  Thrane}]{Romero-Shaw2021}
---. 2021, arXiv e-prints.
\newblock \doarXiv{2108.01284}

\bibitem[{Samsing(2018)}]{Samsing2018d}
Samsing, J. 2018, Physical Review D, 97, 103014,
  \dodoi{10.1103/PhysRevD.97.103014}

\bibitem[{Samsing {et~al.}(2018)Samsing, Askar, \& Giersz}]{Samsing2018c}
Samsing, J., Askar, A., \& Giersz, M. 2018, The Astrophysical Journal, 855,
  124, \dodoi{10.3847/1538-4357/aaab52}

\bibitem[{Samsing {et~al.}(2020{\natexlab{a}})Samsing, D'Orazio, Kremer,
  Rodriguez, \& Askar}]{Samsing2020b}
Samsing, J., D'Orazio, D.~J., Kremer, K., Rodriguez, C.~L., \& Askar, A.
  2020{\natexlab{a}}, Physical Review D, 101, 123010,
  \dodoi{10.1103/PhysRevD.101.123010}

\bibitem[{Samsing {et~al.}(2014)Samsing, MacLeod, \&
  Ramirez-Ruiz}]{Samsing2014}
Samsing, J., MacLeod, M., \& Ramirez-Ruiz, E. 2014, The Astrophysical Journal,
  784, 71, \dodoi{10.1088/0004-637X/784/1/71}

\bibitem[{Samsing \& Ramirez-Ruiz(2017)}]{Samsing2017c}
Samsing, J., \& Ramirez-Ruiz, E. 2017, The Astrophysical Journal Letters, 840,
  L14, \dodoi{10.3847/2041-8213/aa6f0b}

\bibitem[{Samsing {et~al.}(2020{\natexlab{b}})Samsing, Bartos, D'Orazio,
  Haiman, Kocsis, Leigh, Liu, Pessah, \& Tagawa}]{Samsing2020a}
Samsing, J., Bartos, I., D'Orazio, D.~J., {et~al.} 2020{\natexlab{b}}, arXiv
  e-prints.
\newblock \doarXiv{2010.09765}

\bibitem[{Sasaki {et~al.}(2018)Sasaki, Suyama, Tanaka, \&
  Yokoyama}]{Sasaki2018}
Sasaki, M., Suyama, T., Tanaka, T., \& Yokoyama, S. 2018, Classical and Quantum
  Gravity, 35, 063001, \dodoi{10.1088/1361-6382/aaa7b4}

\bibitem[{Silsbee \& Tremaine(2017)}]{Silsbee2017}
Silsbee, K., \& Tremaine, S. 2017, The Astrophysical Journal, 836, 39,
  \dodoi{10.3847/1538-4357/aa5729}

\bibitem[{Spruit(1999)}]{Spruit1999}
Spruit, H.~C. 1999, Astronomy and Astrophysics, 349, 189.
\newblock \doarXiv{9907138}

\bibitem[{Spruit(2002)}]{Spruit2002}
---. 2002, Astronomy {\&} Astrophysics, 381, 923, \dodoi{10.1051/0004-6361}

\bibitem[{Stone {et~al.}(2017)Stone, Metzger, \& Haiman}]{Stone2017}
Stone, N.~C., Metzger, B.~D., \& Haiman, Z. 2017, Monthly Notices of the Royal
  Astronomical Society, 464, 946, \dodoi{10.1093/mnras/stw2260}

\bibitem[{Strader {et~al.}(2012)Strader, Chomiuk, MacCarone, Miller-Jones, \&
  Seth}]{Strader2012}
Strader, J., Chomiuk, L., MacCarone, T.~J., Miller-Jones, J.~C., \& Seth, A.~C.
  2012, Nature, 490, 71, \dodoi{10.1038/nature11490}

\bibitem[{Tagawa {et~al.}(2021)Tagawa, Kocsis, Haiman, Bartos, Omukai, \&
  Samsing}]{Tagawa2021a}
Tagawa, H., Kocsis, B., Haiman, Z., {et~al.} 2021, The Astrophysical Journal,
  907, L20, \dodoi{10.3847/2041-8213/abd4d3}

\bibitem[{Takatsy {et~al.}(2019)Takatsy, B{\'{e}}csy, \& Raffai}]{Takatsy2019}
Takatsy, J., B{\'{e}}csy, B., \& Raffai, P. 2019, Monthly Notices of the Royal
  Astronomical Society, 486, 570, \dodoi{10.1093/mnras/stz820}

\bibitem[{Tiwari {et~al.}(2016)Tiwari, Klimenko, Christensen, Huerta,
  Mohapatra, Gopakumar, Haney, Ajith, McWilliams, Vedovato, Drago, Salemi,
  Prodi, Lazzaro, Tiwari, Mitselmakher, \& {Da Silva}}]{Tiwari2016}
Tiwari, V., Klimenko, S., Christensen, N., {et~al.} 2016, Physical Review D,
  93, 043007, \dodoi{10.1103/PhysRevD.93.043007}

\bibitem[{Usman {et~al.}(2016)Usman, Nitz, Harry, Biwer, Brown, Cabero, Capano,
  Canton, Dent, Fairhurst, Kehl, Keppel, Krishnan, Lenon, Lundgren, Nielsen,
  Pekowsky, Pfeiffer, Saulson, West, \& Willis}]{Usman2016}
Usman, S.~A., Nitz, A.~H., Harry, I.~W., {et~al.} 2016, Classical and Quantum
  Gravity, 33, 215004, \dodoi{10.1088/0264-9381/33/21/215004}

\bibitem[{{Van Der Walt} {et~al.}(2011){Van Der Walt}, Colbert, \&
  Varoquaux}]{numpy2}
{Van Der Walt}, S., Colbert, S.~C., \& Varoquaux, G. 2011, Computing in Science
  and Engineering, 13, 22, \dodoi{10.1109/MCSE.2011.37}

\bibitem[{Virtanen {et~al.}(2020)Virtanen, Gommers, Oliphant, Haberland, Reddy,
  Cournapeau, Burovski, Peterson, Weckesser, Bright, van~der Walt, Brett,
  Wilson, Millman, Mayorov, Nelson, Jones, Kern, Larson, Carey, Polat, Feng,
  Moore, VanderPlas, Laxalde, Perktold, Cimrman, Henriksen, Quintero, Harris,
  Archibald, Ribeiro, Pedregosa, van Mulbregt, Vijaykumar, Bardelli, Rothberg,
  Hilboll, Kloeckner, Scopatz, Lee, Rokem, Woods, Fulton, Masson,
  H{\"{a}}ggstr{\"{o}}m, Fitzgerald, Nicholson, Hagen, Pasechnik, Olivetti,
  Martin, Wieser, Silva, Lenders, Wilhelm, Young, Price, Ingold, Allen, Lee,
  Audren, Probst, Dietrich, Silterra, Webber, Slavi{\v{c}}, Nothman, Buchner,
  Kulick, Sch{\"{o}}nberger, {de Miranda Cardoso}, Reimer, Harrington,
  Rodr{\'{i}}guez, Nunez-Iglesias, Kuczynski, Tritz, Thoma, Newville,
  K{\"{u}}mmerer, Bolingbroke, Tartre, Pak, Smith, Nowaczyk, Shebanov, Pavlyk,
  Brodtkorb, Lee, McGibbon, Feldbauer, Lewis, Tygier, Sievert, Vigna, Peterson,
  More, Pudlik, Oshima, Pingel, Robitaille, Spura, Jones, Cera, Leslie, Zito,
  Krauss, Upadhyay, Halchenko, \& V{\'{a}}zquez-Baeza}]{scipy}
Virtanen, P., Gommers, R., Oliphant, T.~E., {et~al.} 2020, Nature Methods, 17,
  261, \dodoi{10.1038/s41592-019-0686-2}

\bibitem[{Wang {et~al.}(2016)Wang, Spurzem, Aarseth, Giersz, Askar, Berczik,
  Naab, Schadow, \& Kouwenhoven}]{Wang2016}
Wang, L., Spurzem, R., Aarseth, S., {et~al.} 2016, Monthly Notices of the Royal
  Astronomical Society, 458, 1450, \dodoi{10.1093/mnras/stw274}

\bibitem[{Weatherford {et~al.}(2020)Weatherford, Chatterjee, Kremer, \&
  Rasio}]{Weatherford2020}
Weatherford, N.~C., Chatterjee, S., Kremer, K., \& Rasio, F.~A. 2020, The
  Astrophysical Journal, 898, 162, \dodoi{10.3847/1538-4357/ab9f98}

\bibitem[{Wen(2003)}]{Wen2003}
Wen, L. 2003, The Astrophysical Journal, 598, 419, \dodoi{10.1086/378794}

\bibitem[{Wong {et~al.}(2021)Wong, Franciolini, Luca, Baibhav, Berti, Pani,
  Riotto, Ansermet, Geneva, Fisica, \& Universit}]{Wong2021}
Wong, K. W.~K., Franciolini, G., Luca, V.~D., {et~al.} 2021, Physical Review D,
  103, 23026, \dodoi{10.1103/PhysRevD.103.023026}

\bibitem[{Zevin {et~al.}(2019)Zevin, Samsing, Rodriguez, Haster, \&
  Ramirez-Ruiz}]{Zevin2019a}
Zevin, M., Samsing, J., Rodriguez, C., Haster, C.-J., \& Ramirez-Ruiz, E. 2019,
  The Astrophysical Journal, 871, 91, \dodoi{10.3847/1538-4357/aaf6ec}

\bibitem[{Zevin {et~al.}(2021)Zevin, Bavera, Berry, Kalogera, Fragos, Marchant,
  Rodriguez, Antonini, Holz, \& Pankow}]{Zevin2021}
Zevin, M., Bavera, S.~S., Berry, C.~P., {et~al.} 2021, The Astrophysical
  Journal, 910, 152, \dodoi{10.3847/1538-4357/abe40e}

\end{thebibliography}
\bibliographystyle{aasjournal}

\end{document}